\documentclass{aastex631}

\usepackage{multirow} 
\usepackage{xcolor}
\usepackage{amsmath}
\usepackage[figuresleft]{rotating}
\hypersetup{citecolor=violet,filecolor=cyan,urlcolor=blue}
\def\msun{\mathrm{M_{\odot}}}

\begin{document}

\title{Electromagnetic Follow-up to Gravitational Wave Events with the UltraViolet EXplorer (UVEX)}

\author[0000-0002-9225-7756]{Alexander W. Criswell}
\affiliation{Minnesota Institute for Astrophysics, University of Minnesota, Minneapolis, MN 55455, USA}
\affiliation{School of Physics and Astronomy, University of Minnesota, Minneapolis, MN 55455, USA}
\affiliation{Department of Physics and Astronomy, Vanderbilt University, Nashville, TN 37240, USA}
\affiliation{Department of Life and Physical Sciences, Fisk University, Nashville, TN  37208, USA}

\author[0009-0000-9360-4759]{Sydney C. Leggio}
\affiliation{Minnesota Institute for Astrophysics, University of Minnesota, Minneapolis, MN 55455, USA}
\affiliation{School of Physics and Astronomy, University of Minnesota, Minneapolis, MN 55455, USA}

\author[0000-0002-8262-2924]{Michael W. Coughlin}
\affiliation{Minnesota Institute for Astrophysics, University of Minnesota, Minneapolis, MN 55455, USA}
\affiliation{School of Physics and Astronomy, University of Minnesota, Minneapolis, MN 55455, USA}

\author[0000-0001-9898-5597]{Leo P. Singer}
\affiliation{Astroparticle Physics Laboratory, NASA Goddard Space Flight Center, Mail Code 661, Greenbelt, MD 20771, USA}

\author[0000-0002-9108-5059]{R. Weizmann Kiendrebeogo}
\affiliation{Laboratoire de Physique et de Chimie de l’Environnement, Université Joseph KI-ZERBO, Ouagadougou, Burkina Faso}
\affiliation{Artemis, Observatoire de la Côte d’Azur, Université Côte d’Azur, Boulevard de l'Observatoire, F-06304 Nice, France}
\affiliation{School of Physics and Astronomy, University of Minnesota, Minneapolis, MN 55455, USA}

\author[0000-0002-8977-1498]{Igor Andreoni}
\affiliation{Department of Physics and Astronomy, University of North Carolina at Chapel Hill, Chapel Hill, NC 27599-3255, USA}

\author[0009-0008-9546-2035]{Andrew Toivonen}
\affiliation{Minnesota Institute for Astrophysics, University of Minnesota, Minneapolis, MN 55455, USA}
\affiliation{School of Physics and Astronomy, University of Minnesota, Minneapolis, MN 55455, USA}

\author[0000-0001-5857-5622]{Hannah P. Earnshaw}
\affiliation{Division of Physics, Mathematics, and Astronomy, California Institute of Technology, Pasadena, CA 91125, USA}

\author[0000-0003-3703-5154]{Suvi Gezari}
\affiliation{Department of Physics and Astronomy, Johns Hopkins University, 3400 N. Charles Street, Baltimore, MD 21218, USA}
\affiliation{Space Telescope Science Institute, 3700 San Martin Drive, Baltimore, MD 21218, USA}

\author[0000-0002-1984-2932]{Brian W. Grefenstette}
\affiliation{Division of Physics, Mathematics, and Astronomy, California Institute of Technology, Pasadena, CA 91125, USA}

\author[0000-0003-2992-8024]{Fiona A. Harrison}
\affiliation{Division of Physics, Mathematics, and Astronomy, California Institute of Technology, Pasadena, CA 91125, USA}

\author[0000-0002-5619-4938]{Mansi M. Kasliwal}
\affiliation{Division of Physics, Mathematics, and Astronomy, California Institute of Technology, Pasadena, CA 91125, USA}



\begin{abstract}

The Ultraviolet Explorer (UVEX) is expected to fly in 2030 and will have the opportunity --- and the rapid near/far ultraviolet (UV) capabilities --- to glean unprecedented insight into the bright UV emission present in kilonovae like that of AT 170817gfo, the electromagnetic counterpart to binary neutron star merger GW170817. To do so, it will need to perform prompt target-of-opportunity observations following detection of binary neutron star mergers by the LIGO-Virgo-KAGRA gravitational observatories. We present initial simulations to develop UVEX target-of-opportunity strategies for such events and provide the community with detailed initial estimates of the prospects for and characteristics of UVEX target-of-opportunity observations following gravitational-wave events, considering fiducial scenarios for the fifth and sixth LIGO-Virgo-KAGRA observing runs. Additionally, in light of the relatively few binary neutron star mergers observed since GW170817, we consider variant target-of-opportunity strategies for UVEX to maximize scientific gain in the case of a lowered binary neutron star merger rate.
\end{abstract}



\section{Introduction}
\subsection{UVEX: The UltraViolet EXplorer}\label{sec:uvex}
As of February 13th, 2024, the UltraViolet EXplorer (UVEX; PI Fiona Harrison) has been selected by NASA as its next Astrophysics Medium-Class Explorer mission with a launch date of 2030 \citep{fisher_new_2024}. UVEX will be a space telescope with wide-field, co-aligned imaging in the near- and far-ultraviolet (NUV/FUV) band with a 12.25 deg$^2$ field of view. It will additionally possess moderate-resolution ($R\geq1000$) spectroscopic capabilities across a broad bandpass covering much of the FUV and NUV \citep{kulkarni_science_2023}. In addition to the rapid target-of-opportunity (ToO) capabilities considered in this work, UVEX will serve as a joint FUV/NUV deep survey instrument, providing a crucial multi-band UV counterpart to the optical and infrared surveys of the coming era \citep[Rubin, Euclid, and Roman;][]{ivezic_lsst:_2019, laureijs_euclid_2011,akeson_the_2019}. These capabilities allow UVEX to cover a missing sector of the observational landscape of the 2030s: a deep, multi-band UV mission with ample sky coverage and rapid ToO response times. The Hubble Space Telescope, while of course highly sensitive, possesses a narrow field of view (0.002 deg$^2$) and slow ($>2$ weeks) ToO response times; the Neil Gehrels Swift Observatory \citep[\textit{Swift};][]{gehrels_the_2004} only covers the NUV; and the Ultraviolet Transient Astronomical Satellite \citep[ULTRASAT; ][]{sagiv_science_2014,shvartzvald_ultrasat:_2024}, while effective for rapid ToOs and imaging in the NUV with a planned ToO response time of $<15$ min from ToO trigger and a FOV of 204 deg$^2$, does not provide the depth, spectral capabilities, or FUV coverage of UVEX. For detailed comparison of the relevant capabilities of current and planned UV observatories to UVEX and typical requirements for follow-up to gravitational wave (GW) events, see Table~\ref{tab:uvex_in_context}. For further details as to the planned capabilities and broad science portfolio of UVEX, including its planned surveys and time-domain studies beyond those considered here, refer to \citet{kulkarni_science_2023}.

\begin{table}
    \centering
    \begin{tabular}{cccccc}
         \hline\hline
         & Swift & Hubble & ULTRASAT & UVEX & O6 BNS/KNe\\
         \hline
         UV Band(s) & NUV & NUV & NUV & NUV+FUV &NUV+FUV\\
         FOV & 0.08 deg$^2$ & 0.002 deg$^2$  & 204 deg$^2$ & 12.25 deg$^2$ & 10s -- 100s of deg$^2$ \\
         Depth (AB mag)& 24 & 28.75 & 22.5 & 24.7 (NUV)/24.6 (FUV) & 19.6--25.0 (NUV at 300 Mpc)\\
         ToO Response Time & $<150$ s & $>2$ weeks & $<15$ min & $3$ hrs (average) & $\sim$6 hrs\\
         \hline
    \end{tabular}
    \caption{UVEX capabilities compared to other current and upcoming UV observatories and typical expected values for LVK-detected BNS mergers and their KN counterparts in O6. The ``FOV'' values given for O6 BNS/KNe are the expected sky localizations for well-localized BNS in O6 \citep{petrov_data-driven_2022}. The range of KN apparent magnitudes corresponds to the minimum/maximum peak NUV luminosity for KN emission models considered in \citet{kulkarni_science_2023}, observed at a distance of 300 Mpc (the equivalent quote for the fainter FUV emission is 19.5--27.2 AB mags). The ToO response time listed for BNS mergers is the expected timescale after which it will likely no longer be possible distinguish between emission models for the KN UV emission; see e.g. Figure 14 of \citet{kulkarni_science_2023}. The depths quoted for ULTRASAT and UVEX are limiting magnitudes for a $5\sigma$ detection in a standard 900 s dwell. The depth quoted for Hubble is the current limiting magnitude of WFC3/UVIS at 2500\r{A} as given by Figure 3 of the Hubble Space Telescope Primer for Cycle 33 \citep{peeples_hst_2025}. The depth quoted for \textit{Swift} is the $5\sigma$ limiting magnitude of the UVOT instrument for a 1000 s exposure with the white light filter \citep{gehrels_the_2004}. The ToO response time for \textit{Swift} is based on updated capabilities presented in \citep{tohuvavohu_swiftly_2024}. Remaining quotes drawn from  \citet{gehrels_the_2004}, \citet{peeples_hst_2025}, \citet{shvartzvald_ultrasat:_2024}, \citet{kulkarni_science_2023}, and \href{https://www.uvex.caltech.edu/page/for-astronomers}{the UVEX Mission website}.}
    \label{tab:uvex_in_context}
\end{table}

\subsection{Multimessenger Astronomy}\label{sec:mma}
A central goal of UVEX is to leverage the capabilities described above to perform rapid electromagnetic (EM) follow-up to gravitational wave (GW) events, so as to capture the crucial early-time UV emission of kilonovae (KNe).\footnote{Singular: kilonova (KN).} KNe are the electromagnetically-bright transient counterparts to binary neutron star (BNS) and neutron star - black hole (NSBH) mergers, formed when matter comprising some fraction of the component neutron star(s) is ejected during the late inspiral and merger \citep{li_transient_1998}. The optical/near-infrared (nIR) KN emission is driven by radioactive decay of heavy elements formed in the ejecta through rapid neutron capture ($r$-process) nucleosynthesis \citep{metzger_electromagnetic_2010a}; KNe also feature bright UV emission at early times \citep{evans_swift_2017}, although the underlying factors driving this emission have yet to be determined \citep[see e.g. discussion in ][]{kulkarni_science_2023}.\footnote{``Kilonova" has historically referred in a technical sense to the optical/nIR emission of such a counterpart; following observation with \textit{Swift} of early-time UV excess in KN AT2017gfo \citep{evans_swift_2017}, the term is at times now taken to be inclusive of UV emission. As this work primarily considers the UV emission that will be observed by UVEX, we adopt the UV/optical/nIR convention for simplicity's sake.} It is worth noting that BNS/NSBH mergers can also produce short gamma-ray bursts (sGRBs) and long-lasting radio emission, the former likely powered by accretion onto the merger remnant, and the latter believed to arise from a jet structure; refer to \citet{metzger_kilonovae_2019} for a review of KN theory and observations across the EM spectrum. 

One of the most powerful and scientifically fruitful means by which one can achieve KN observations is through prompt follow-up with EM observatories of BNS or NSBH mergers detected through GWs. Indeed, the BNS merger GW170817 \citep{abbott_gw170817:_2017} and its EM counterpart, KN AT 2017gfo \citep{abbott_multi-messenger_2017,soares-santos_the_2017,cowperthwaite_the_2017,nicholl_the_2017,chornock_the_2017,margutti_the_2017,alexander_the_2017}, led to a profound and far-reaching scientific yield. This first --- and, so far, only --- multimessenger event of its kind resulted in insights across many aspects of astrophysics, including cosmology \citep[e.g.,][]{abbott_a_2017,guidorzi_improved_2017,hjorth_the_2017,hotokezaka_a_2019,dietrich_multimessenger_2020a,wang_multimessenger_2021,bulla_multi-messenger_2022,palmese_standard_2024}, the nature of the remnant object \citep[e.g.,][]{smartt_a_2017,abbott_estimating_2017, abbott_gravitational_2017, yu_a_2018, gill_when_2019a,  margalit_the_2019, murguia-berthier_the_2021}, constraints on the dense nuclear equation of state \citep[EoS;][]{coughlin_multi-messenger_2019,raaijmakers_constraining_2020,miller_the_2021,dietrich_multimessenger_2020,essick_nonparametric_2020,capano_stringent_2020,raaijmakers_constraints_2021,al-mamun_combining_2021,biswas_impact_2021,pang_nuclear-physics_2021,breschi_at2017gfo:_2021,legred_impact_2021,nicholl_tight_2021,huth_constraining_2022}, the origin of heavy elements through $r$-process enrichment \citep[e.g.,][]{chornock_the_2017,coulter_swope_2017,cowperthwaite_the_2017,kasen_origin_2017,kasliwal_illuminating_2017,kilpatrick_electromagnetic_2017,pian_spectroscopic_2017,rosswog_detectability_2017,smartt_a_2017,rosswog_the_2018,watson_identification_2019,kasliwal_spitzer_2022}, and beyond; see \citet{nakar_the_2020} and \citet{margutti_first_2021} for a review. In addition to the rich astrophysics of the EM counterpart itself, EM information regarding the distance and redshift of the host galaxy breaks important degeneracies for the analysis of the GW signal, further allowing for improved characterization of the latter \citep[e.g.,][]{abbott_gw170817:_2017,guidorzi_improved_2017,gianfagna_joint_2023}. Further joint EM-GW observations of neutron-star-containing mergers and their KNe could provide unparalleled, independent constraints on the Hubble constant \citep[e.g.,][]{bulla_multi-messenger_2022, coughlin_measuring_2020, coughlin_standardizing_2020}, solidify our understanding of KN dynamics and the dense nuclear EoS \citep[e.g.,][]{nicholl_tight_2021,breschi_bayesian_2024}, and more.

With this rich scientific bounty in mind, the motivation for pursuing EM follow-up observations to GW events --- especially those that contain a NS and can therefore produce a KN --- is clear. Since GW170817, several additional neutron-star-containing mergers have been detected via GWs \citep{abbott_gw190425:_2020,abbott_observation_2021}. However, no additional KNe counterparts have been located. While the absence of an observed EM counterpart can still provide some information \citep[e.g.,][]{ahumada_searching_2024, kasliwal_kilonova_2020}, the prospect of repeating the scientific windfall of GW170817 --- and then surpassing it via observations of an entire population of KNe --- remains a tantalizing one.

\subsection{Challenges for Electromagnetic Follow-up}\label{sec:challenges}
Numerous thorough searches were performed for the BNS merger GW190425 \citep{abbott_gw190425:_2020} \citep[e.g.,][]{antier_the_2020, coughlin_growth_2019, gompertz_searching_2020, hosseinzadeh_follow-up_2019, saleem_on_2020, song_viewing_2019,paek_gravitational-wave_2024}, as well as the NSBH mergers GW200105 and GW200115 \citep{abbott_observation_2021} \citep[e.g.,][]{page_swift-xrt_2020,antier_grandma_2020,anand_optical_2021, dichiara_constraints_2021} and GW events in O4 \citep{ahumada_searching_2024}. Part of the challenge lies in the broad sky area that any such follow-up endeavor must search. GW detectors do not have high angular resolution to individual sources, and while a network of 3+ detectors allows for significant improvements in this respect, GW localization regions --- that is, the area of the sky covered by, e.g., the 90\% credible level (C.L.) error region for the GW-inferred localization --- are expected to remain large (many 10s or 100s of square degrees) for the immediate future \citep{petrov_data-driven_2022,kiendrebeogo_updated_2023a}. Tiling a GW localization region spanning such a broad area of the sky is a difficult task for most modern observatories. Adding this difficulty, light curve predictions from KN models for these events are still quite uncertain. This is largely due to the fact that we do not yet understand the dominant mechanism powering the KN at early times. Possibilities include a predominantly nucleosynthesis-powered KN \citep{li_transient_1998, metzger_electromagnetic_2010a,hotokezaka_radioactive_2020,banerjee_simulations_2020,klion_the_2021,banerjee_diversity_2024}, a nucleosynthesis-powered KN with additional contributions from $\beta$-decay of free neutrons \citep{metzger_neutron-powered_2015,gottlieb_electromagnetic_2020}, and a shock-powered KN \citep{piro_evidence_2018}. Depending on which of these prescriptions is used, the corresponding light-curve predictions can vary by more than 4 mag over the first few hours after merger \citep[see, e.g., Fig. 43 of][]{kulkarni_science_2023}. When this uncertainty is joined by the intrinsic GW uncertainties on the inclination $\iota$ and luminosity distance $D_L$ of the BNS/NSBH merger in question, the depth at which a given telescope needs to observe each field to be able to actually detect a potential KN becomes difficult to estimate. One could, of course, take a highly conservative approach and observe each field at extreme depth, but doing so requires large per-field exposure times. 

Time is, however, of the essence. Not only are KNe fast-evolving and fade quickly \citep[as seen for AT 2017gfo in, e.g.,][]{cowperthwaite_the_2017}, but many open questions as to the physical nature of KNe and the origin of heavy elements can \textit{only} be answered by observing the KN during the rapid evolution of its light curve at early times. As noted above, different underlying mechanisms can cause drastically different KN luminosities in the first few hours. After 5--10 hrs, however, the magnitude difference between (e.g.) shock-powered vs. nucleosynthesis-powered KN models shrinks dramatically, likely to within observational uncertainty of one another (e.g., Fig. 43 of \citet{kulkarni_science_2023} and surrounding discussion). The exact nature of the early-time evolution of the KN lightcurve is ambiguous, as the first observations of KN AT 2017gfo were performed 15 hrs after the merger \citep{abbott_multi-messenger_2017,soares-santos_the_2017,cowperthwaite_the_2017,nicholl_the_2017,chornock_the_2017,margutti_the_2017,alexander_the_2017}. Taken together, these considerations both complicate and lend additional impetus to EM-GW follow-up observations, requiring them to be deep, broad, \textit{and} rapid. This quandary is the driving motivation for employing wide-field survey telescopes in the pursuit of EM follow-up to GW events. These observatories are designed with wide fields of view and the ability to quickly move from pointing to pointing for efficient survey operations. The primary challenge then becomes how one can achieve the necessary depth for successful followup of a given KN counterpart in the short time before it fades \textit{without} sacrificing coverage of the GW localization region. 
% We consider here an approach which limits the events on which one triggers ToO follow-up observations to only those events that are sufficiently nearby/bright that they will likely be detectable given some predetermined maximum acceptable exposure time. This work seeks to estimate the prospects for such observations with UVEX, and further develop potential ToO strategies for EM follow-up to GW events during UVEX's mission.

\subsection{UVEX Target-of-Opportunity Strategy}
The preliminary UVEX ToO strategy for EM-GW follow-up of BNS mergers, as presented in \citet{kulkarni_science_2023}, is designed to fulfill two scientific goals. The first, for UVEX to perform follow-up of a sample of BNS mergers detected in GWs; the second, to do so with sufficient alacrity as to capture the crucial early-time evolution of their KN counterparts. This latter consideration amounts to obtaining the first epoch of observations within 6 hours of merger; subject to an average ToO response time of $<3$ hrs, this leaves on average a maximum of 3 hours to tile the 90\% C.L. GW localization region. Following this initial epoch, UVEX would continue to repeat its tiling pattern over the localization area for a total of 24 hours to obtain a multi-epoch FUV + NUV light curve. UVEX will likely fly concurrent to O5 or O6, each of which is expected to observe a large sample BNS mergers; however the majority of these mergers are not anticipated to be ``golden" events like that of GW170817. Instead, detected BNS mergers are expected to span a broad range of inferred distances and sky localization areas \citep{petrov_data-driven_2022,kiendrebeogo_updated_2023a}. Pursuing follow-up to all detected BNS events will not be feasible for several reasons, including the fact that UVEX has numerous other science goals; only a fraction of UVEX's baseline mission is dedicated to EM-GW ToO observations. More importantly, it will not be possible for UVEX to fully tile large localization regions with sufficient depth to detect distant KNe \textit{and} do so quickly enough to capture their early-time evolution. As such, the task of defining an EM-GW ToO strategy for UVEX becomes that of selection: on which mergers do we trigger ToO observations so as to maximize the total scientific gain of our follow-up observations, given these limitations?

Towards this end, it is necessary to define a set of criteria for ToO selection, based on the information available in a realistic EM-GW follow-up setting. From low-latency LVK analyses of a given GW signal \citep[e.g.,][]{chaudhary_low-latency_2024}, we will have access to estimates of the luminosity distance and sky localization for a given BNS candidate; this information, in conjunction with the imaging capabilities of the UVEX instrument itself and a (conservative) estimate of the KN luminosity, allow us to evaluate candidate ToO triggers along the following axes: 1) the total sky area covered the 90\% C.L. GW localization region, 2) the fraction of the 90\% C.L. GW localization region that UVEX is able to tile within 3 hours, and 3) that each UVEX pointing in the localization region can reach sufficient per-tile exposure time so as to reach a fiducial KN source FUV magnitude depth, given the UV background and GW-inferred distance estimate for that pointing and current modelling uncertainties for the UV KN luminosity.

\subsection{UVEX EM-GW Selection Criteria}
The UVEX EM-GW ToO selection criteria presented in \citet{kulkarni_science_2023} were defined such that UVEX could achieve ToO observations for follow-up of $\geq 20$ BNS mergers, assuming the GWTC-2 rate of $320^{+490}_{-240}\ \mathrm{Gpc}^{-3}\,\mathrm{yr}^{-1}$ \citep{abbott_gwtc-2:_2021}, the \textit{BNS-broad} observing scenarios (see \S\ref{sec:obs_scenarios}), and a five-detector network consisting of LIGO Hanford, LIGO Livingston, Virgo, KAGRA, and LIGO India at design (A+) sensitivity. These criteria, as presented in the UVEX proposal and concept study report (CSR), are as follows: 1) the 90\% C.L. GW localization region must be $\leq100$ deg$^2$, 2) UVEX must be able to cover $\geq99\%$ of the GW localization probability within 3 hours, and 3) that each UVEX pointing must reach a source FUV magnitude depth of -12.1 AB mag, given the UV background and GW-inferred mean distance estimate for that pointing. The choice of FUV source magnitude is a conservative one such that UVEX will be able to detect and characterize the KN counterpart regardless of current modelling uncertainties as to the KN peak magnitude. See Appendix E2 of \citet{kulkarni_science_2023} for further details as to the specific light curve models considered. The final set of UVEX ToO selection criteria are summarized in Table~\ref{tab:uvex_criteria}. These criteria were estimated in the UVEX proposal to yield 20 (35) ToO triggers in 1.5 years of O5 (O6), given the assumptions outlined above. Further details as to the simulations and results leading to these criteria as performed for UVEX proposal
can be found in \citet{kulkarni_science_2023}. In this work, we provide pertinent details of these estimates' underlying LVK observing scenarios (\S\ref{sec:obs_scenarios}), detail the analysis framework and codebase used to determine the UVEX EM-GW selection strategy (\S\ref{sec:uvex_followup}), present a method for estimating the strategy-dependent success rate of such observations (\S\ref{sec:success_estimates}), provide updated estimates of the associated prospects for UVEX EM-GW ToOs based on the latest LVK observing scenarios (\S\ref{sec:prospects}), and lend additional consideration to strategies which can maximise the scientific success of UVEX EM-GW follow-up in the case of a lower astrophysical BNS rate (\S\ref{sec:new_strats}).

\begin{table}
    \centering
    \begin{tabular}{ll}
    \hline\hline
    Selection Criteria  & Value \\ 
    \hline 
    Area of GW 90\% C.L. Localization & $\leq100$ deg$^2$\\
    Total Time to Tile & $\leq 3$ hr (10800 s)\\
    Per-tile Exposure Time & $\leq 3$ hr (10800 s)\\
    Tiling Coverage & $\geq99$\%\\
    FUV Depth (Source Magnitude) & -12.1 AB mag\\
    \hline
    \end{tabular}
    \caption{UVEX ToO selection criteria as established in the UVEX proposal and CSR. The tiling coverage is the minimum percentage of the GW 90\% C.L. localization that can be tiled in 3 hrs. The FUV depth is the AB source magnitude achievable by UVEX in each pointing, assuming the mean GW distance estimate. Note that these criteria can be relaxed to ensure UVEX's ability to achieve impactful KN science under adverse conditions.}
    \label{tab:uvex_criteria}
\end{table}
\newpage
\section{Observing Scenarios}\label{sec:obs_scenarios}
Realistic estimates of the prospects for EM-GW follow-up with UVEX require well-founded predictions for the GW landscape of the coming years. This work uses the GW observing scenarios presented in \citet{kiendrebeogo_updated_2023a}. \citet{kiendrebeogo_updated_2023a} consider two models for the underlying astrophysical BNS/NSBH/BBH mass/spin distributions: the \textit{LRR} model, which serves as a comparison point to previous studies \citep[see e.g.,][]{abbott_prospects_2020a,petrov_data-driven_2022}; and the \textit{PDB/GWTC-3} model, which instead draws on the population inference model of \citet{farah_bridging_2022} as fit to the GWTC-3 catalogue \citep{abbott_gwtc-3:_2023}. In contrast to the \textit{LRR} model, which parameterizes and draws from independent distributions for each source classification (i.e., BNS/NSBH/BBH), the \textit{PDB/GWTC-3} model treats the entire compact binary mass spectrum holistically with a single function consisting of a broken power law with a dip in the lower mass gap at $2.72-6.13\msun$.\footnote{Hence, \textit{PDB}: Power law + Dip + Break.} Further details can be found in \citet{farah_bridging_2022,abbott_gwtc-3:_2023,kiendrebeogo_updated_2023a}. Additionally, some of the original considerations for UVEX used a variation of the \citet{petrov_data-driven_2022} scenarios. These simulations, produced for the UVEX proposal prior to the advent of the updated \citet{kiendrebeogo_updated_2023a} observing scenarios, follow the methods of \citet{petrov_data-driven_2022} save that they employ a uniform neutron star mass distribution\footnote{The \citet{petrov_data-driven_2022} and \textit{LRR} models assume normally distributed neutron star masses with a mean of $1.33\msun$ and standard deviation of $0.09\msun$.} on $[1,2]\,\msun$, aiming to reflect the broader neutron star mass distribution supported by the events of GWTC-3 \citep{abbott_gwtc-3:_2023} and as such are referred to hereafter as the \textit{BNS-broad} scenarios. The observing scenarios employed in this work use the \textit{PDB/GWTC-3} model of \citet{kiendrebeogo_updated_2023a}. All results presented in this work assume the GWTC-3 \citep{abbott_gwtc-3:_2023} astrophysical BNS merger rate of $210^{+240}_{-120}$ $\mathrm{Gpc}^{-3}\,\mathrm{yr}^{-1}$.

Each observing scenario is comprised of a large number of simulated GW waveforms with source parameter distributions as described above. These waveforms are then injected into simulated detector noise and recovered using a matched filter search. The simulated event rates for BNS/NSBH/BBHs used in the \citet{petrov_data-driven_2022} and \citet{kiendrebeogo_updated_2023a} scenarios are much larger than the rates for such events determined by LVK observations \citep[e.g.,][]{abbott_gwtc-3:_2023}, so as to reduce the impact of Monte Carlo error; the simulated rate can be straightforwardly scaled to an estimated rate through the procedure discussed in \S\ref{sec:uvexfollowup_postproc}. Recovered signals with a $SNR>10$ are then characterized with the rapid localization software {\tt BAYESTAR} \citep{singer_rapid_2016,singer_going_2016, singer_supplement:_2016} to determine posterior estimates of the event luminosity distance and GW localization on the sky. The detected event catalogues, corresponding GW localization skymaps, and posterior summary statistics produced through this process are the primary data products used for the work presented here. Further details can be found in \citet{petrov_data-driven_2022} and \citet{kiendrebeogo_updated_2023a}.

\section{{\tt uvex-followup}}\label{sec:uvex_followup}
The {\tt uvex-followup} code was developed for the UVEX proposal and science paper \citep{kulkarni_science_2023}, in order to rapidly process the large observing scenario datasets of \citet{petrov_data-driven_2022} and \citet{kiendrebeogo_updated_2023a} and compute UVEX follow-up prospects as realized for a set of user-defined ToO selection criteria. {\tt uvex-followup} is open source and can be found at \href{https://github.com/criswellalexander/uvex-followup}{https://github.com/criswellalexander/uvex-followup}. The code as presented here is an updated version of the infrastructure used for \citet{kulkarni_science_2023}, and operates as follows:

\subsection{Preprocessing}\label{sec:uvexfollowup_preproc}
First, the observing scenario dataset is filtered to only those events with a total 90\% C.L. sky localization area less than or equal to a user-specified maximum area (e.g. in Table~\ref{tab:uvexfollowup_config}, 100 sq. deg.). This allows us to remove events with broad sky localizations that are unlikely to prove to be suitable follow-up candidates. The down-selected dataset is then split into a user-specified number of batches to allow for parallel computation of the following steps.

For each remaining event, we calculate the necessary UVEX exposure time in each sky pixel within the localization region to reach a SNR of 5 for a user-specified fiducial FUV source magnitude (e.g. in Table~\ref{tab:uvexfollowup_config}, -12.1 AB mags.). The exposure time can be computed for an apparent magnitude corresponding to either the mean or 90\% confidence upper limit of the GW luminosity distance estimate. The results presented in this work use the mean luminosity distance estimate. Exposure times are calculated with the UVEX exposure time calculator (ETC) within {\tt uvex-mission}.\footnote{{\tt uvex-mission} is under development by Brian Grefenstette and the UVEX science team. It is currently housed within a private repository on GitHub, but is planned to be made available to the public before UVEX launches. The work presented here uses {\tt uvex-mission} version v0.11-158 (current as of October 16th, 2024).} The UVEX ETC accounts for relevant aspects of the UVEX instrument (field of view, FUV/NUV bandpasses, CMOS detector noise contributions, etc.), as well as the UV background contribution at the sky location in question. It important to note that the results presented in this work are conditioned on our current understanding of UVEX performance, and may change as knowledge of the telescope performance improves. Due to constraints within our current observation scheduling algorithm (which is not yet capable of scheduling UVEX observations with adaptive exposure times; see the following section), we take the per-tile exposure time to be the maximum per-pixel exposure time across the entire localization region. If this final exposure time exceeds a user-defined maximum exposure time, that event is discarded. If the calculated exposure time is below some user-defined minimum, it is set to that minimum value. All remaining events are then re-batched and prepared as input for the observation scheduling software.

\subsection{Scheduling}
A simulated observing plan for each event is procured through {\tt dorado-scheduling}. {\tt dorado-scheduling} was developed for Dorado, a concept for a similar mission in smallsat format that preceded UVEX; the code is open source and available at \href{https://github.com/nasa/dorado-scheduling}{https://github.com/nasa/dorado-scheduling}. It leverages Mixed-Integer Linear Programming to find the globally optimal UVEX\footnote{{\tt dorado-scheduling} supports several missions, including UVEX and the titular Dorado.} observing plan for tiling a GW localization region, given a per-tile exposure time, subject to the constraints of the telescope motion and total allowable observing time. The scheduling code used for the simulations that supported the Dorado concept study did adaptively optimize the exposure time for each individual field based on SNR modeling across the entire field of view. However, this capability is not yet supported for UVEX. We are working to implement the adaptive exposure time scheduling algorithm for UVEX and other missions; these efforts will be described in an upcoming manuscript \citep{singer_optimal_2025}. For further details on {\tt dorado-scheduling} and its underlying algorithm, see the {\tt dorado-scheduling} documentation.\footnote{\href{https://dorado-scheduling.readthedocs.io/}{https://dorado-scheduling.readthedocs.io/}}

\subsection{Postprocessing}\label{sec:uvexfollowup_postproc}
With observing plans in hand, {\tt uvex-followup} then computes the total tiling coverage achieved for each event. Any event for which the observing plan does not achieve a user-specified threshold coverage of the GW localization (e.g. in Table~\ref{tab:uvexfollowup_config}, 99\%) is discarded. This can occur for several reasons; the most common culprits are 1) events at large distances with broad sky localizations, and 2) events for which a significant portion of the localization region is outside of the UVEX field of regard (FoR) due to Sun/Moon/Earth exclusion. 

Any events that remain in the dataset after this procedure, having fulfilled all the selection criteria, are considered selected ToO triggers. The proportion of selected events to the total number of events in the observing scenario simulation is then normalized to a user-specified astrophysical rate (e.g. in Table~\ref{tab:uvexfollowup_config}, $210^{+240}_{-120}\ \mathrm{Gpc}^{-3}\,\mathrm{yr}^{-1}$). For a given user-specified observing run overlap duration (e.g. in Table~\ref{tab:uvexfollowup_config}, 1.5 yr), estimates are then given for the expected number of UVEX ToO triggers. Following \citet{petrov_data-driven_2022} and \citet{kiendrebeogo_updated_2023a}, the uncertainty of these estimates accounts for both the log-normal error of the astrophysical rate measurement and uncertainty in the observed number of events due to Poisson statistics, but does not include the Monte Carlo error of the observing scenario simulations themselves. Statistics for the distribution of exposure times and that of the number of UVEX pointings are also computed. Finally, summary plots are produced. All outputs are saved, along with a file containing the selected events and their characteristics, to a directory of the user's choice.

\section{Estimated Success Rates}\label{sec:success_estimates}
For a given choice of ToO selection strategy and observing scenario, we estimate the success rate and corresponding counterpart detection rates as follows. A triggered ToO observation is considered successful if it 1) observes the pixel that contains the true (simulated) source location $\mathbf{n}_{\mathrm{true}}$, and 2) the observation reaches sufficient depth as to detect the counterpart in the first epoch of observations. As a primary goal of UVEX EM-GW follow-up is to capture the early-time behavior of the KN light curve, we do not account for the possibility of achieving detection by stacking observations across multiple subsequent epochs. The former consideration is purely binary, i.e.,
\begin{equation}\label{eq:p_success_loc}
    p_{\mathrm{loc}}(\mathcal{S}) = \begin{cases}
        1 & \mathbf{n}_{\mathrm{true}} \in \mathrm{observed\ fields}\\
        0 & \mathbf{n}_{\mathrm{true}} \notin \mathrm{observed\ fields}.\\
    \end{cases}
\end{equation}
The latter, however, is somewhat more involved. At a base level, condition (2) requires that the limiting magnitude of our observation exceeds the true apparent magnitude of the counterpart (i.e., $m_{\mathrm{lim, obs}} \geq m_{\mathrm{KN}}$). The limiting magnitude of our observations, as described in \S\ref{sec:uvex_followup}, is such that UVEX can observe a counterpart with assumed absolute magnitude $\hat{M}_{\mathrm{AB}}$ at the point-estimate distance $\hat{D}_L$ given by the GW distance posterior mean, such that
\begin{equation}
    m_{\mathrm{lim, obs}} = \hat{M}_{\mathrm{AB}} + 5\log_{10}(\hat{D}_L) - 5.
\end{equation}
Additional complicating factors (extinction, UV background, etc.) are accounted for within the UVEX ETC. The true apparent magnitude of the counterpart, $m_{\mathrm{KN}}$, is given in terms of the true intrinsic KN absolute magnitude and the actual (simulated) source distance:
\begin{equation}
    m_{\mathrm{KN}} = M_{\mathrm{AB,true}} + 5\log_{10}(D_{L,\mathrm{true}}) - 5.
\end{equation}
While $D_{L,\mathrm{true}}$ is known for events contained within the observing scenarios simulations, the intrinsic KN UV luminosity is as-of-yet relatively unconstrained. We assume that our ignorance as to the intrinsic KN peak magnitude can be captured with a uniform prior in FUV absolute magnitude, such that
\begin{equation}\label{eq:abs_kn_mag_prior}
    \pi(M_{\mathrm{AB,true}}) \sim \mathcal{U}(-17.9,-10.2)\, \mathrm{AB\ mags}.
\end{equation}
The prior bounds are chosen so as to encapsulate the full extent of KN FUV emission models considered in \citet{kulkarni_science_2023}; see Fig. 43 and surrounding discussion therein. We emphasize that this distribution is solely used to estimate the probability of success for the ToO strategies considered in this work as applied to a given observing scenario. The strategies in this work assume a fixed point estimate of the intrinsic KN luminosity with which to compute necessary exposure times and observing schedules. We do not at this juncture have a good understanding of the true distribution of UV KN luminosities, in part because of uncertainty as to the underlying KN UV emission mechanism(s). As such, the choice of a uniform distribution approximates the current broad uncertainty in the absolute magnitudes of the (simulated) KNe in our simulations. In contrast, a probabilistic approach to the ToO selection strategy itself would benefit from assuming a prior that less heavily weights brighter events such as a Gaussian or half-Gaussian distribution. Such an approach is an important aspect of further work on EM-GW follow-up with UVEX \citep[see discussion in \S\ref{sec:discussion_conclusion};][]{singer_optimal_2025}. 

We can now write an expression for the probability of success in terms of depth, $p_{\mathrm{depth}}(\mathcal{S})$:
\begin{equation}\label{eq:p_success_depth}
    p_{\mathrm{depth}}(\mathcal{S}) = \int_{-17.9}^{-10.2} \mathcal{F}(M_{\mathrm{AB,true}};\,\hat{M}_{\mathrm{AB}},\hat{D}_L,D_{L,\mathrm{true}})\pi(M_{\mathrm{AB,true}})\,dM_{\mathrm{AB,true}},
\end{equation}
where
\begin{equation}
    \mathcal{F}(M_{\mathrm{AB,true}};\,\hat{M}_{\mathrm{AB}},\hat{D}_L,D_{L,\mathrm{true}}) = \begin{cases}
        1 & \hat{M}_{\mathrm{AB}} + 5\log_{10}(\hat{D}_L) - 5 \geq M_{\mathrm{AB,true}} + 5\log_{10}(D_{L,\mathrm{true}}) - 5\\
        0 & \mathrm{otherwise.}
    \end{cases}
\end{equation}
From here, it is straightforward to compute the total estimated probability of success for a given ToO observation as a  combination of Eqs.~\eqref{eq:p_success_loc} and~\eqref{eq:p_success_depth}:
\begin{equation}
    p_{\mathrm{tot}}(\mathcal{S}) = p_{\mathrm{loc}}(\mathcal{S})\times p_{\mathrm{depth}}(\mathcal{S}).
\end{equation}
We apply this formalism to all selected ToO targets within each considered strategy/simulation. The overall success rate for a given scenario is computed by taking the mean of the values of $p_{\mathrm{tot}}(\mathcal{S})$ for all selected events contained within. We additionally estimate the rate of KN counterpart detection (as opposed to ToO triggers) for each strategy/scenario by including the overall success rate of a given strategy as an additional efficiency factor of the Poisson process.

\section{Updated Prospects for UVEX EM-GW Follow-up}\label{sec:prospects}
The prospects presented here for UVEX EM-GW follow-up build on the procedure developed for the UVEX proposal and selection process, bringing it in line with the current post-selection UVEX timeline and mission details --- although these remain subject to change --- as well as our improved understanding of the BNS rate and mass distribution. As the precise details of the LVK observing run timeline and detector sensitivity in the coming years remain uncertain, we consider two fiducial observing scenarios. The ``O5" scenario considers 1.5 years of overlap with a LVK detector network consisting of the LIGO Hanford, LIGO Livingston, Virgo, and KAGRA at their respective projected O5 sensitivities. The ``O6" scenario considers the same configuration at the projected sensitivities for O6 and a fiducial 1.5 year overlap. These prospects are estimated by applying our UVEX EM-GW ToO selection criteria as developed for the UVEX proposal to the updated scenarios of \citet{kiendrebeogo_updated_2023a} with the GWTC-3 astrophysical BNS rate estimate of $210^{+240}_{-120}\ \mathrm{Gpc}^{-3}\,\mathrm{yr}^{-1}$ \citep{abbott_gwtc-3:_2023}. The {\tt uvex-followup} configuration used for both the O5 and O6 simulations is given in Table~\ref{tab:uvexfollowup_config}. These updated estimates retain the UVEX ToO selection strategy developed for the UVEX proposal, and can therefore be taken as a modernization of the quotes presented in \citet{kulkarni_science_2023}.\footnote{i.e., 20 ToO triggers for O5 and 35 for O6; the \citep{kulkarni_science_2023} quotes did not include the detailed treatment of uncertainty now present within {\tt uvex-followup} (as described in \S\ref{sec:uvexfollowup_postproc}).} We consider an expanded set of ToO selection strategies in \S\ref{sec:new_strats}.

For the  O5 (O6) simulation, with the fiducial observing strategy detailed in Table~\ref{tab:uvexfollowup_config} and an overlap of 1.5 years of concurrent observations, UVEX is estimated to trigger EM-GW ToO observations for $8.9^{+11.1}_{-5.0}$ ($11.3^{+14.0}_{-6.3}$) BNS mergers. The success rate of the fiducial observing strategy as applied to the O5 (O6) scenarios is 76.3\% (77.5\%), corresponding to an estimated $6.8^{+8.4}_{-3.8}$ ($8.8^{+10.9}_{-4.9}$) KN counterpart detections. The overall distribution of these ToO triggers with respect to their GW distance estimates and sky localizations for O5 and O6 can be found in top panels of Fig.~\ref{fig:selection_multi}. It is expected that the LVK BNS rate measurement will change following O4; to allow the estimates presented in this work to be robust to rate updates in the coming years, we provide a simple tool to convert these quotes to the corresponding quote with a BNS rate of the user's choice. This tool can be found at \href{https://github.com/criswellalexander/uvex-rate-updater}{https://github.com/criswellalexander/uvex-rate-updater}.

\begin{table}[hb]
    \centering
    \begin{tabular}{ll}
    \hline\hline
    Configuration Parameter  & Setting \\ 
    \hline 
    Maximum Localization Area & 100 deg$^2$\\
    Maximum Tiling Time & 3 hr (10800 s)\\
    Minimum Per-tile Exposure Time & 500 s\\
    Maximum Per-tile Exposure Time & 3 hr (10800 s)\\
    Minimum Coverage & $99$\%\\
    Fiducial FUV Source Magnitude & -12.1 AB mag\\
    GW Distance Estimate & mean\\
    BNS Rate &$210^{+240}_{-120}\ \mathrm{Gpc}^{-3}\,\mathrm{yr}^{-1}$\\
    Observing Run Overlap & 1.5 yr\\
    \hline
    \end{tabular}
    \caption{Configuration settings for {\tt uvex-followup} used for the fiducial O5 and O6 UVEX prospects simulations. Each parameter is described in \S\ref{sec:uvex_followup}. These settings are identical to those used in the UVEX proposal and science paper \citep{kulkarni_science_2023}, save for the BNS rate.}
    \label{tab:uvexfollowup_config}
\end{table}

\section{Expanded Strategies for UVEX Follow-up}\label{sec:new_strats}
In the event that the dearth of BNS mergers in O4 continues and is indicative of a relatively lower BNS rate than had previously been assumed, UVEX may need to alter its EM-GW ToO strategy accordingly. We therefore present several potential variant strategies for EM-GW ToO selection with UVEX, and discuss the advantages and disadvantages of each. We consider three potential adaptive axes, relaxing the stringency of our analysis with respect to the maximum considered GW localization area, the minimum acceptable percent coverage by UVEX of the localization region, and the assumed intrinsic KN luminosity. For each case considered, simulations were performed with {\tt uvex-followup} as applied to the observing scenarios of \citet{kiendrebeogo_updated_2023a}. The UVEX EM-GW ToO trigger count estimates, success rates, and estimated number of counterpart detections for each variant strategy as applied to the O5 and O6 observing scenarios are found in Table~\ref{tab:uvex_success_rates}.

\subsection{Maximum Area}
As discussed in \S\ref{sec:uvexfollowup_preproc}, {\tt uvex-followup} automatically rejects events with a GW 90\% localization area greater than a user-defined maximum area (default 100 deg$^2$). This cut is largely motivated from a computational perspective; it was considered unlikely for a GW localization of greater than 100 deg$^2$ to pass the other, more-computationally-intensive steps in the selection workflow. To ensure that this criterion was not needlessly restrictive, we considered increased maximum area cuts of 150 and 200 deg$^2$. Expanding the allowed localization areas in this way resulted in no deviation from the default values produced with a maximum area cut of 100 deg$^2$. We conclude that --- in isolation --- adjusting the maximum area cut beyond 100 deg$^2$ has minimal effect on the number triggered ToO observations, and that this computationally-expedient approach is robust to within the Monte Carlo error of the observing scenarios simulations. It is worth noting that allowing for a longer total epoch duration or adaptive exposure time could increase the accessible total localization areas; we leave such considerations to future work (see \S\ref{sec:discussion_conclusion}). Variation of this parameter was omitted in following simulations.\footnote{Save for the least conservative strategy, M-13.1+CT90 (see description in \S\ref{sec:strats_kn_mag}). For completeness, we consider this strategy with a maximum area cut of 200 deg$^2$. We again see no deviation from the number of events with an area cut of 100 deg$^2$.} 

\subsection{Coverage Threshold}\label{sec:strats_cov_thresh}
The percent coverage threshold defines the minimum allowed coverage by UVEX of the GW localization region. It is given as the total integrated probability tiled by the observing plan; e.g., 99\% coverage equates to a 99\% chance that the counterpart is present within the observed fields. Higher coverage thresholds yield a higher confidence that the UVEX ToO observing plan will include the KN counterpart; however, a stringent threshold may reject events with a high probability of success, but which are missing some small area of the localization region due to sun exclusion, time constraints, etc.. It is worth noting that the threshold defines a \textit{minimum} level of coverage, and therefore most selected events will in fact encapsulate a greater total probability than given by the threshold. The (quite conservative) fiducial coverage threshold employed for the UVEX proposal process was 99\%. We consider two variant strategies --- dubbed CT95 and CT90 --- in which this threshold is relaxed to 95\% and 90\%, respectively, and consider the impact on the overall ToO count and success rate. Doing so yields a small increase in ToO triggers with a negligible change in success rate: adopting the CT90 strategy leads to $3.2^{+3.9}_{-1.8}$ and $3.4^{+4.3}_{-2.0}$ additional triggers for the O5 and O6 scenarios, for a total of $12.1^{+15.0}_{-6.8}$ and $14.7^{+18.3}_{-8.3}$ triggers, respectively. The success rate of the CT90 strategy as applied to the O5 scenario is 0.3\% higher than that of the fiducial case and is only reduced by 0.2\% for the O6 scenario, yielding a total of $9.3^{+11.5}_{-5.2}$ ($11.4^{+14.1}_{-6.4}$) detected counterparts for O5 (O6). The relative ToO trigger counts and success rates for coverage thresholds of 99\%, 95\%, and 90\% can be found in Table~\ref{tab:uvex_success_rates}. An event selection plot for the CT90 strategy can be found in the bottom left panel of Fig.~\ref{fig:selection_multi}. In general, relaxing this requirement results better retention of nearby/well-localized events that fall marginally outside of the UVEX FoR; compare, e.g., the proportion of events lost to FoR exclusion in the O6 simulation with this strategy as opposed to with the fiducial strategy, as seen in Fig.~\ref{fig:selection_multi}.

% \begin{figure}
%     \centering
%     \includegraphics[width=0.9\linewidth]{uvex_event_selection_O6_cov90_defaults.pdf}
%     \caption{UVEX ToO selection for the O6 simulation with a 90\% threshold for minimum coverage by UVEX of the GW localization. Points correspond to individual simulated events from the \citet{kiendrebeogo_updated_2023a} O6 observing scenario, so the total number of points is not one-to-one with the expected number of observed BNS mergers (although it is proportional); scaling from the simulated event rate to an astrophysical BNS rate of $210^{+240}_{-120}\ \mathrm{Gpc}^{-3}\,\mathrm{yr}^{-1}$ \citep{abbott_gwtc-3:_2023} yields $14.7^{+18.3}_{-8.3}$ UVEX ToO triggers over 1.5 years of O6. Events are plotted by their mean GW distance estimate and 90\% C.L. GW localization areas. Points in grey are not selected due to the overall threshold of $\leq100$ deg$^2$ GW 90\% C.L. localization area. Events with $\leq100$ deg$^2$ localizations that are not selected due falling outside of the UVEX FoR or exceeding the maximum epoch time constraint of 3 hrs are marked as black and orange $\times$s, respectively. Selected events are color-coded by per-tile exposure time.}
%     \label{fig:uvex_cov90_O6}
% \end{figure}

\subsection{Assumed Intrinsic KN Magnitude}\label{sec:strats_kn_mag}
As discussed in \S\ref{sec:challenges}, the expected distribution of intrinsic KN magnitudes is essentially unconstrained. The fiducial approach, by assuming a extremely conservative KN absolute FUV magnitude of -12.1 AB mags, ensures that the ToO is successful regardless of what the underlying mechanism is for KN UV emission (provided the true counterpart is in the tiled region). This is of course advantageous, especially when one is in a position to pick and choose follow-up targets from a large population of BNS mergers. If scarcity is a factor, however, the scientific return of ToO observations will likely be increased by following up more events at a reduced depth. We therefore consider a brighter assumed KN absolute magnitude of -13.1 AB mag. This strategy is referred to hereafter as M-13.1. Relaxing our assumption in this respect still allows UVEX to probe most UV emission mechanisms of interest while significantly reducing the required per-tile exposure time. Moreover, any non-detections that would occur as a result of this choice could still prove informative \citep[as in, e.g.,][]{coughlin_implications_2020a}.

\begin{figure*}
    \centering
    % Top row
    \includegraphics[width=0.47\textwidth]{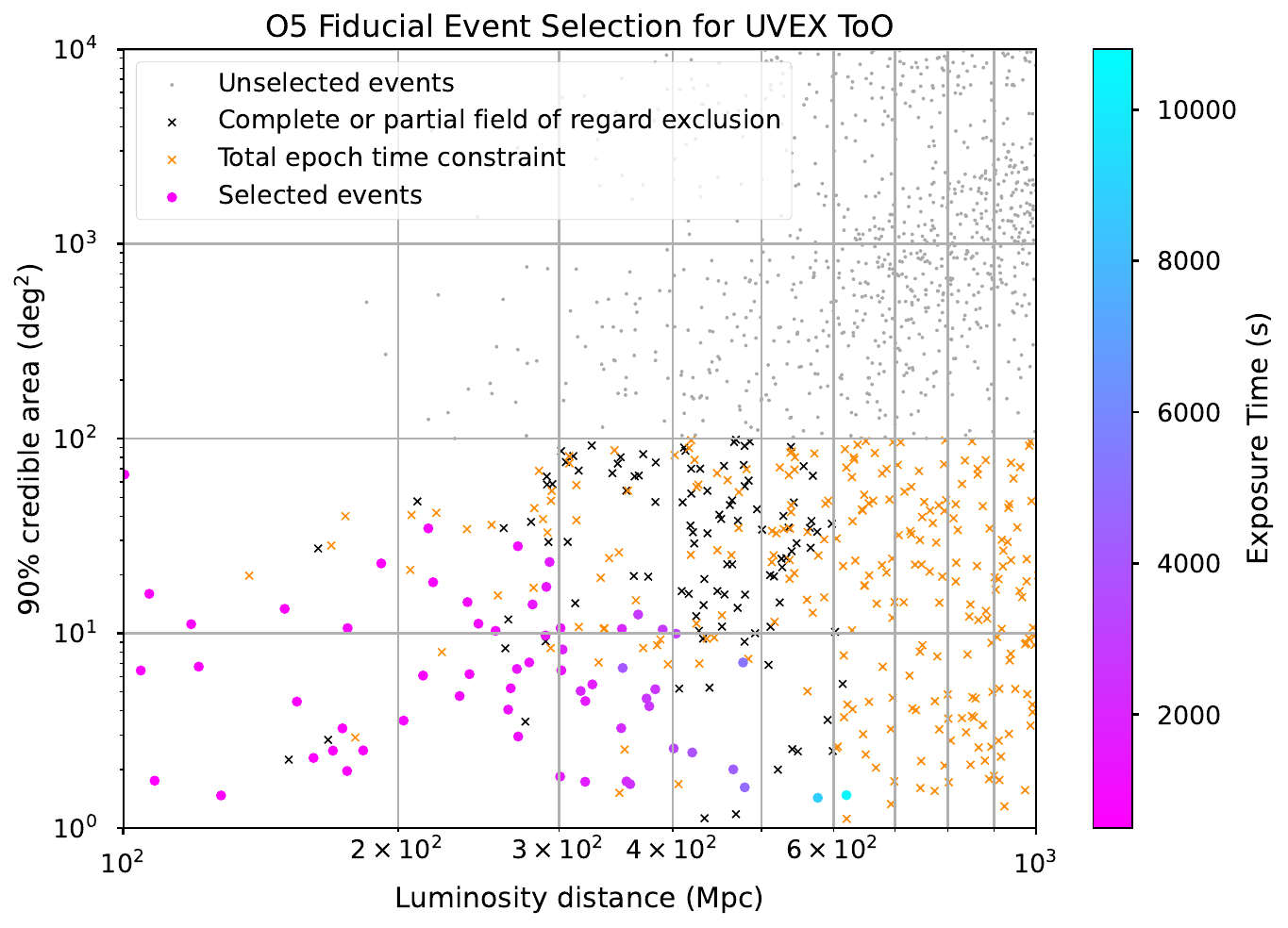}
    \includegraphics[width=0.47\textwidth]{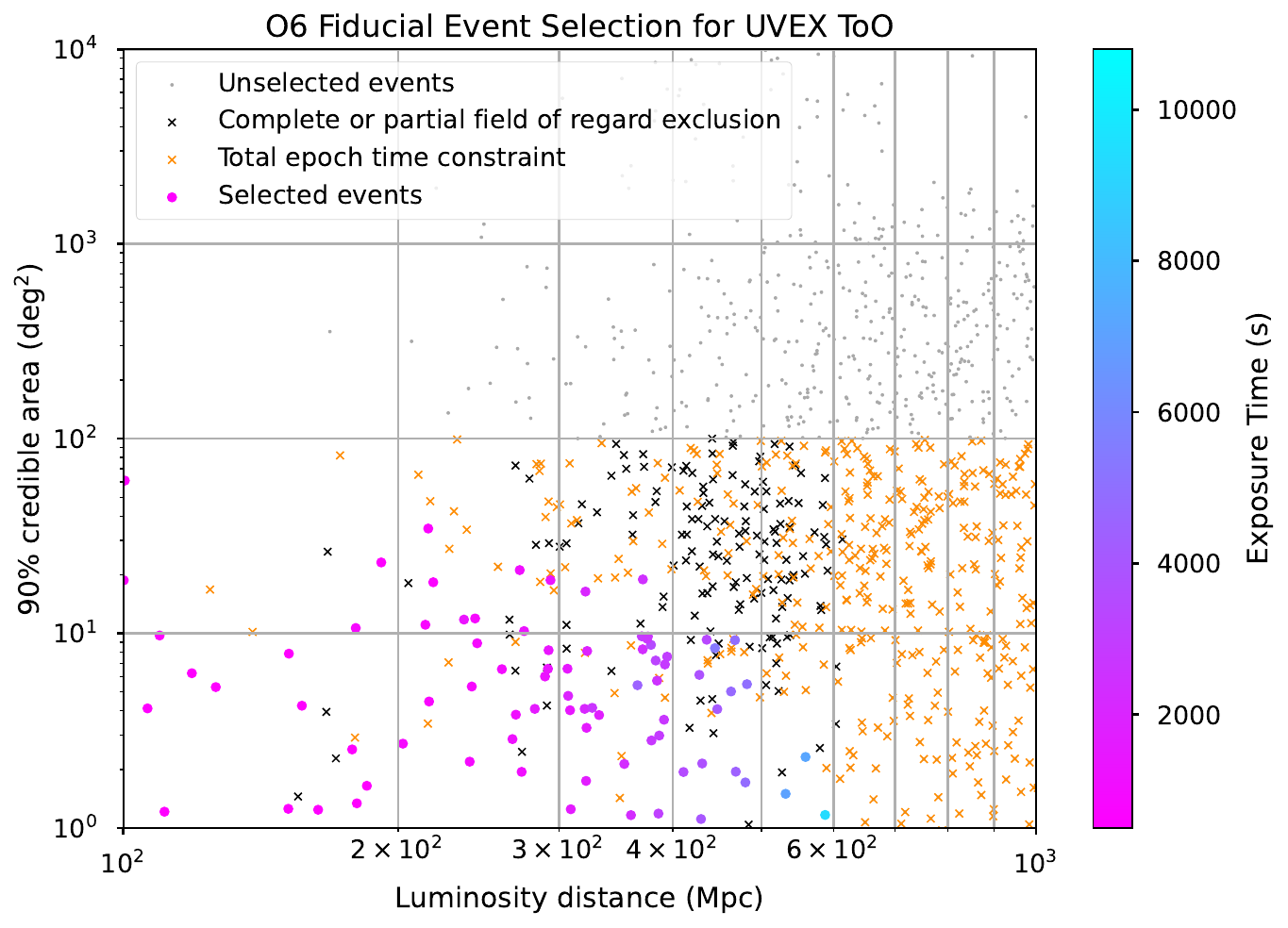}
    
    % Bottom row
    \includegraphics[width=0.47\textwidth]{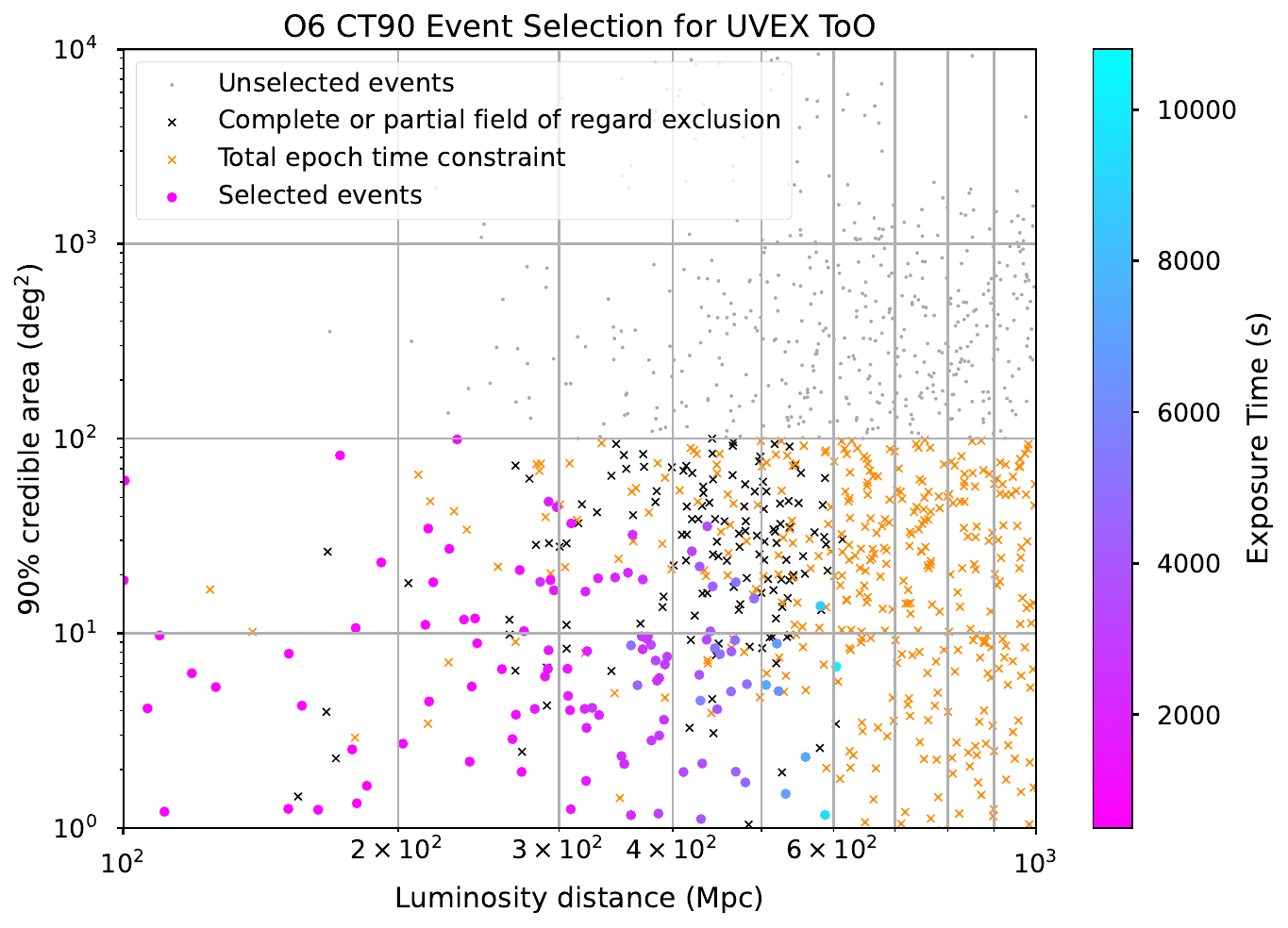}
    \includegraphics[width=0.47\textwidth]{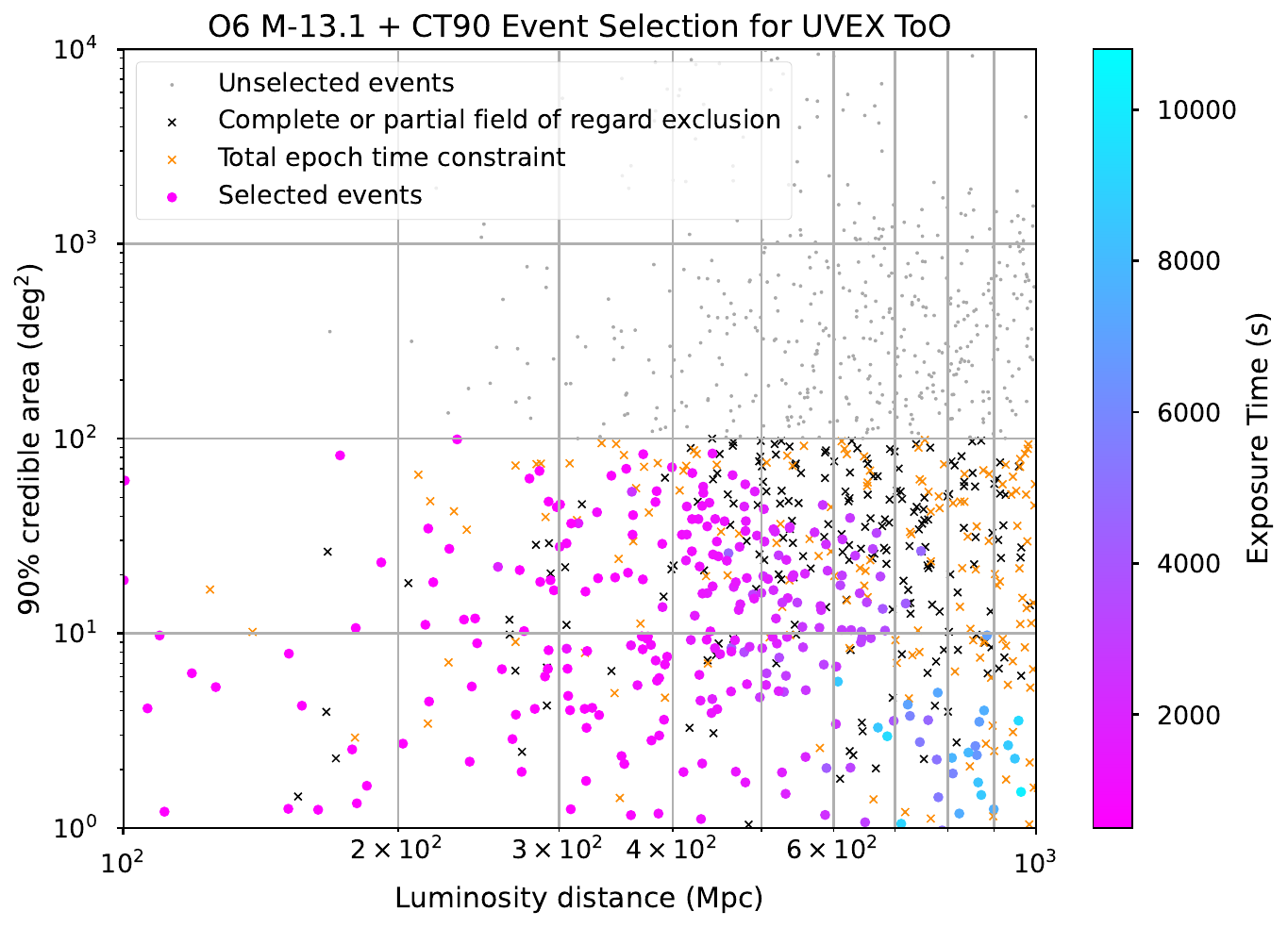}

    \caption{UVEX ToO selection summary plots for selected simulations discussed in \S\ref{sec:prospects}--\ref{sec:new_strats}. \textbf{Top:} event selection for the fiducial strategy (\S\ref{sec:prospects}) as applied to the O5 and O6 observing scenarios. \textbf{Bottom:} event selection for selected expanded ToO strategies. The bottom left panel shows the CT90 strategy (\S\ref{sec:strats_cov_thresh}), which reduces the minimum required UVEX coverage of the GW localization region to 90\%. The bottom right panel shows the M-13.1+CT90 strategy (\S\ref{sec:strats_kn_mag}), which additionally assumes a less-conservative value of the intrinsic KN luminosity for exposure time calculations. Points correspond to individual simulated events from the \citet{kiendrebeogo_updated_2023a} observing scenarios, so the total number of points is not one-to-one with the expected number of observed BNS mergers (although it is proportional); scaling from the simulated event rate to an astrophysical BNS rate of $210^{+240}_{-120}\ \mathrm{Gpc}^{-3}\,\mathrm{yr}^{-1}$ \citep{abbott_gwtc-3:_2023} yields the UVEX BNS ToO trigger rates given in Table~\ref{tab:uvex_success_rates}. Events are plotted by their mean GW distance estimate and 90\% C.L. GW localization areas. Points in grey are not selected due to the overall threshold of $\leq100$ deg$^2$ GW 90\% C.L. localization area. Events with $\leq100$ deg$^2$ localizations that are not selected due falling outside of the UVEX FoR or exceeding the maximum epoch time constraint of 3 hrs are marked as black and orange $\times$s, respectively. Selected events are color-coded by per-tile exposure time.}
    \label{fig:selection_multi}
\end{figure*}

Perhaps unsurprisingly, this strategy carries the most significant effect on the number of UVEX EM-GW ToO triggers, almost doubling the ToO trigger count compared to the fiducial strategy. The M-13.1 variation alone results in an estimated $6.8^{+8.3}_{-3.8}$ and $11.2^{+13.9}_{-6.3}$ additional ToO triggers for O5 and O6, for a total of $15.7^{+19.4}_{-8.8}$ and $22.5^{+27.9}_{-12.6}$ triggers, respectively. We also consider a joint strategy, dubbed M-13.1 + CT90, which couples M-13.1 with a relaxed coverage threshold of 90\%. This approach yields an estimated $15.0^{+18.6}_{-8.4}$ and $21.8^{+27.0}_{-12.2}$ additional triggers for O5 and O6, for a total of $23.9^{+29.7}_{-13.4}$ and $33.1^{+41.0}_{-18.5}$ triggers, respectively. By and large, the additional triggers result from the increased distances at which this strategy will trigger ToOs, as can be seen in Fig.~\ref{fig:selection_multi}. While adopting the M-13.1 + CT90 strategy does result in a substantially reduced success rate (63.5\%  and 63.6\% for the O5 and O6 scenarios), it nonetheless yields the highest estimate for successful counterpart detections: $15.2^{+18.8}_{-8.6}$ ($21.1^{+26.1}_{-11.9}$) for O5 (O6). The trigger counts, success rates, and estimated counterpart detections for each strategy can be found in Table~\ref{tab:uvex_success_rates}.

% \begin{figure}
%     \centering
%     \includegraphics[width=0.9\linewidth]{uvex_event_selection_O6_cov90_mag13.pdf}
%     \caption{UVEX ToO selection for the O6 simulation with a 90\% threshold for minimum coverage by UVEX of the GW localization region and an assumed intrinsic KN magnitude of -13.1 AB mags. Points correspond to individual simulated events from the \citet{kiendrebeogo_updated_2023a} O6 observing scenario, so the total number of points is not one-to-one with the expected number of observed BNS mergers (although it is proportional); scaling from the simulated event rate to an astrophysical BNS rate of $210^{+240}_{-120}\ \mathrm{Gpc}^{-3}\,\mathrm{yr}^{-1}$ \citep{abbott_gwtc-3:_2023} yields $14.7^{+18.3}_{-8.3}$ UVEX ToO triggers over 1.5 years of O6. Events are plotted by their mean GW distance estimate and 90\% C.L. GW localization areas. Points in grey are not selected due to the overall threshold of $\leq100$ deg$^2$ GW 90\% C.L. localization area. Events with $\leq100$ deg$^2$ localizations that are not selected due falling outside of the UVEX FoR or exceeding the maximum epoch time constraint of 3 hrs are marked as black and orange $\times$s, respectively. Selected events are color-coded by per-tile exposure time.}
%     \label{fig:uvex_KNmag_cov90_O6}
% \end{figure}

\setlength{\extrarowheight}{2pt}
\begin{table}
    \renewcommand{\arraystretch}{1.35}
    \centering
    \begin{tabular}{lccc}
    \hline\hline
    \multirow{2}{*}{Strategy}  & \multicolumn{3}{c}{Observing Run: O5 (O6)} \\ 
        \cline{2-4}  
    & BNS ToO Triggers & Success Rate & Expected Successful Observations \\ 
    \hline 
    \textbf{Fiducial} & $8.9^{+11.1}_{-5.0}$ ($11.3^{+14.0}_{-6.3}$) & 76.3\% (77.5\%) & $6.8^{+8.4}_{-3.8}$ ($8.8^{+10.9}_{-4.9}$) \\
    \hline
    A150 & $8.9^{+11.1}_{-5.0}$ ($11.3^{+14.0}_{-6.3}$) &  76.3\% (77.5\%) & $6.8^{+8.4}_{-3.8}$ ($8.8^{+10.9}_{-4.9}$) \\
    A200 & $8.9^{+11.1}_{-5.0}$ ($11.3^{+14.0}_{-6.3}$) & 76.3\% (77.5\%) & $6.8^{+8.4}_{-3.8}$ ($8.8^{+10.9}_{-4.9}$) \\
    \hline
    CT95 & $10.7^{+13.2}_{-6.0}$ ($13.1^{+16.3}_{-7.3}$) & 76.7\% (77.2\%) & $8.2^{+10.2}_{-4.6}$ ($10.1^{+12.5}_{-5.7}$) \\
    \textbf{CT90} & $12.1^{+15.0}_{-6.8}$ ($14.7^{+18.3}_{-8.3}$) & 76.9\% (77.3\%) & $9.3^{+11.5}_{-5.2}$ ($11.4^{+14.1}_{-6.4}$) \\
    \hline
    M-13.1 & $15.7^{+19.4}_{-8.8}$ ($22.5^{+27.9}_{-12.6}$) & 78.6\% (80.4\%) & $12.3^{+15.3}_{-6.9}$ ($18.1^{+22.4}_{-10.2}$) \\
    M-13.1 + CT95 & $21.4^{+26.5}_{-12.0}$ ($28.9^{+35.9}_{-16.2}$) & 63.6\% (70.0\%) & $13.6^{+16.9}_{-7.7}$ ($20.2^{+25.1}_{-11.4}$) \\
    \textbf{M-13.1 + CT90} & $23.9^{+29.7}_{-13.4}$ ($33.1^{+41.0}_{-18.5}$) & 63.5\% (63.6\%) & $15.2^{+18.8}_{-8.6}$ ($21.1^{+26.1}_{-11.9}$) \\
    \hline\hline
    \end{tabular}
    \caption{Summary of ToO rates, success rates and, and estimated number of successful observations for all UVEX ToO selection strategies considered in this work. Quotes are given as ``quote for O5 (quote for O6)". The fiducial strategy corresponds to the original UVEX ToO selection strategy. A150 and A200 include events with GW localization 90\% C.L. areas up to 150 and 200 deg$^2$, respectively (all other strategies retain a 100 deg$^2$ area cut). CT95 and CT90 relax the requirement for UVEX coverage of the 90\% C.L. localization area to 95\%/90\% from the fiducial 99\%. The M-13.1 strategy assumes an intrinsic KN absolute magnitude of -13.1 AB mags (c.f. the fiducial assumption of -12.1 AB mags). M-13.1+CT95 and M-13.1+CT90 combine the M-13.1 strategy with the CT95/CT90 strategies. }
    \label{tab:uvex_success_rates}
\end{table}

\section{Discussion and Future Work}\label{sec:discussion_conclusion}
We present a detailed description of the {\tt uvex-followup} code used to estimate prospects for EM follow-up to BNS mergers detected in GWs with UVEX, as well as updated prospects for ToO observations of KN counterparts with UVEX concurrent to LVK observing runs O5 and O6 ($8.9^{+11.1}_{-5.0}$ and $11.3^{+14.0}_{-6.3}$ triggers, respectively). In light of a likely-reduced BNS rate following a dearth of BNS detections in O4 to-date, we consider expanded strategies beyond those considered in the UVEX proposal and selection process. We find that relaxing certain stringent assumptions for ToO selection --- namely, the minimum \% coverage of the GW localization and the assumed intrinsic KN magnitude --- can help ensure that the scientific return of EM-GW ToO observations with UVEX is robust to a lower rate of BNS detections, increasing the number of UVEX ToO triggers by 168.5\% and 192.9\% (for O5 and O6, respectively) while only reducing the success rate by $\sim$12.8\% and $\sim$13.9\% (for an assumed KN absolute magnitude of -13.1 AB mags and a coverage threshold of 90\%). Conditioned on the observing scenarios of \citet{kiendrebeogo_updated_2023a}, the GWTC-3 BNS rate of $210^{+240}_{-120}\ \mathrm{Gpc}^{-3}\,\mathrm{yr}^{-1}$, and an assumption of uniform uncertainty in the intrinsic KN peak magnitude as given in Eq.~\eqref{eq:abs_kn_mag_prior}, we estimate that the fiducial UVEX EM-GW ToO strategy yields $6.8^{+8.4}_{-3.8}$ ($8.8^{+10.9}_{-4.9}$)counterpart detections in O5 (O6), whereas the M-13.1 + CT90 variant strategy more than doubles this figure, yielding $15.2^{+18.8}_{-8.6}$ ($21.1^{+26.1}_{-11.9}$) UVEX-detected KN counterparts. We emphasize that these results are based on early estimates of the UVEX instrument performance, which may change as the understanding of the telescope performance matures. Additionally, we provide a tool for easily updating these results with a new estimate of the BNS rate.

It is worth noting that while the approach taken in this work relies on a series of binary criteria, the actual reality of the KN counterpart landscape will be much more nuanced. The calculations performed in this work leverage only the GW point estimate of the merger distance, rather than the full luminosity distance posterior distribution. Moreover, they do not capture the full variation in UV backgrounds, foregrounds and extinction across the a given GW localization (absent the capability for adaptive exposure times, we instead use the largest required exposure time across the field for every pointing). A promising direction of future work is to recast the question of UVEX ToO triggers probabilistically, accounting for the full joint distance and sky localization posterior, incorporating adaptive exposure times, and treating our uncertainty as to the intrinsic KN magnitude with a broad prior distribution. Such a method would not only be able to employ the optimal exposure time for each field within the GW localization region, but also allow UVEX ToO observations to be triggered based entirely on how likely they are to succeed, marginalizing over the uncertainty of the GW observation and KN luminosity. Efforts towards such an approach are ongoing and will be presented in a future work \citep{singer_optimal_2025}. It will also be important to consider in future both the advantages and drawbacks of extending the total epoch duration, potentially providing additional depth/coverage at the cost of a more sparsely-sampled lightcurve.

Additionally, this work and prior estimates \citep{kulkarni_science_2023} consider only UVEX ToO prospects for BNS mergers. A small fraction\footnote{10-20\%; see \citet{biscoveanu_population_2023} and references therein.} of NSBH mergers are also expected to produce EM-bright counterparts; as such, prospects for UVEX ToO observations following GW-detected NSBH mergers should be explored in future. Finally, we have restrained ourselves to the questions regarding the rate of and procedures for UVEX EM-GW ToO observations of KNe; it will be valuable to consider what constraints can be placed on astrophysical parameters of interest (namely, the underlying mechanism of early-time UV emission in KNe), how those constraints depend on the size and nature of the sample, and how such considerations may impact the overall UVEX EM-GW ToO strategy.

\section*{Data Availability}
The version of {\tt uvex-followup} used for this study can be found on Zenodo \citep{criswell_criswellalexander/uvex-followup:_2024}. The data products created for this work are available on Zenodo \citep{leggio_data_2024}. The observing scenarios used in this work are publicly available; see \citet{kiendrebeogo_updated_2023a} and references therein.

% In the bibliography the format for data or code follows this format: \\

% \noindent author year, title, version, publisher, prefix:identifier\\

% \citet{2015ApJ...805...23C} provides a example of how the citation in the
% article references the external code at
% \doi{10.5281/zenodo.15991}.  Unfortunately, bibtex does
% not have specific bibtex entries for these types of references so the
% ``@misc'' type should be used.  The Repository tutorial explains how to
% code the ``@misc'' type correctly.  The most recent aasjournal.bst file,
% available with \aastex\ v6, will output bibtex ``@misc'' type properly.

% Also note that the akcnowlodgment environment does not support long amounts of text. If you have a lot of people and institutions to acknowledge, do not use this command. Instead, create a new
\section*{Acknowledgments}
Portions of this manuscript were adapted from the doctoral dissertation of AWC, \textit{Astrophysical Inferences from Multimessenger Ensembles}. AWC acknowledges support by NSF grant no. 2125764. MWC acknowledges support from the National Science Foundation with grant numbers PHY-2308862 and PHY-2117997. The authors acknowledge the use of computing resources provided by the Minnesota Supercomputing Institute at the University of Minnesota. This work used the Delta CPU at University of Illinois at Urbana-Champaign National Center for Supercomputing Applications through allocation AST200029 from the Advanced Cyberinfrastructure Coordination Ecosystem: Services \& Support (ACCESS) program, which is supported by U.S. National Science Foundation grants no. 2138259,  2138286, 2138307, 2137603, and 2138296.



\software{{\tt uvex-followup} (\href{https://github.com/criswellalexander/uvex-followup}{https://github.com/criswellalexander/uvex-followup}), {\tt dorado-scheduling} (\href{https://github.com/nasa/dorado-scheduling}{https://github.com/nasa/dorado-scheduling}), {\tt uvex-mission} (Grefenstette et al., private repository), {\tt BAYESTAR} \citep{singer_rapid_2016},  Matplotlib \citep{hunter_matplotlib:_2007}, Numpy \citep{harris_array_2020a}, Scipy \citep{virtanen_scipy_2020a}, Python \citep{vanrossum_python_2009}, Healpy \citep{zonca_healpy:_2019}, Pandas \citep{mckinney_data_2010a}, and Astropy \citep{robitaille_astropy:_2013,theastropycollaboration_the_2018,astropycollaboration_the_2022a}.
          }


% \newpage

\appendix

\section{Histograms}
\begin{figure*}
    \centering
    \includegraphics[width=0.31\textwidth]{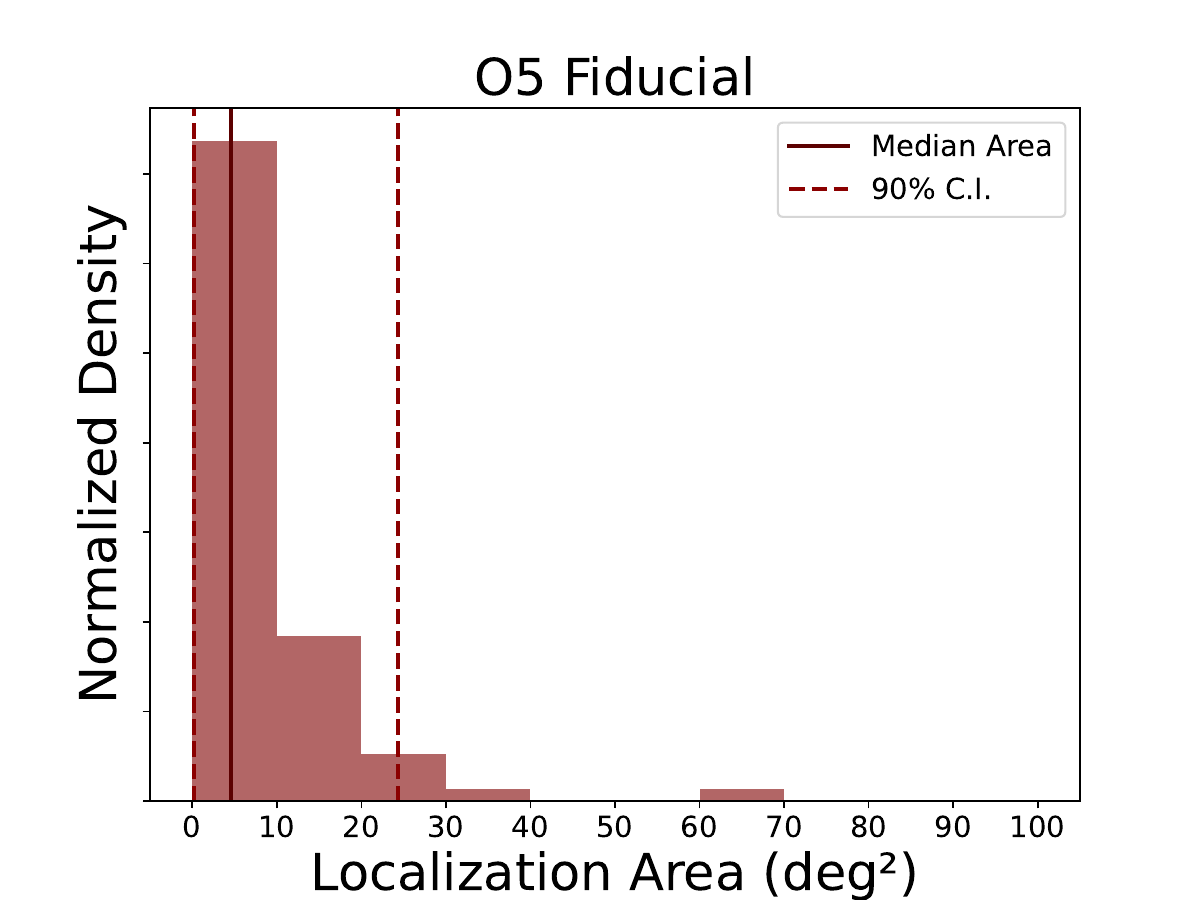}
    \includegraphics[width=0.31\textwidth]{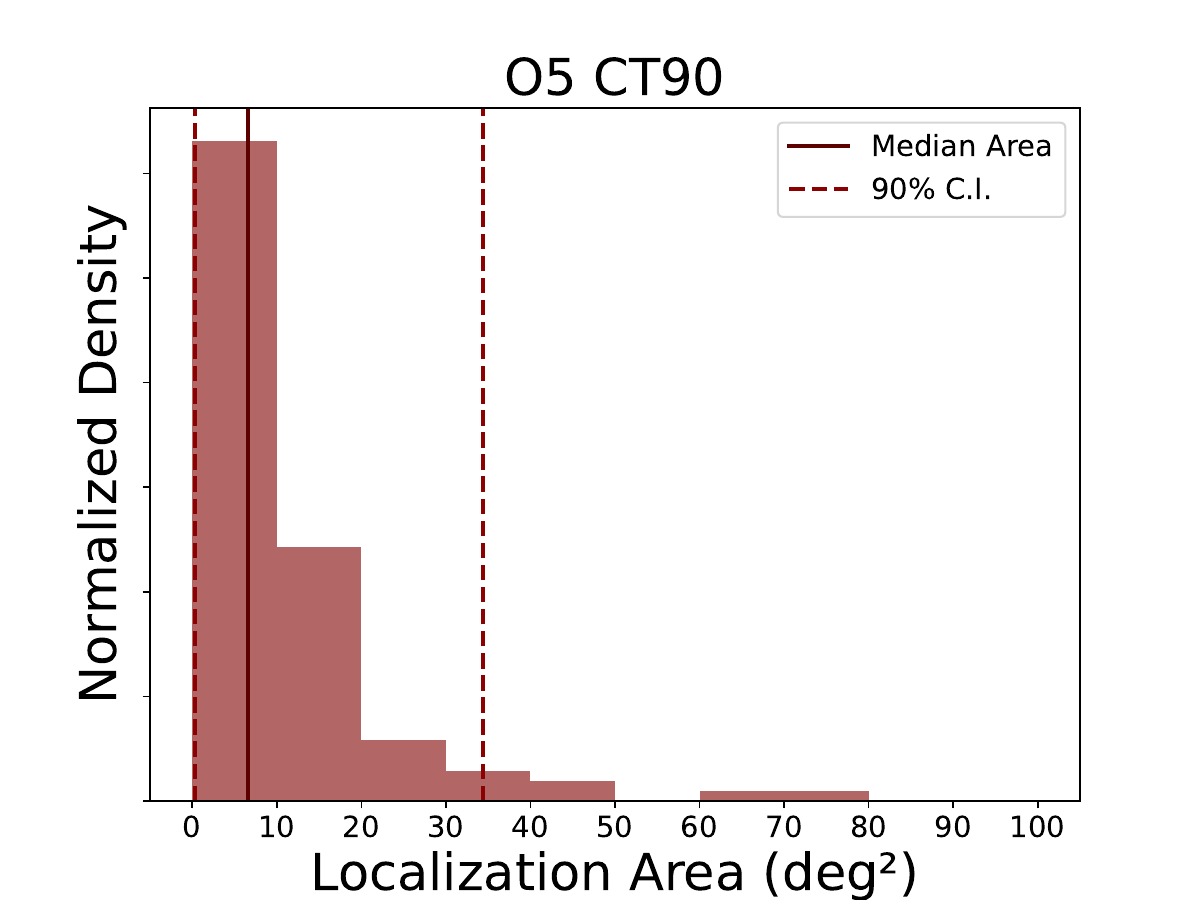}
    \includegraphics[width=0.31\textwidth]{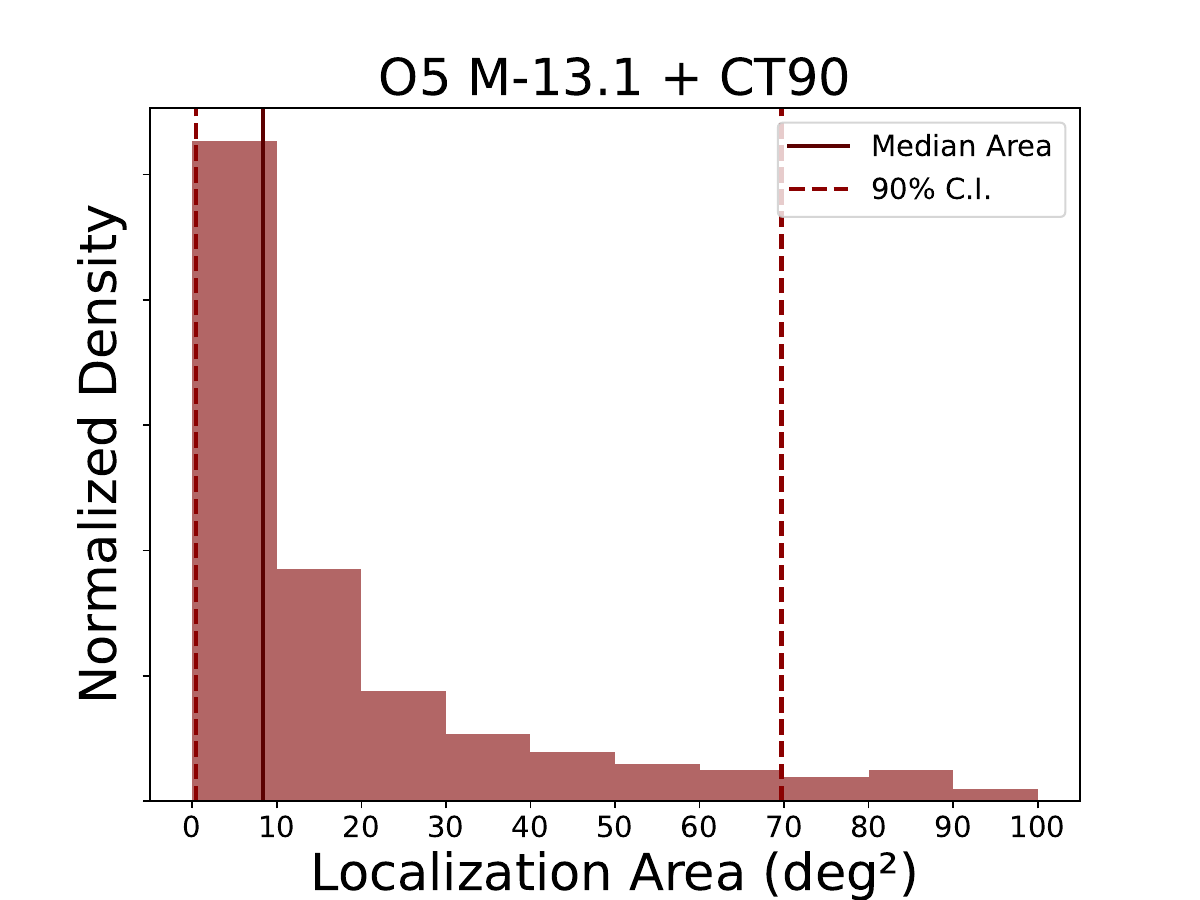}
   
    \includegraphics[width=0.30\textwidth]{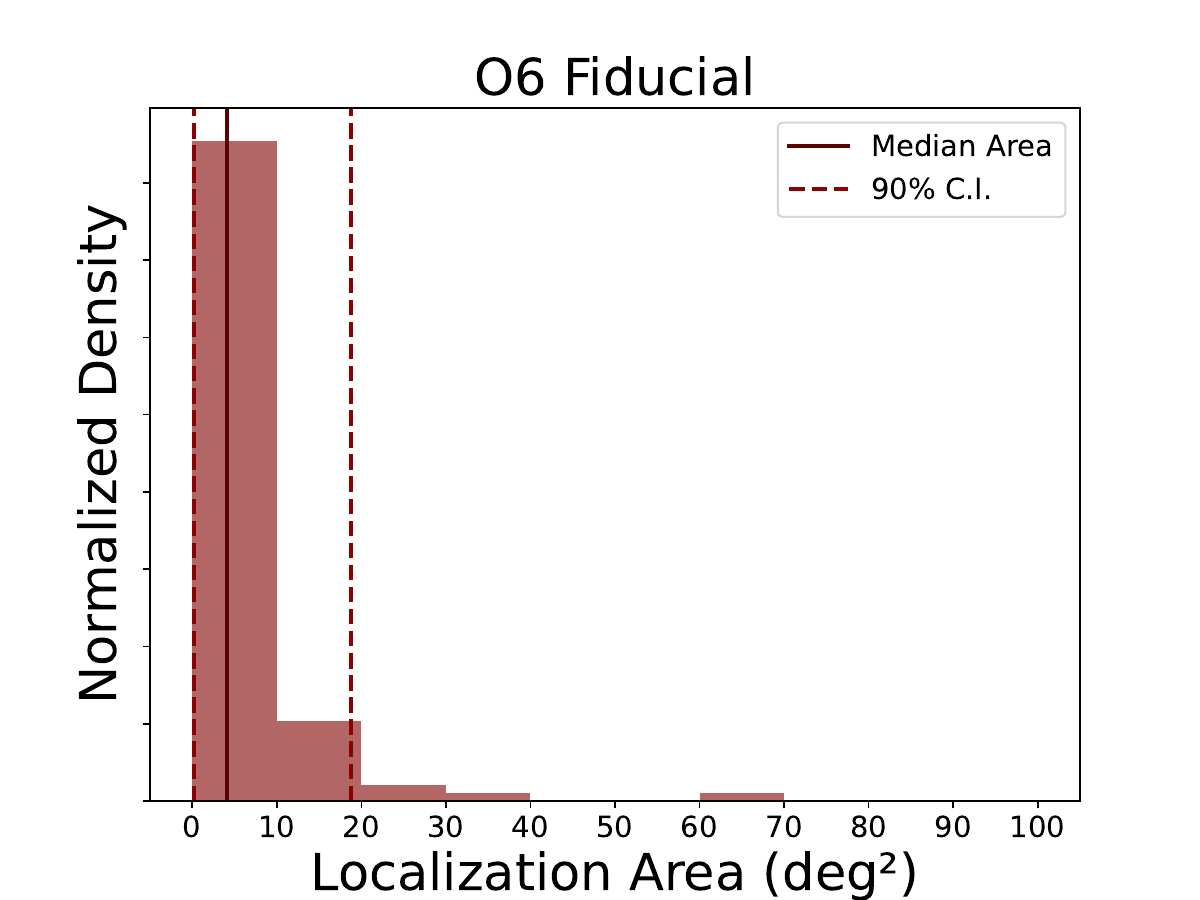}
    \includegraphics[width=0.30\textwidth]{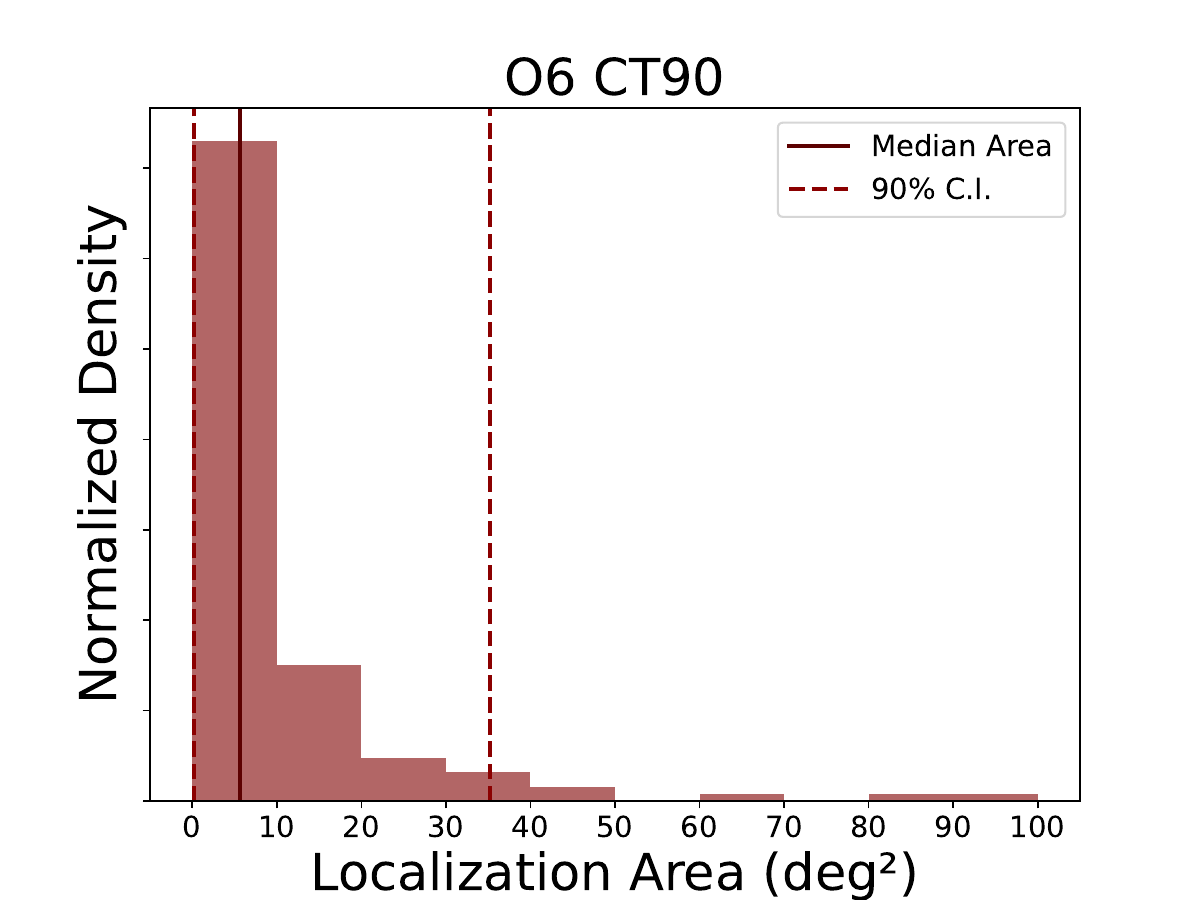}
    \includegraphics[width=0.30\textwidth]{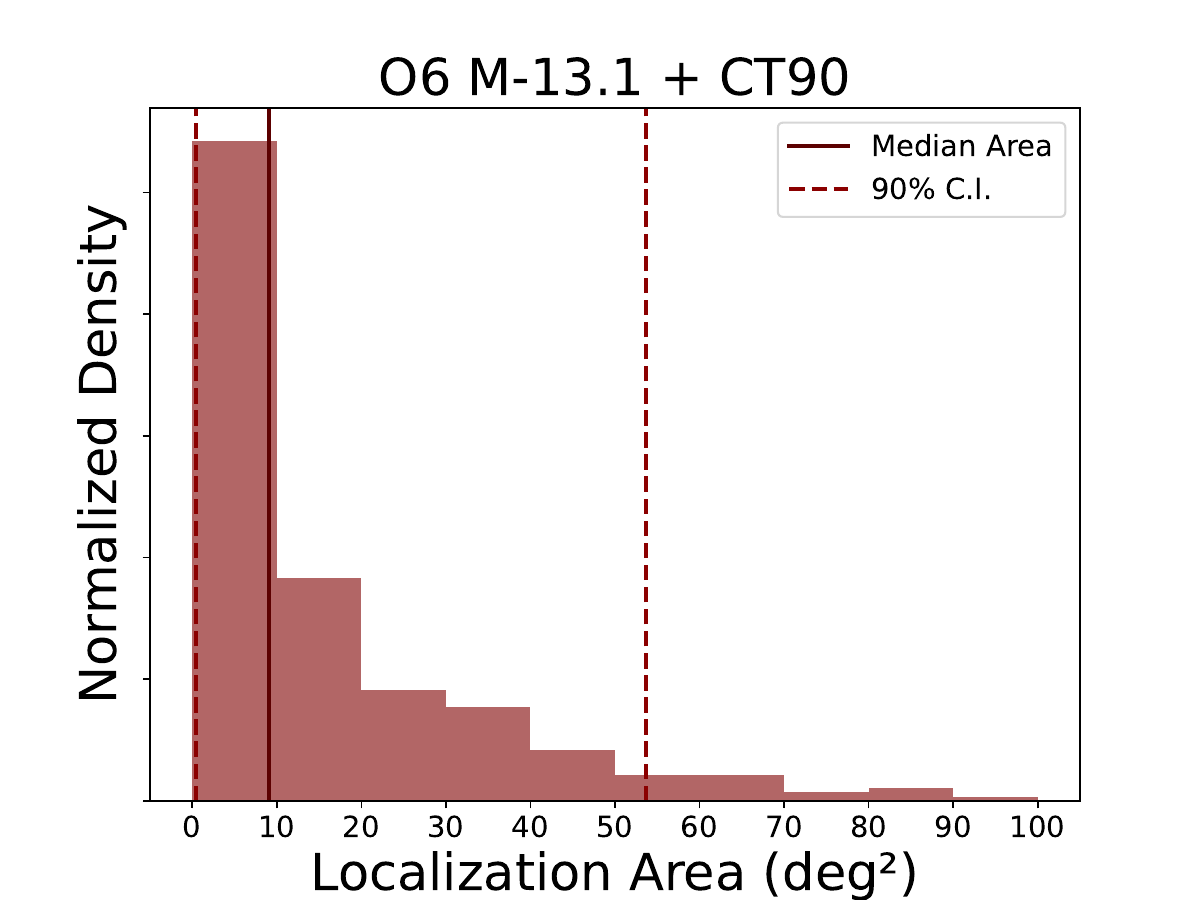}
   
    \caption{Localization area distributions for UVEX BNS ToO triggers in the simulations of O5 (top) and O6 (bottom). Distributions are shown for the fiducial UVEX EM-GW strategy, the variant strategy with a 90\% coverage threshold and the variant strategy with a 90\% coverage threshold and an assumed KN absolute magnitude of -13.1 AB mags for both the O5 and O6 simulations. The median area and 90\% credible intervals (C.I.) are indicated with solid and dashed lines, respectively. Mean and C.I. values for the LVK O5 simulations are as follows: $4.68^{+19.64}_{-4.47}$ $\text{deg}^2$ for fiducial, $6.68^{+27.78}_{-6.47}$ $\text{deg}^2$ for CT90, and $8.40^{+61.21}_{-7.99}$ $\text{deg}^2$ for M-13.1 + CT90. Mean and C.I. values for the LVK O6 simulations are as follows: $4.21^{+14.6}_{-4.08}$ $\text{deg}^2$ for fiducial, $5.56^{+29.54}_{-5.43}$ $\text{deg}^2$ for CT90, and $9.19^{+44.54}_{-8.87}$ $\text{deg}^2$ for M-13.1 + CT90.}
    \label{fig:uvex_areas}
\end{figure*}

\begin{figure*}
    \centering
    \includegraphics[width=0.30\textwidth]{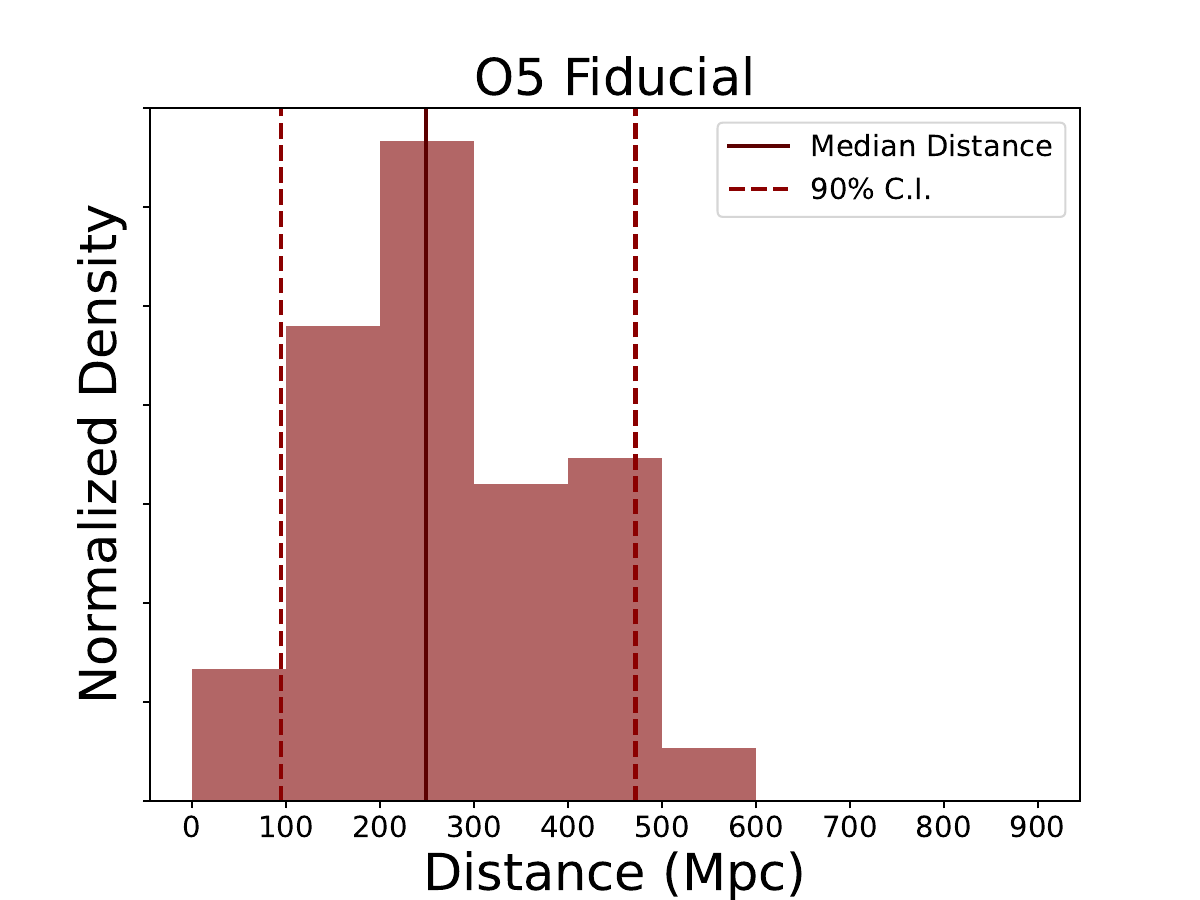}
    \includegraphics[width=0.30\textwidth]{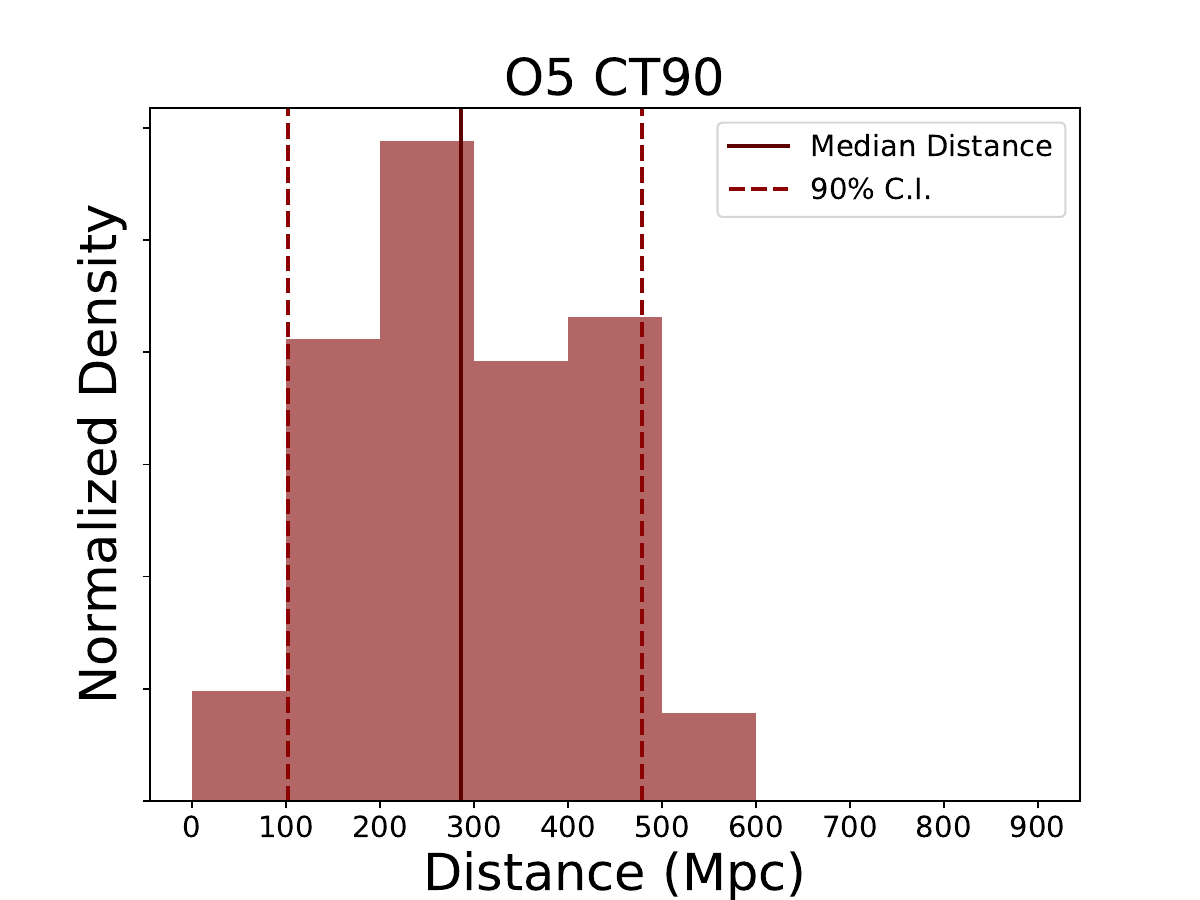}
    \includegraphics[width=0.30\textwidth]{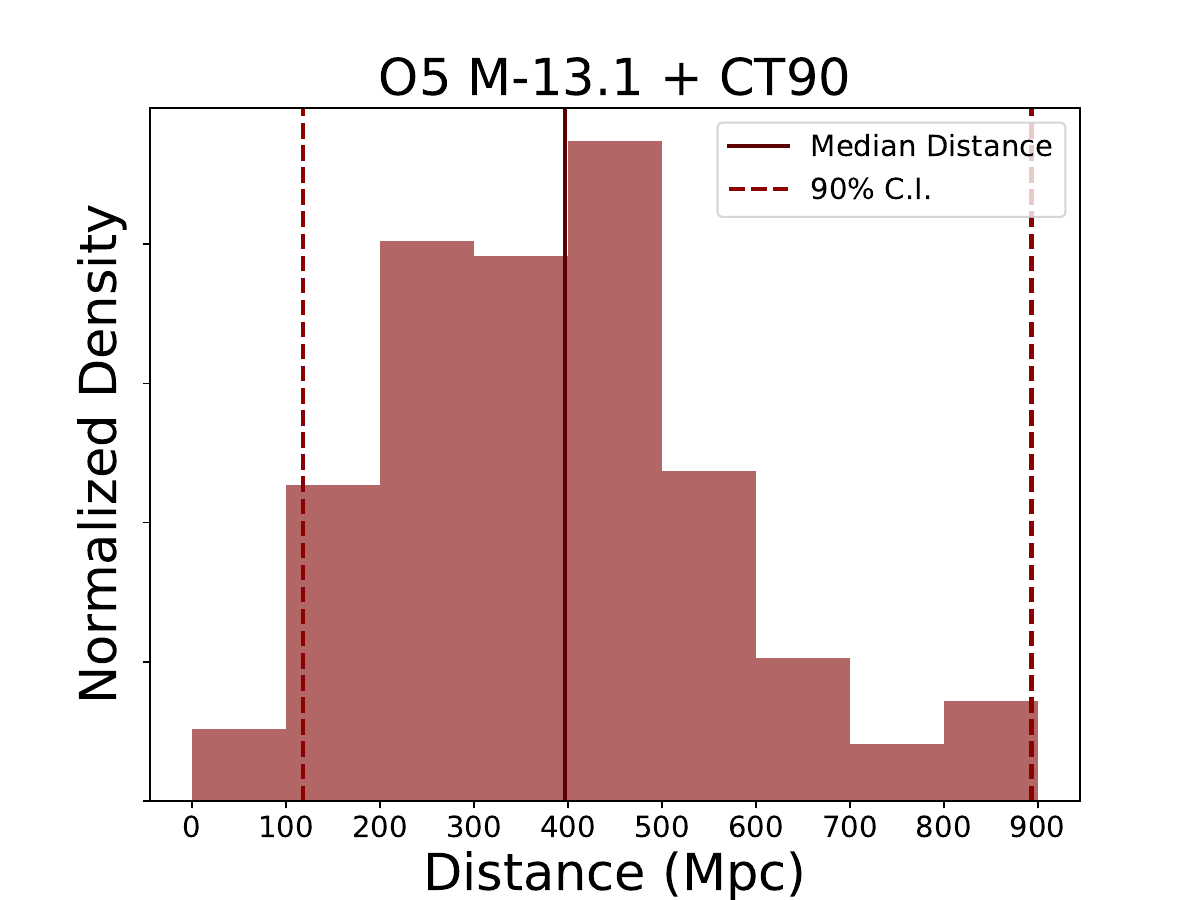}
    \includegraphics[width=0.30\textwidth]{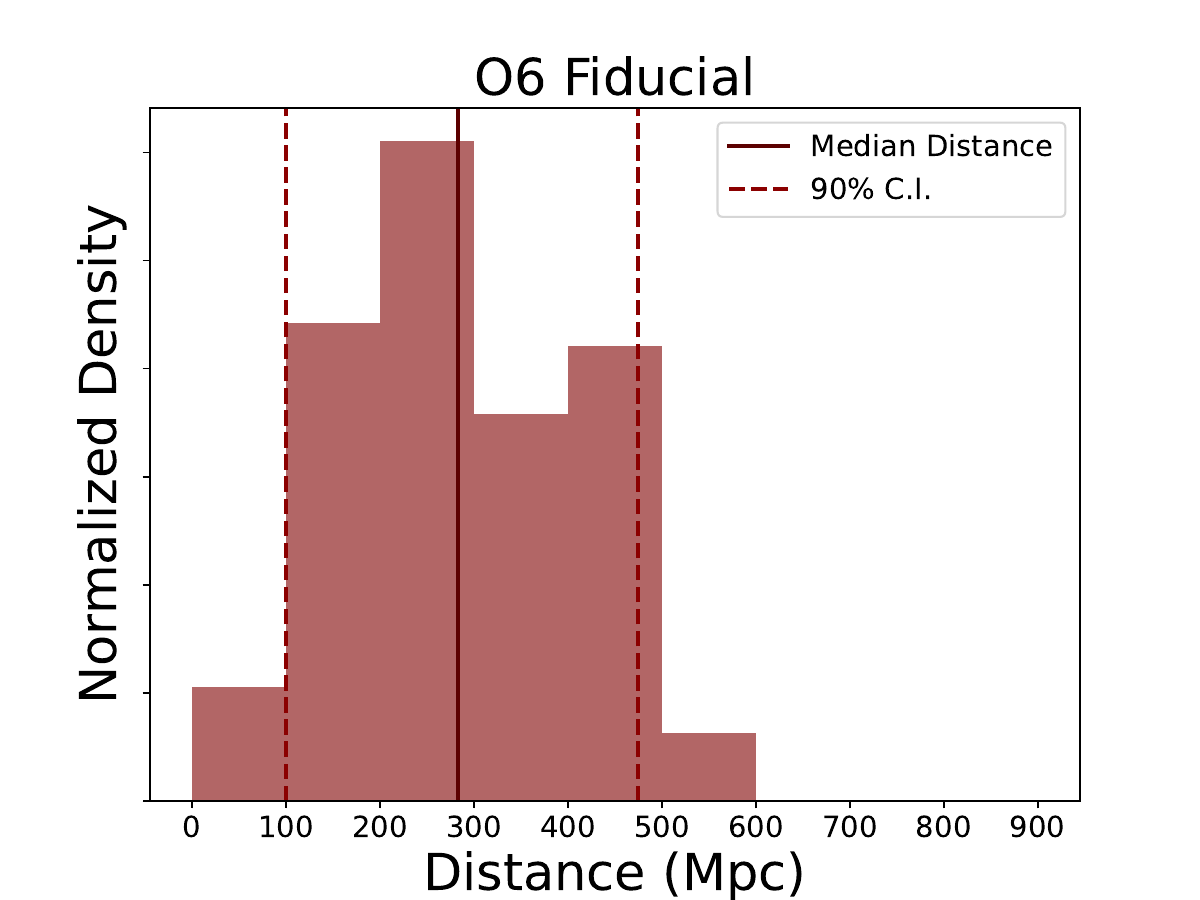}
    \includegraphics[width=0.30\textwidth]{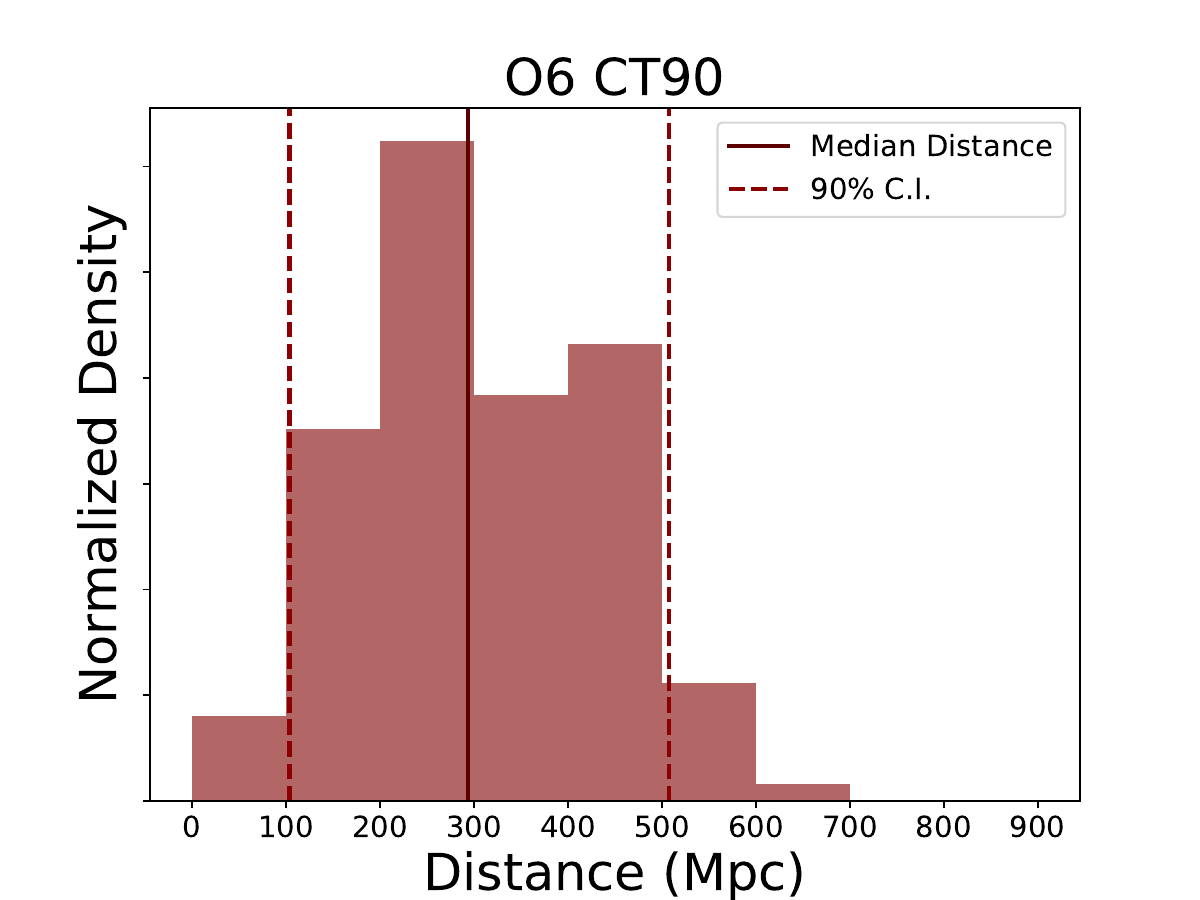}
    \includegraphics[width=0.30\textwidth]{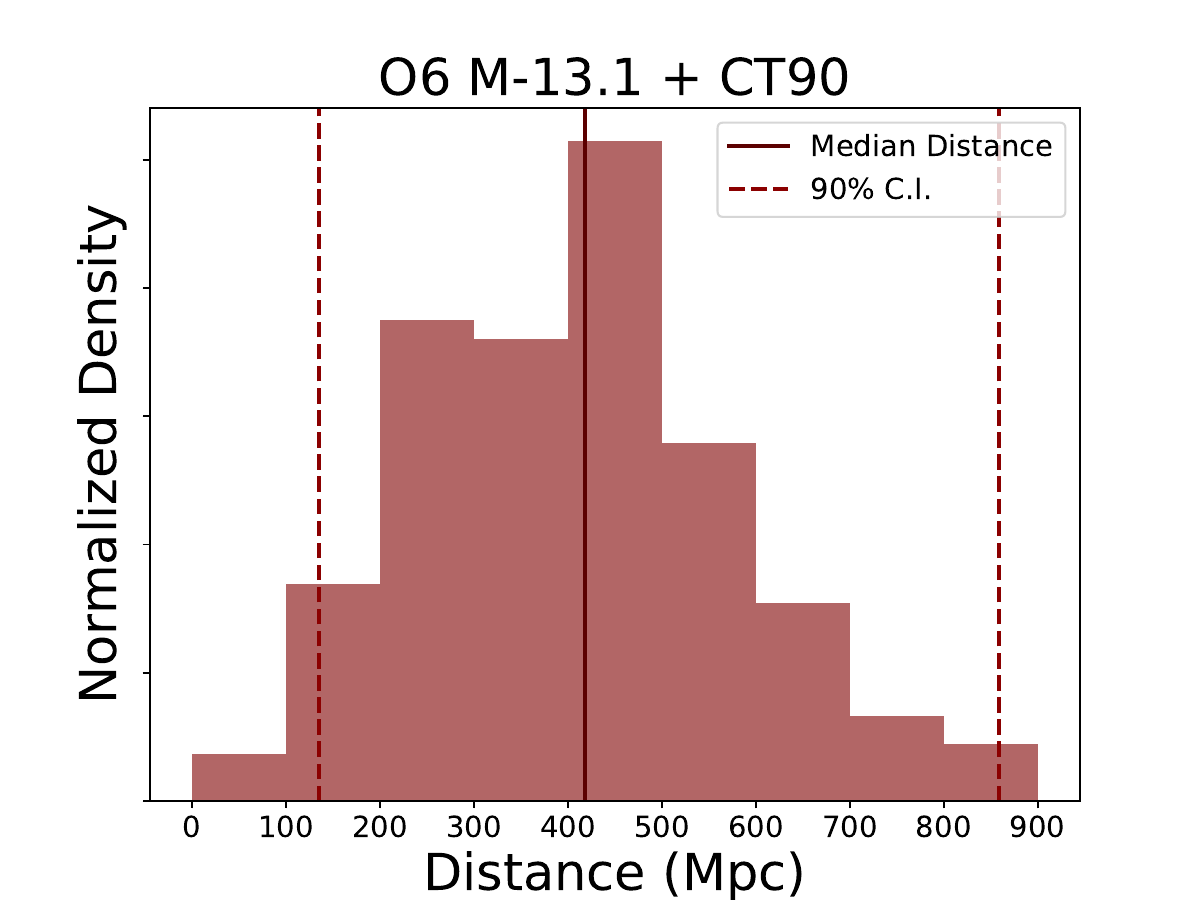}
   
    \caption{Event distance distributions for UVEX BNS ToO triggers in the simulations of O5 (top) and O6 (bottom). Distributions are shown for the fiducial UVEX EM-GW strategy, the variant strategy with a 90\% coverage threshold and the variant strategy with a 90\% coverage threshold and an assumed KN absolute magnitude of -13.1 AB mags for both the O5 and O6 simulations. The median distance and 90\% C.I. are indicated with solid and dashed lines, respectively. Mean and C.I. values for the LVK O5 simulations are as follows: $248.9^{+222.5}_{-154.9}$ Mpc for fiducial, $286.7^{+192.1}_{-185.2}$ Mpc for CT90, and $397.3^{+496.6}_{-279.4}$ Mpc for M-13.1 + CT90. Mean and C.I. values for the LVK O6 simulations are as follows: $283.0^{+190.5}_{-183.2}$ Mpc for fiducial, $293.6^{+214.0}_{-189.5}$ Mpc for CT90, and $418.1^{+441.5}_{-283.5}$ Mpc for M-13.1 + CT90.} 
    \label{fig:uvex_distances}
\end{figure*}

\begin{figure*}
    \centering
    \includegraphics[width=0.32\textwidth]{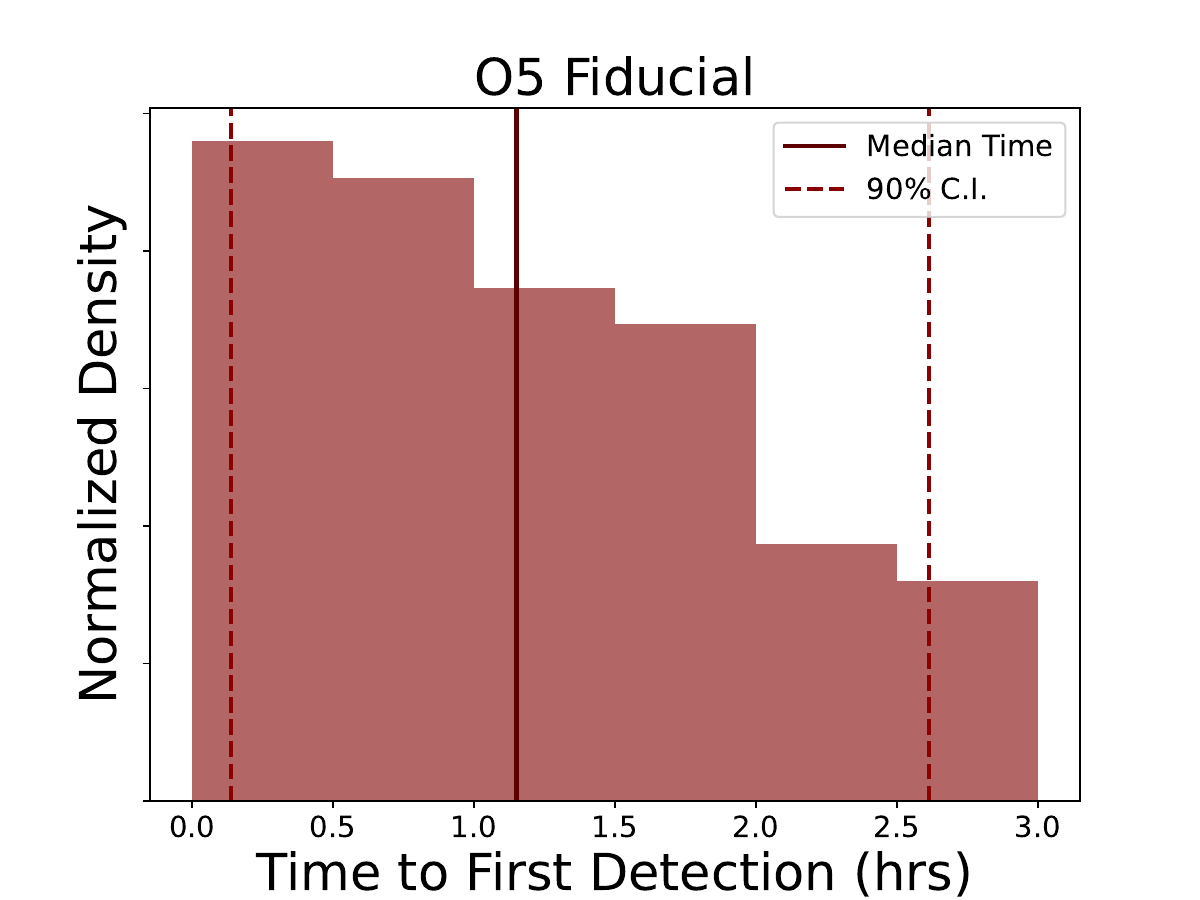}
    \includegraphics[width=0.32\textwidth]{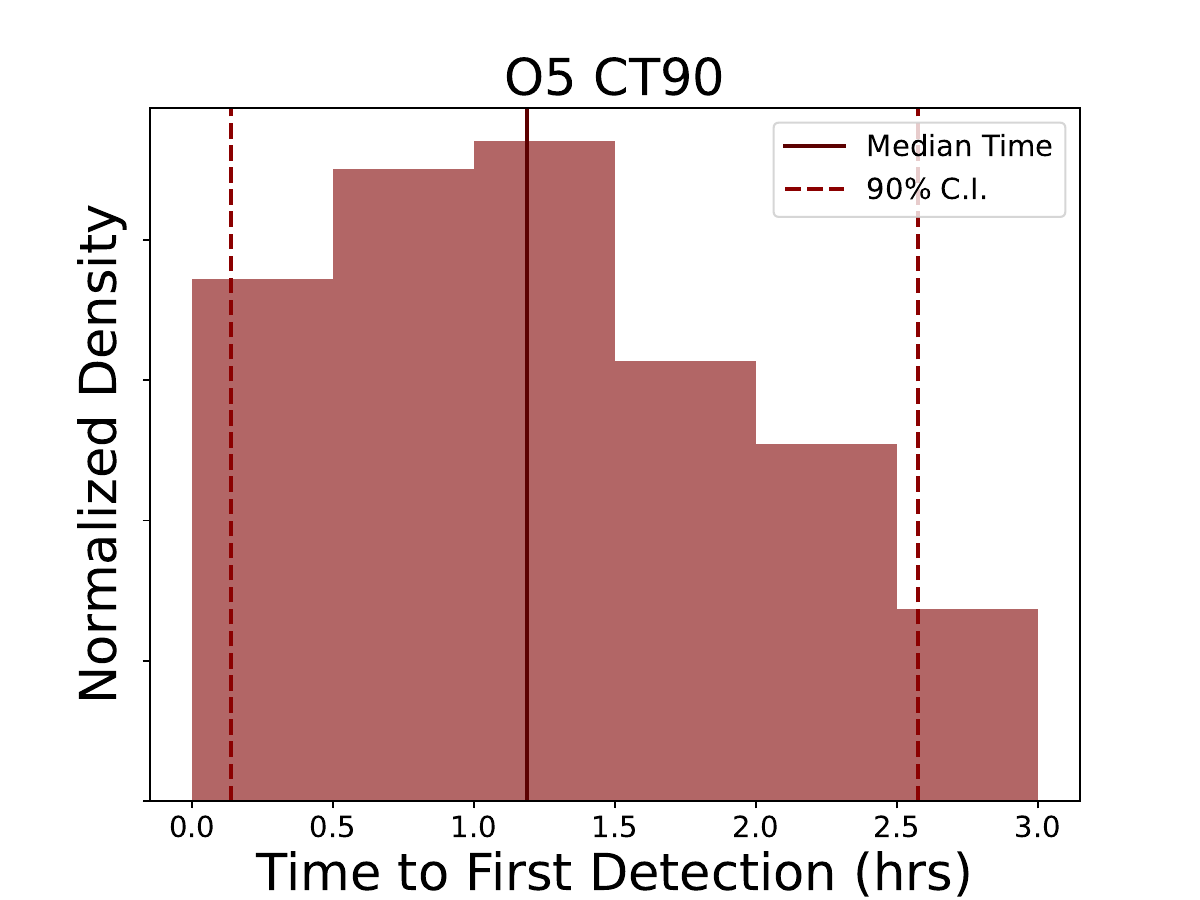}
    \includegraphics[width=0.32\textwidth]{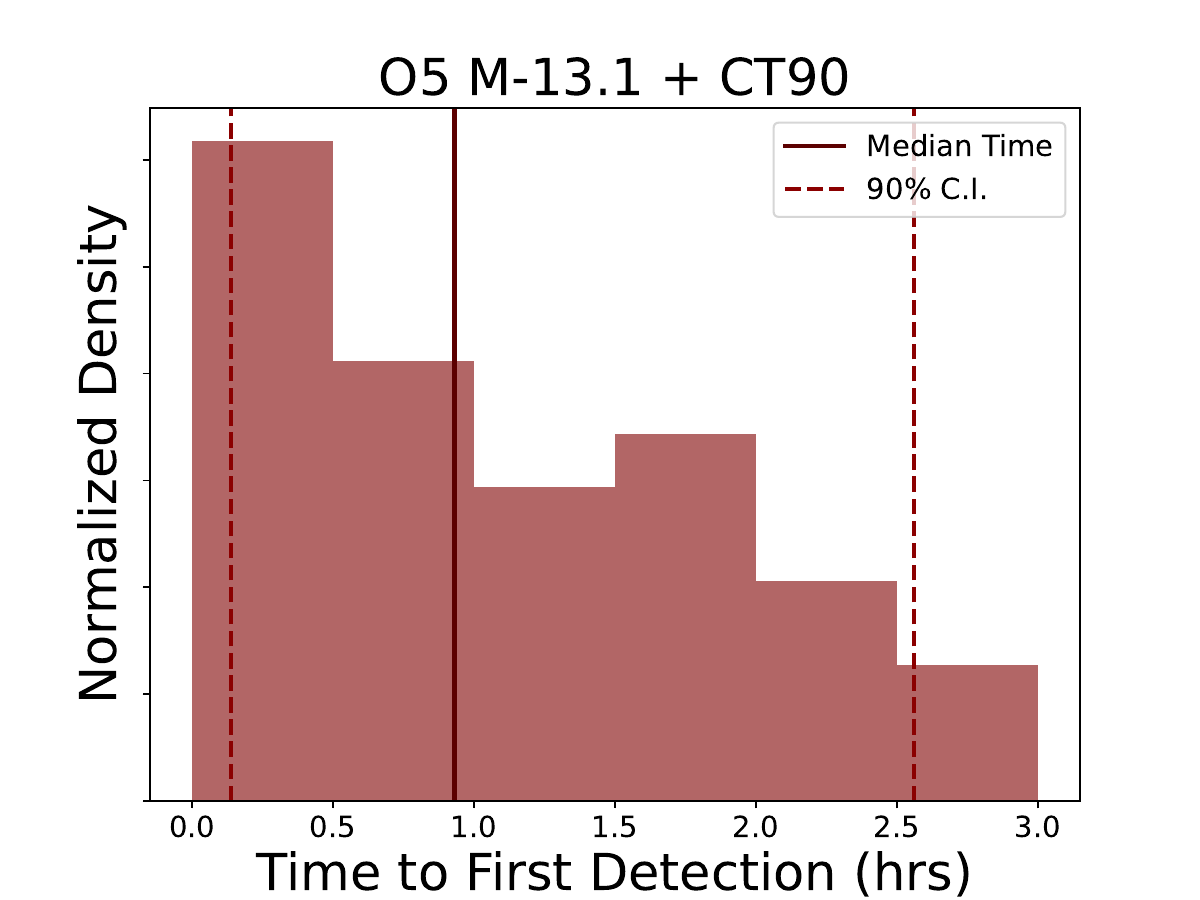}
    \includegraphics[width=0.32\textwidth]{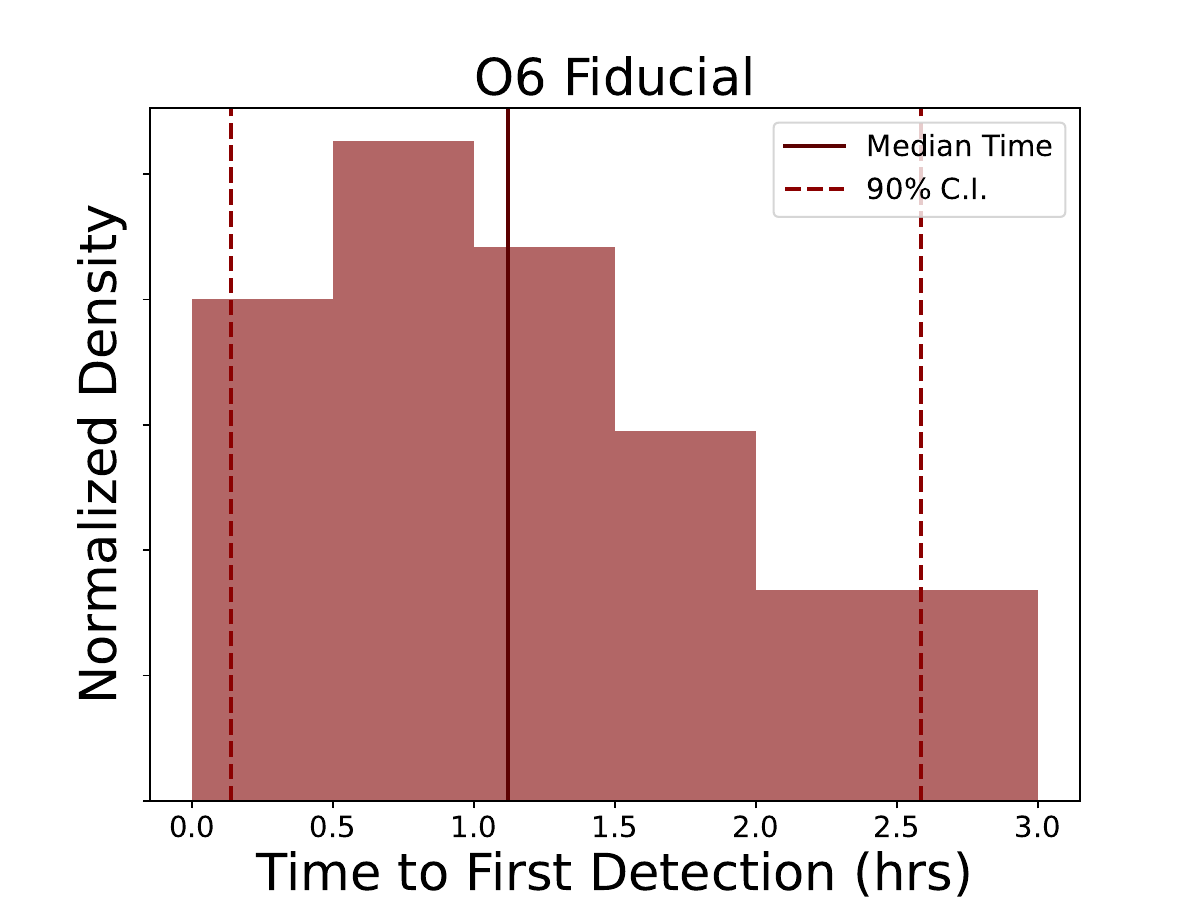}
    \includegraphics[width=0.32\textwidth]{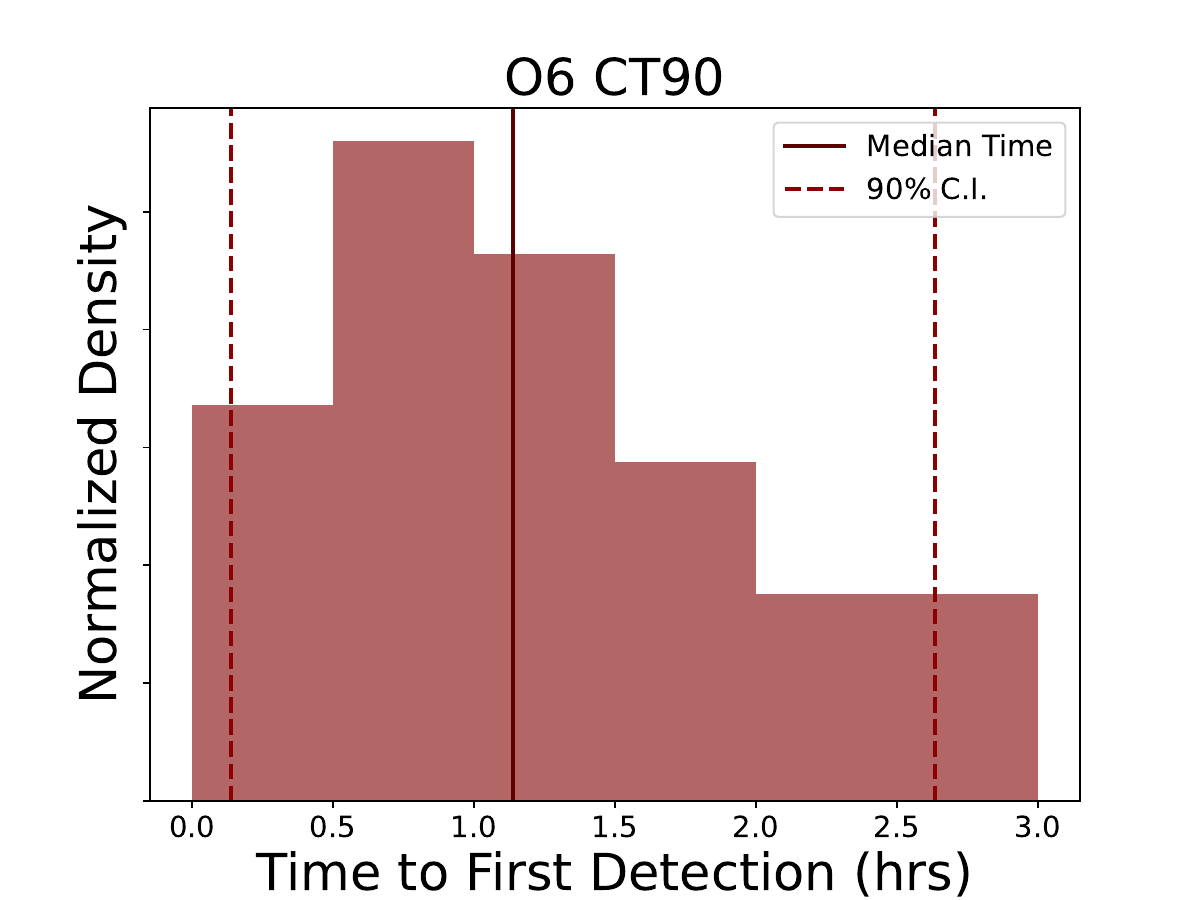}
    \includegraphics[width=0.32\textwidth]{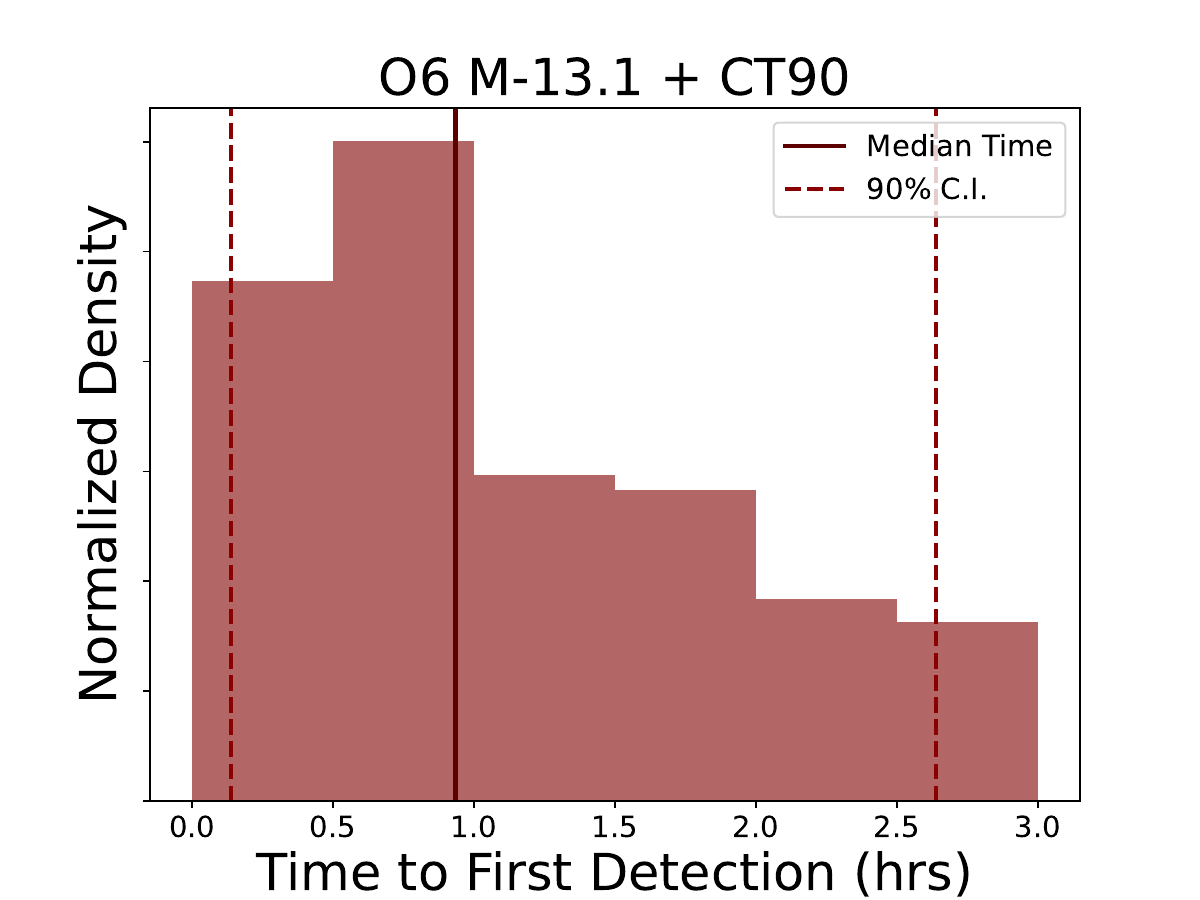}
   
    \caption{Time to first detection distributions for for UVEX BNS ToO triggers in the simulations of O5 (top) and O6 (bottom). Note that $t=0$ denotes the beginning of observations, \textit{not} the time of merger. The UVEX ToO response time requirement is $<3$ hrs from receipt of the trigger to the start of observations. Distributions are shown for the fiducial UVEX EM-GW strategy, the variant strategy with a 90\% coverage threshold and the variant strategy with a 90\% coverage threshold and an assumed KN absolute magnitude of -13.1 AB mags for both the O5 and O6 simulations. The median time and 90\% C.I. are indicated with solid and dashed lines, respectively. Mean and C.I. values for the LVK O5 simulations are as follows: $1.15^{+1.46}_{-1.01}$ hours for fiducial, $1.19^{+1.38}_{-1.05}$ hours for CT90, and $0.93^{+1.63}_{-0.79}$ hours for M-13.1 + CT90. Mean and C.I. values for the LVK O6 simulations are as follows: $1.12^{+1.47}_{-0.98}$ hours for fiducial, $1.14^{+1.49}_{-1.01}$ hours for CT90, and $0.93^{+1.71}_{-0.79}$ hours for M-13.1 + CT90.} 
    \label{fig:uvex_times}
\end{figure*}

We include for reference histograms of 90\% C.L. GW localization areas (Fig.~\ref{fig:uvex_areas}), event distances (Fig.~\ref{fig:uvex_distances}),  and time to first detection (Fig.~\ref{fig:uvex_times}) for events selected for ToO triggers with the fiducial UVEX EM-GW strategy, the variant CT90 strategy with a 90\% coverage threshold (discussed in \S\ref{sec:strats_cov_thresh}), and the variant M-13.1+CT90 strategy with a 90\% coverage threshold and an assumed KN absolute magnitude of -13.1 AB mags (discussed in \S\ref{sec:strats_kn_mag}) for both the O5 and O6 simulations. 


\clearpage
\bibliography{UVEX_EMGW.bib,Code_Packages.bib}{}

\begin{thebibliography}{}
\expandafter\ifx\csname natexlab\endcsname\relax\def\natexlab#1{#1}\fi
\providecommand{\url}[1]{\href{#1}{#1}}
\providecommand{\dodoi}[1]{doi:~\href{http://doi.org/#1}{\nolinkurl{#1}}}
\providecommand{\doeprint}[1]{\href{http://ascl.net/#1}{\nolinkurl{http://ascl.net/#1}}}
\providecommand{\doarXiv}[1]{\href{https://arxiv.org/abs/#1}{\nolinkurl{https://arxiv.org/abs/#1}}}

\bibitem[{Abbott {et~al.}(2017{\natexlab{a}})Abbott, Abbott, Abbott, Acernese, Ackley, Adams, Adams, Addesso, Adhikari, Adya, Affeldt, Afrough, Agarwal, Agathos, Agatsuma, Aggarwal, Aguiar, Aiello, Ain, Ajith, Allen, Allen, Allocca, Altin, Amato, Ananyeva, Anderson, Anderson, Angelova, Antier, Appert, Arai, Araya, Areeda, Arnaud, Arun, Ascenzi, Ashton, Ast, Aston, Astone, Atallah, Aufmuth, Aulbert, AultONeal, Austin, {Avila-Alvarez}, Babak, Bacon, Bader, Bae, Bailes, Baker, Baldaccini, Ballardin, Ballmer, Banagiri, Barayoga, Barclay, Barish, Barker, Barkett, Barone, Barr, Barsotti, Barsuglia, Barta, Barthelmy, Bartlett, Bartos, Bassiri, Basti, Batch, Bawaj, Bayley, Bazzan, B{\'e}csy, Beer, Bejger, Belahcene, Bell, Berger, Bergmann, Bernuzzi, Bero, Berry, Bersanetti, Bertolini, Betzwieser, Bhagwat, Bhandare, Bilenko, Billingsley, Billman, Birch, Birney, Birnholtz, Biscans, Biscoveanu, Bisht, Bitossi, Biwer, Bizouard, Blackburn, Blackman, Blair, Blair, Blair, Bloemen, Bock, Bode, Boer, Bogaert, Bohe, Bondu,
  Bonilla, Bonnand, Boom, Bork, Boschi, Bose, Bossie, Bouffanais, Bozzi, Bradaschia, Brady, Branchesi, Brau, Briant, Brillet, Brinkmann, Brisson, Brockill, Broida, Brooks, Brown, Brown, Brunett, Buchanan, Buikema, Bulik, Bulten, Buonanno, Buskulic, Buy, Byer, Cabero, Cadonati, Cagnoli, Cahillane, Calder{\'o}n~Bustillo, Callister, Calloni, Camp, Canepa, Canizares, Cannon, Cao, Cao, Capano, Capocasa, Carbognani, Caride, Carney, Carullo, Casanueva~Diaz, Casentini, Caudill, Cavagli{\`a}, Cavalier, Cavalieri, Cella, Cepeda, {Cerd{\'a}-Dur{\'a}n}, Cerretani, Cesarini, Chamberlin, Chan, Chao, Charlton, Chase, {Chassande-Mottin}, Chatterjee, Chatziioannou, Cheeseboro, Chen, Chen, Chen, Cheng, Chia, Chincarini, Chiummo, Chmiel, Cho, Cho, Chow, Christensen, Chu, Chua, Chua, Chung, Chung, Ciani, Ciolfi, Cirelli, Cirone, Clara, Clark, Clearwater, Cleva, Cocchieri, Coccia, Cohadon, Cohen, Colla, Collette, Cominsky, Constancio, Conti, Cooper, Corban, Corbitt, {Cordero-Carri{\'o}n}, Corley, Cornish, Corsi, Cortese, Costa,
  Coughlin, Coughlin, Coulon, Countryman, Couvares, Covas, Cowan, Coward, Cowart, Coyne, Coyne, Creighton, Creighton, Cripe, Crowder, Cullen, Cumming, Cunningham, Cuoco, Dal~Canton, D{\'a}lya, Danilishin, D'Antonio, Danzmann, Dasgupta, Da~Silva~Costa, Dattilo, Dave, Davier, Davis, Daw, Day, De, DeBra, Degallaix, De~Laurentis, Del{\'e}glise, Del~Pozzo, Demos, Denker, Dent, De~Pietri, Dergachev, De~Rosa, DeRosa, De~Rossi, DeSalvo, {de Varona}, Devenson, Dhurandhar, D{\'i}az, Dietrich, Di~Fiore, Di~Giovanni, Di~Girolamo, Di~Lieto, Di~Pace, Di~Palma, Di~Renzo, Doctor, Dolique, Donovan, Dooley, Doravari, Dorrington, Douglas, Dovale~{\'A}lvarez, Downes, Drago, Dreissigacker, Driggers, Du, Ducrot, Dudi, Dupej, Dwyer, Edo, Edwards, Effler, Eggenstein, Ehrens, Eichholz, Eikenberry, Eisenstein, Essick, Estevez, Etienne, Etzel, Evans, Evans, Factourovich, Fafone, Fair, Fairhurst, Fan, Farinon, Farr, Farr, {Fauchon-Jones}, Favata, Fays, Fee, Fehrmann, Feicht, Fejer, {Fernandez-Galiana}, Ferrante, Ferreira, Ferrini,
  Fidecaro, Finstad, Fiori, Fiorucci, Fishbach, Fisher, {Fitz-Axen}, Flaminio, Fletcher, Fong, Font, Forsyth, Forsyth, Fournier, Frasca, Frasconi, Frei, Freise, Frey, Frey, Fries, Fritschel, Frolov, Fulda, Fyffe, Gabbard, Gadre, Gaebel, Gair, Gammaitoni, Ganija, Gaonkar, {Garcia-Quiros}, Garufi, Gateley, Gaudio, Gaur, Gayathri, Gehrels, Gemme, Genin, Gennai, George, George, Gergely, Germain, Ghonge, Ghosh, Ghosh, Ghosh, Giaime, Giardina, Giazotto, Gill, Glover, Goetz, Goetz, Gomes, Goncharov, Gonz{\'a}lez, Gonzalez~Castro, Gopakumar, Gorodetsky, Gossan, Gosselin, Gouaty, Grado, Graef, Granata, Grant, Gras, Gray, Greco, Green, Gretarsson, Groot, Grote, Grunewald, Gruning, Guidi, Guo, Gupta, Gupta, Gushwa, Gustafson, Gustafson, Halim, Hall, Hall, Hamilton, Hammond, Haney, Hanke, Hanks, Hanna, Hannam, Hannuksela, Hanson, Hardwick, Harms, Harry, Harry, Hart, Haster, Haughian, Healy, Heidmann, Heintze, Heitmann, Hello, Hemming, Hendry, Heng, Hennig, Heptonstall, Heurs, Hild, Hinderer, Ho, Hoak, Hofman, Holt, Holz,
  Hopkins, Horst, Hough, Houston, Howell, Hreibi, Hu, Huerta, Huet, Hughey, Husa, Huttner, {Huynh-Dinh}, Indik, Inta, Intini, Isa, Isac, Isi, Iyer, Izumi, Jacqmin, Jani, Jaranowski, Jawahar, {Jim{\'e}nez-Forteza}, Johnson, {Johnson-McDaniel}, Jones, Jones, Jonker, Ju, Junker, Kalaghatgi, Kalogera, Kamai, Kandhasamy, Kang, Kanner, Kapadia, Karki, Karvinen, Kasprzack, Kastaun, Katolik, Katsavounidis, Katzman, Kaufer, Kawabe, K{\'e}f{\'e}lian, Keitel, Kemball, Kennedy, Kent, Key, Khalili, Khan, Khan, Khan, Khazanov, Kijbunchoo, Kim, Kim, Kim, Kim, Kim, Kim, Kimbrell, King, King, {Kinley-Hanlon}, Kirchhoff, Kissel, Kleybolte, Klimenko, Knowles, Koch, Koehlenbeck, Koley, Kondrashov, Kontos, Korobko, Korth, Kowalska, Kozak, Kr{\"a}mer, Kringel, Krishnan, Kr{\'o}lak, Kuehn, Kumar, Kumar, Kumar, Kuo, Kutynia, Kwang, Lackey, Lai, Landry, Lang, Lange, Lantz, Lanza, Larson, {Lartaux-Vollard}, Lasky, Laxen, Lazzarini, Lazzaro, Leaci, Leavey, Lee, Lee, Lee, Lee, Lee, Lehmann, Lenon, Leon, Leonardi, Leroy, Letendre, Levin,
  Li, Linker, Littenberg, Liu, Liu, Lo, Lockerbie, London, Lord, Lorenzini, Loriette, Lormand, Losurdo, Lough, Lousto, Lovelace, L{\"u}ck, Lumaca, Lundgren, Lynch, Ma, Macas, Macfoy, Machenschalk, MacInnis, Macleod, Maga{\~n}a~Hernandez, {Maga{\~n}a-Sandoval}, Maga{\~n}a~Zertuche, Magee, Majorana, Maksimovic, Man, Mandic, Mangano, Mansell, Manske, Mantovani, Marchesoni, Marion, M{\'a}rka, M{\'a}rka, Markakis, Markosyan, Markowitz, Maros, Marquina, Marsh, Martelli, Martellini, Martin, Martin, Martynov, Marx, Mason, Massera, Masserot, Massinger, {Masso-Reid}, Mastrogiovanni, Matas, Matichard, Matone, Mavalvala, Mazumder, McCarthy, McClelland, McCormick, McCuller, McGuire, McIntyre, McIver, McManus, McNeill, McRae, McWilliams, Meacher, Meadors, Mehmet, Meidam, {Mejuto-Villa}, Melatos, Mendell, Mercer, Merilh, Merzougui, Meshkov, Messenger, Messick, Metzdorff, Meyers, Miao, Michel, Middleton, Mikhailov, Milano, Miller, Miller, Miller, Millhouse, {Milovich-Goff}, Minazzoli, Minenkov, Ming, Mishra, Mitra,
  Mitrofanov, Mitselmakher, Mittleman, Moffa, Moggi, Mogushi, Mohan, Mohapatra, Molina, Montani, Moore, Moraru, Moreno, Morisaki, Morriss, Mours, {Mow-Lowry}, Mueller, Muir, Mukherjee, Mukherjee, Mukherjee, Mukund, Mullavey, Munch, Mu{\~n}iz, Muratore, Murray, Nagar, Napier, Nardecchia, Naticchioni, Nayak, Neilson, Nelemans, Nelson, Nery, Neunzert, Nevin, Newport, Newton, Ng, Nguyen, Nguyen, Nichols, Nielsen, Nissanke, Nitz, Noack, Nocera, Nolting, North, Nuttall, Oberling, O'Dea, Ogin, Oh, Oh, Ohme, Okada, Oliver, Oppermann, Oram, O'Reilly, Ormiston, Ortega, O'Shaughnessy, Ossokine, Ottaway, Overmier, Owen, Pace, Page, Page, Pai, Pai, Palamos, Palashov, Palomba, {Pal-Singh}, Pan, Pan, Pang, Pang, Pankow, Pannarale, Pant, Paoletti, Paoli, Papa, Parida, Parker, Pascucci, Pasqualetti, Passaquieti, Passuello, Patil, Patricelli, Pearlstone, Pedraza, Pedurand, Pekowsky, Pele, Penn, Perez, Perreca, Perri, Pfeiffer, Phelps, Piccinni, Pichot, Piergiovanni, Pierro, Pillant, Pinard, Pinto, Pirello, Pitkin, Poe,
  Poggiani, Popolizio, Porter, Post, Powell, Prasad, Pratt, Pratten, Predoi, Prestegard, Prijatelj, Principe, Privitera, Prix, Prodi, Prokhorov, Puncken, Punturo, Puppo, P{\"u}rrer, Qi, Quetschke, Quintero, {Quitzow-James}, Raab, Rabeling, Radkins, Raffai, Raja, Rajan, Rajbhandari, Rakhmanov, Ramirez, {Ramos-Buades}, Rapagnani, Raymond, Razzano, Read, Regimbau, Rei, Reid, Reitze, Ren, Reyes, Ricci, Ricker, Rieger, Riles, Rizzo, Robertson, Robie, Robinet, Rocchi, Rolland, Rollins, Roma, Romano, Romano, Romel, Romie, Rosi{\'n}ska, Ross, Rowan, R{\"u}diger, Ruggi, Rutins, Ryan, Sachdev, Sadecki, Sadeghian, Sakellariadou, Salconi, Saleem, Salemi, Samajdar, Sammut, Sampson, Sanchez, Sanchez, {Sanchis-Gual}, Sandberg, Sanders, Sassolas, Sathyaprakash, Saulson, Sauter, Savage, Sawadsky, Schale, Scheel, Scheuer, Schmidt, Schmidt, Schnabel, Schofield, Sch{\"o}nbeck, Schreiber, Schuette, Schulte, Schutz, Schwalbe, Scott, Scott, Seidel, Sellers, Sengupta, Sentenac, Sequino, Sergeev, Shaddock, Shaffer, Shah, Shahriar,
  Shaner, Shao, Shapiro, Shawhan, Sheperd, Shoemaker, Shoemaker, Siellez, Siemens, Sieniawska, Sigg, Silva, Singer, Singh, Singhal, Sintes, Slagmolen, Smith, Smith, Smith, Somala, Son, Sonnenberg, Sorazu, Sorrentino, Souradeep, Spencer, Srivastava, Staats, Staley, Steinke, Steinlechner, Steinlechner, Steinmeyer, Stevenson, Stone, Stops, Strain, Stratta, Strigin, Strunk, Sturani, Stuver, Summerscales, Sun, Sunil, Suresh, Sutton, Swinkels, Szczepa{\'n}czyk, Tacca, Tait, Talbot, Talukder, Tanner, T{\'a}pai, Taracchini, Tasson, Taylor, Taylor, Tewari, Theeg, Thies, Thomas, Thomas, Thomas, Thorne, Thorne, Thrane, Tiwari, Tiwari, Tokmakov, Toland, Tonelli, Tornasi, {Torres-Forn{\'e}}, Torrie, T{\"o}yr{\"a}, Travasso, Traylor, Trinastic, Tringali, Trozzo, Tsang, Tse, Tso, Tsukada, Tsuna, Tuyenbayev, Ueno, Ugolini, Unnikrishnan, Urban, Usman, Vahlbruch, Vajente, Valdes, Vallisneri, {van Bakel}, {van Beuzekom}, {van den Brand}, Van Den~Broeck, {Vander-Hyde}, {van der Schaaf}, {van Heijningen}, {van Veggel}, Vardaro,
  Varma, Vass, Vas{\'u}th, Vecchio, Vedovato, Veitch, Veitch, Venkateswara, Venugopalan, Verkindt, Vetrano, Vicer{\'e}, Viets, Vinciguerra, Vine, Vinet, Vitale, Vo, Vocca, Vorvick, Vyatchanin, Wade, Wade, Wade, Walet, Walker, Wallace, Walsh, Wang, Wang, Wang, Wang, Wang, Ward, Warner, Was, Watchi, Weaver, Wei, Weinert, Weinstein, Weiss, Wen, Wessel, We{\ss}els, Westerweck, Westphal, Wette, Whelan, Whitcomb, Whiting, Whittle, Wilken, Williams, Williams, Williamson, Willis, Willke, Wimmer, Winkler, Wipf, Wittel, Woan, Woehler, Wofford, Wong, Worden, Wright, Wu, Wysocki, Xiao, Yamamoto, Yancey, Yang, Yap, Yazback, Yu, Yu, Yvert, Zadro{\.z}ny, Zanolin, Zelenova, Zendri, Zevin, Zhang, Zhang, Zhang, Zhang, Zhao, Zhou, Zhou, Zhu, Zhu, Zimmerman, Zucker, \& Zweizig}]{abbott_gw170817:_2017}
Abbott, B.~P., Abbott, R., Abbott, T.~D., {et~al.} 2017{\natexlab{a}}, Physical Review Letters, 119, 161101, \dodoi{10.1103/PhysRevLett.119.161101}

\bibitem[{Abbott {et~al.}(2017{\natexlab{b}})Abbott, Abbott, Abbott, Acernese, Ackley, Adams, Adams, Addesso, Adhikari, Adya, Affeldt, Afrough, Agarwal, Agathos, Agatsuma, Aggarwal, Aguiar, Aiello, Ain, Ajith, Allen, Allen, Allocca, Altin, Amato, Ananyeva, Anderson, Anderson, Angelova, Antier, Appert, Arai, Araya, Areeda, Arnaud, Arun, Ascenzi, Ashton, Ast, Aston, Astone, Atallah, Aufmuth, Aulbert, AultONeal, Austin, {Avila-Alvarez}, Babak, Bacon, Bader, Bae, Baker, Baldaccini, Ballardin, Ballmer, Banagiri, Barayoga, Barclay, Barish, Barker, Barkett, Barone, Barr, Barsotti, Barsuglia, Barta, Barthelmy, Bartlett, Bartos, Bassiri, Basti, Batch, Bawaj, Bayley, Bazzan, B{\'e}csy, Beer, Bejger, Belahcene, Bell, Berger, Bergmann, Bero, Berry, Bersanetti, Bertolini, Betzwieser, Bhagwat, Bhandare, Bilenko, Billingsley, Billman, Birch, Birney, Birnholtz, Biscans, Biscoveanu, Bisht, Bitossi, Biwer, Bizouard, Blackburn, Blackman, Blair, Blair, Blair, Bloemen, Bock, Bode, Boer, Bogaert, Bohe, Bondu, Bonilla, Bonnand,
  Boom, Bork, Boschi, Bose, Bossie, Bouffanais, Bozzi, Bradaschia, Brady, Branchesi, Brau, Briant, Brillet, Brinkmann, Brisson, Brockill, Broida, Brooks, Brown, Brown, Brunett, Buchanan, Buikema, Bulik, Bulten, Buonanno, Buskulic, Buy, Byer, Cabero, Cadonati, Cagnoli, Cahillane, Bustillo, Callister, Calloni, Camp, Canepa, Canizares, Cannon, Cao, Cao, Capano, Capocasa, Carbognani, Caride, Carney, Diaz, Casentini, Caudill, Cavagli{\`a}, Cavalier, Cavalieri, Cella, Cepeda, {Cerd{\'a}-Dur{\'a}n}, Cerretani, Cesarini, Chamberlin, Chan, Chao, Charlton, Chase, {Chassande-Mottin}, Chatterjee, Chatziioannou, Cheeseboro, Chen, Chen, Chen, Cheng, Chia, Chincarini, Chiummo, Chmiel, Cho, Cho, Chow, Christensen, Chu, Chua, Chua, Chung, Chung, Ciani, Ciolfi, Cirelli, Cirone, Clara, Clark, Clearwater, Cleva, Cocchieri, Coccia, Cohadon, Cohen, Colla, Collette, Cominsky, Jr, Conti, Cooper, Corban, Corbitt, {Cordero-Carri{\'o}n}, Corley, Cornish, Corsi, Cortese, Costa, Coughlin, Coughlin, Coulon, Countryman, Couvares, Covas,
  Cowan, Coward, Cowart, Coyne, Coyne, Creighton, Creighton, Cripe, Crowder, Cullen, Cumming, Cunningham, Cuoco, Canton, D{\'a}lya, Danilishin, D'Antonio, Danzmann, Dasgupta, Costa, Dattilo, Dave, Davier, Davis, Daw, Day, De, DeBra, Degallaix, Laurentis, Del{\'e}glise, Pozzo, Demos, Denker, Dent, Pietri, Dergachev, Rosa, DeRosa, Rossi, DeSalvo, de~Varona, Devenson, Dhurandhar, D{\'i}az, Fiore, Giovanni, Girolamo, Lieto, Pace, Palma, Renzo, Doctor, Dolique, Donovan, Dooley, Doravari, Dorrington, Douglas, {\'A}lvarez, Downes, Drago, Dreissigacker, Driggers, Du, Ducrot, Dupej, Dwyer, Edo, Edwards, Effler, Ehrens, Eichholz, Eikenberry, Eisenstein, Essick, Estevez, Etienne, Etzel, Evans, Evans, Factourovich, Fafone, Fair, Fairhurst, Fan, Farinon, Farr, Farr, {Fauchon-Jones}, Favata, Fays, Fee, Fehrmann, Feicht, Fejer, {Fernandez-Galiana}, Ferrante, Ferreira, Ferrini, Fidecaro, Finstad, Fiori, Fiorucci, Fishbach, Fisher, {Fitz-Axen}, Flaminio, Fletcher, Fong, Font, Forsyth, Forsyth, Fournier, Frasca, Frasconi,
  Frei, Freise, Frey, Frey, Fries, Fritschel, Frolov, Fulda, Fyffe, Gabbard, Gadre, Gaebel, Gair, Gammaitoni, Ganija, Gaonkar, {Garcia-Quiros}, Garufi, Gateley, Gaudio, Gaur, Gayathri, Gehrels, Gemme, Genin, Gennai, George, George, Gergely, Germain, Ghonge, Ghosh, Ghosh, Ghosh, Giaime, Giardina, Giazotto, Gill, Glover, Goetz, Goetz, Gomes, Goncharov, Gonz{\'a}lez, Castro, Gopakumar, Gorodetsky, Gossan, Gosselin, Gouaty, Grado, Graef, Granata, Grant, Gras, Gray, Greco, Green, Gretarsson, Griswold, Groot, Grote, Grunewald, Gruning, Guidi, Guo, Gupta, Gupta, Gushwa, Gustafson, Gustafson, Halim, Hall, Hall, Hamilton, Hammond, Haney, Hanke, Hanks, Hanna, Hannam, Hannuksela, Hanson, Hardwick, Harms, Harry, Harry, Hart, Haster, Haughian, Healy, Heidmann, Heintze, Heitmann, Hello, Hemming, Hendry, Heng, Hennig, Heptonstall, Heurs, Hild, Hinderer, Hoak, Hofman, Holt, Holz, Hopkins, Horst, Hough, Houston, Howell, Hreibi, Hu, Huerta, Huet, Hughey, Husa, Huttner, {Huynh-Dinh}, Indik, Inta, Intini, Isa, Isac, Isi, Iyer,
  Izumi, Jacqmin, Jani, Jaranowski, Jawahar, {Jim{\'e}nez-Forteza}, Johnson, Jones, Jones, Jonker, Ju, Junker, Kalaghatgi, Kalogera, Kamai, Kandhasamy, Kang, Kanner, Kapadia, Karki, Karvinen, Kasprzack, Katolik, Katsavounidis, Katzman, Kaufer, Kawabe, K{\'e}f{\'e}lian, Keitel, Kemball, Kennedy, Kent, Key, Khalili, Khan, Khan, Khan, Khazanov, Kijbunchoo, Kim, Kim, Kim, Kim, Kim, Kim, Kimbrell, King, King, {Kinley-Hanlon}, Kirchhoff, Kissel, Kleybolte, Klimenko, Knowles, Koch, Koehlenbeck, Koley, Kondrashov, Kontos, Korobko, Korth, Kowalska, Kozak, Kr{\"a}mer, Kringel, Krishnan, Kr{\'o}lak, Kuehn, Kumar, Kumar, Kumar, Kuo, Kutynia, Kwang, Lackey, Lai, Landry, Lang, Lange, Lantz, Lanza, Larson, {Lartaux-Vollard}, Lasky, Laxen, Lazzarini, Lazzaro, Leaci, Leavey, Lee, Lee, Lee, Lee, Lee, Lehmann, Lenon, Leonardi, Leroy, Letendre, Levin, Li, Linker, Littenberg, Liu, Lo, Lockerbie, London, Lord, Lorenzini, Loriette, Lormand, Losurdo, Lough, Lousto, Lovelace, L{\"u}ck, Lumaca, Lundgren, Lynch, Ma, Macas, Macfoy,
  Machenschalk, MacInnis, Macleod, Hernandez, {Maga{\~n}a-Sandoval}, Zertuche, Magee, Majorana, Maksimovic, Man, Mandic, Mangano, Mansell, Manske, Mantovani, Marchesoni, Marion, M{\'a}rka, M{\'a}rka, Markakis, Markosyan, Markowitz, Maros, Marquina, Marsh, Martelli, Martellini, Martin, Martin, Martynov, Mason, Massera, Masserot, Massinger, {Masso-Reid}, Mastrogiovanni, Matas, Matichard, Matone, Mavalvala, Mazumder, McCarthy, McClelland, McCormick, McCuller, McGuire, McIntyre, McIver, McManus, McNeill, McRae, McWilliams, Meacher, Meadors, Mehmet, Meidam, {Mejuto-Villa}, Melatos, Mendell, Mercer, Merilh, Merzougui, Meshkov, Messenger, Messick, Metzdorff, Meyers, Miao, Michel, Middleton, Mikhailov, Milano, Miller, Miller, Miller, Millhouse, {Milovich-Goff}, Minazzoli, Minenkov, Ming, Mishra, Mitra, Mitrofanov, Mitselmakher, Mittleman, Moffa, Moggi, Mogushi, Mohan, Mohapatra, Montani, Moore, Moraru, Moreno, Morriss, Mours, {Mow-Lowry}, Mueller, Muir, Mukherjee, Mukherjee, Mukherjee, Mukund, Mullavey, Munch,
  Mu{\~n}iz, Muratore, Murray, Napier, Nardecchia, Naticchioni, Nayak, Neilson, Nelemans, Nelson, Nery, Neunzert, Nevin, Newport, Newton, Ng, Nguyen, Nguyen, Nichols, Nielsen, Nissanke, Nitz, Noack, Nocera, Nolting, North, Nuttall, Oberling, O'Dea, Ogin, Oh, Oh, Ohme, Okada, Oliver, Oppermann, Oram, O'Reilly, Ormiston, Ortega, O'Shaughnessy, Ossokine, Ottaway, Overmier, Owen, Pace, Page, Page, Pai, Pai, Palamos, Palashov, Palomba, {Pal-Singh}, Pan, Pan, Pang, Pang, Pankow, Pannarale, Pant, Paoletti, Paoli, Papa, Parida, Parker, Pascucci, Pasqualetti, Passaquieti, Passuello, Patil, Patricelli, Pearlstone, Pedraza, Pedurand, Pekowsky, Pele, Penn, Perez, Perreca, Perri, Pfeiffer, Phelps, Piccinni, Pichot, Piergiovanni, Pierro, Pillant, Pinard, Pinto, Pirello, Pitkin, Poe, Poggiani, Popolizio, Porter, Post, Powell, Prasad, Pratt, Pratten, Predoi, Prestegard, Price, Prijatelj, Principe, Privitera, Prodi, Prokhorov, Puncken, Punturo, Puppo, P{\"u}rrer, Qi, Quetschke, Quintero, {Quitzow-James}, Raab, Rabeling,
  Radkins, Raffai, Raja, Rajan, Rajbhandari, Rakhmanov, Ramirez, {Ramos-Buades}, Rapagnani, Raymond, Razzano, Read, Regimbau, Rei, Reid, Reitze, Ren, Reyes, Ricci, Ricker, Rieger, Riles, Rizzo, Robertson, Robie, Robinet, Rocchi, Rolland, Rollins, Roma, Romano, Romel, Romie, Rosi{\'n}ska, Ross, Rowan, R{\"u}diger, Ruggi, Rutins, Ryan, Sachdev, Sadecki, Sadeghian, Sakellariadou, Salconi, Saleem, Salemi, Samajdar, Sammut, Sampson, Sanchez, Sanchez, {Sanchis-Gual}, Sandberg, Sanders, Sassolas, Sathyaprakash, Saulson, Sauter, Savage, Sawadsky, Schale, Scheel, Scheuer, Schmidt, Schmidt, Schnabel, Schofield, Sch{\"o}nbeck, Schreiber, Schuette, Schulte, Schutz, Schwalbe, Scott, Scott, Seidel, Sellers, Sengupta, Sentenac, Sequino, Sergeev, Shaddock, Shaffer, Shah, Shahriar, Shaner, Shao, Shapiro, Shawhan, Sheperd, Shoemaker, Shoemaker, Siellez, Siemens, Sieniawska, Sigg, Silva, Singer, Singh, Singhal, Sintes, Slagmolen, Smith, Smith, Smith, Somala, Son, Sonnenberg, Sorazu, Sorrentino, Souradeep, Spencer, Srivastava,
  Staats, Staley, Steinke, Steinlechner, Steinlechner, Steinmeyer, Stevenson, Stone, Stops, Strain, Stratta, Strigin, Strunk, Sturani, Stuver, Summerscales, Sun, Sunil, Suresh, Sutton, Swinkels, Szczepa{\'n}czyk, Tacca, Tait, Talbot, Talukder, Tanner, T{\'a}pai, Taracchini, Tasson, Taylor, Taylor, Tewari, Theeg, Thies, Thomas, Thomas, Thomas, Thorne, Thorne, Thrane, Tiwari, Tiwari, Tokmakov, Toland, Tonelli, Tornasi, {Torres-Forn{\'e}}, Torrie, T{\"o}yr{\"a}, Travasso, Traylor, Trinastic, Tringali, Trozzo, Tsang, Tse, Tso, Tsukada, Tsuna, Tuyenbayev, Ueno, Ugolini, Unnikrishnan, Urban, Usman, Vahlbruch, Vajente, Valdes, van Bakel, van Beuzekom, van~den Brand, Broeck, {Vander-Hyde}, van~der Schaaf, van Heijningen, van Veggel, Vardaro, Varma, Vass, Vas{\'u}th, Vecchio, Vedovato, Veitch, Veitch, Venkateswara, Venugopalan, Verkindt, Vetrano, Vicer{\'e}, Viets, Vinciguerra, Vine, Vinet, Vitale, Vo, Vocca, Vorvick, Vyatchanin, Wade, Wade, Wade, Walet, Walker, Wallace, Walsh, Wang, Wang, Wang, Wang, Wang, Ward,
  Warner, Was, Watchi, Weaver, Wei, Weinert, Weinstein, Weiss, Wen, Wessel, Wessels, Westerweck, Westphal, Wette, Whelan, Whitcomb, Whiting, Whittle, Wilken, Williams, Williams, Williamson, Willis, Willke, Wimmer, Winkler, Wipf, Wittel, Woan, Woehler, Wofford, Wong, Worden, Wright, Wu, Wysocki, Xiao, Yamamoto, Yancey, Yang, Yap, Yazback, Yu, Yu, Yvert, Zadro{\textbackslash}.zny, Zanolin, Zelenova, Zendri, Zevin, Zhang, Zhang, Zhang, Zhang, Zhao, Zhou, Zhou, Zhu, Zhu, Zimmerman, Zucker, Zweizig, {Wilson-Hodge}, Bissaldi, Blackburn, Briggs, Burns, Cleveland, Connaughton, Gibby, Giles, Goldstein, Hamburg, Jenke, Hui, Kippen, Kocevski, McBreen, Meegan, Paciesas, Poolakkil, Preece, Racusin, Roberts, Stanbro, Veres, von Kienlin, Savchenko, Ferrigno, Kuulkers, Bazzano, Bozzo, Brandt, Chenevez, Courvoisier, Diehl, Domingo, Hanlon, Jourdain, Laurent, Lebrun, Lutovinov, {Martin-Carrillo}, Mereghetti, Natalucci, Rodi, Roques, Sunyaev, Ubertini, Aartsen, Ackermann, Adams, Aguilar, Ahlers, Ahrens, Samarai, Altmann,
  Andeen, Anderson, Ansseau, Anton, Arg{\"u}elles, Auffenberg, Axani, Bagherpour, Bai, Barron, Barwick, Baum, Bay, Beatty, Tjus, Bernardini, Besson, Binder, Bindig, Blaufuss, Blot, Bohm, B{\"o}rner, Bos, Bose, B{\"o}ser, Botner, Bourbeau, Bourbeau, Bradascio, Braun, Brayeur, Brenzke, Bretz, Bron, {Brostean-Kaiser}, Burgman, Carver, Casey, Casier, Cheung, Chirkin, Christov, Clark, Classen, Coenders, Collin, Conrad, Cowen, Cross, Day, de~Andr{\'e}, Clercq, DeLaunay, Dembinski, Ridder, Desiati, de~Vries, de~Wasseige, de~With, DeYoung, {D{\'i}az-V{\'e}lez}, di~Lorenzo, Dujmovic, Dumm, Dunkman, Dvorak, Eberhardt, Ehrhardt, Eichmann, Eller, Evenson, Fahey, Fazely, Felde, Filimonov, Finley, Flis, Franckowiak, Friedman, Fuchs, Gaisser, Gallagher, Gerhardt, Ghorbani, Giang, Glauch, Gl{\"u}senkamp, Goldschmidt, Gonzalez, Grant, Griffith, Haack, Hallgren, Halzen, Hanson, Hebecker, Heereman, Helbing, Hellauer, Hickford, Hignight, Hill, Hoffman, Hoffmann, {Hokanson-Fasig}, Hoshina, Huang, Huber, Hultqvist, H{\"u}nnefeld,
  In, Ishihara, Jacobi, Japaridze, Jeong, Jero, Jones, Kalaczynski, Kang, Kappes, Karg, Karle, Kauer, Keivani, Kelley, Kheirandish, Kim, Kim, Kintscher, Kiryluk, Kittler, Klein, Kohnen, Koirala, Kolanoski, K{\"o}pke, Kopper, Kopper, Koschinsky, Koskinen, Kowalski, Krings, Kroll, Kr{\"u}ckl, Kunnen, Kunwar, Kurahashi, Kuwabara, Kyriacou, Labare, Lanfranchi, Larson, Lauber, {Lesiak-Bzdak}, Leuermann, Liu, Lu, L{\"u}nemann, Luszczak, Madsen, Maggi, Mahn, Mancina, Maruyama, Mase, Maunu, McNally, Meagher, Medici, Meier, Menne, Merino, Meures, Miarecki, Micallef, Moment{\'e}, Montaruli, Moore, Moulai, Nahnhauer, Nakarmi, Naumann, Neer, Niederhausen, Nowicki, Nygren, Pollmann, Olivas, O'Murchadha, Palczewski, Pandya, Pankova, Peiffer, Pepper, de~los Heros, Pieloth, Pinat, Price, Przybylski, Raab, R{\"a}del, Rameez, Rawlins, Rea, Reimann, Relethford, Relich, Resconi, Rhode, Richman, Robertson, Rongen, Rott, Ruhe, Ryckbosch, Rysewyk, S{\"a}lzer, Herrera, Sandrock, Sandroos, Santander, Sarkar, Sarkar, Satalecka,
  Schlunder, Schmidt, Schneider, Schoenen, Sch{\"o}neberg, Schumacher, Seckel, Seunarine, Soedingrekso, Soldin, Song, Spiczak, Spiering, Stachurska, Stamatikos, Stanev, Stasik, Stettner, Steuer, Stezelberger, Stokstad, St{\"o}ssl, Strotjohann, Stuttard, Sullivan, Sutherland, Taboada, Tatar, Tenholt, {Ter-Antonyan}, Terliuk, Te{\v s}i{\'c}, Tilav, Toale, Tobin, Toscano, Tosi, Tselengidou, Tung, Turcati, Turley, Ty, Unger, Usner, Vandenbroucke, Driessche, van Eijndhoven, Vanheule, van Santen, Vehring, Vogel, Vraeghe, Walck, Wallace, Wallraff, Wandler, Wandkowsky, Waza, Weaver, Weiss, Wendt, Werthebach, Whelan, Wiebe, Wiebusch, Wille, Williams, Wills, Wolf, Wood, Woolsey, Woschnagg, Xu, Xu, Xu, Yanez, Yodh, Yoshida, Yuan, Zoll, Balasubramanian, Mate, Bhalerao, Bhattacharya, Vibhute, Dewangan, Rao, Vadawale, Svinkin, Hurley, Aptekar, Frederiks, Golenetskii, Kozlova, Lysenko, Oleynik, Tsvetkova, Ulanov, Cline, Li, Xiong, Zhang, Lu, Song, Cao, Chang, Chen, Chen, Chen, Chen, Chen, Chen, Cui, Cui, Deng, Dong, Du, Fu,
  Gao, Gao, Gao, Ge, Gu, Guan, Guo, Han, Hu, Huang, Huo, Jia, Jiang, Jiang, Jin, Jin, Li, Li, Li, Li, Li, Li, Li, Li, Li, Li, Li, Liang, Liao, Liu, Liu, Liu, Liu, Liu, Liu, Liu, Lu, Lu, Luo, Ma, Meng, Nang, Nie, Ou, Qu, Sai, Sun, Tan, Tao, Tao, Tuo, Wang, Wang, Wang, Wang, Wang, Wen, Wu, Wu, Xiao, Xu, Xu, Yan, Yang, Yang, Yang, Zhang, Zhang, Zhang, Zhang, Zhang, Zhang, Zhang, Zhang, Zhang, Zhang, Zhang, Zhang, Zhang, Zhang, Zhang, Zhang, Zhang, Zhang, Zhao, Zhao, Zhao, Zheng, Zhu, Zhu, Zou, Albert, Andr{\'e}, Anghinolfi, Ardid, Aubert, Aublin, Avgitas, Baret, {Barrios-Mart{\'i}}, Basa, Belhorma, Bertin, Biagi, Bormuth, Bourret, Bouwhuis, Br{\^a}nza{\c s}, Bruijn, Brunner, Busto, Capone, Caramete, Carr, Celli, Moursli, Chiarusi, Circella, Coelho, Coleiro, Coniglione, Costantini, Coyle, Creusot, D{\'i}az, Deschamps, Bonis, Distefano, Palma, Domi, Donzaud, Dornic, Drouhin, Eberl, Bojaddaini, Khayati, Els{\"a}sser, Enzenh{\"o}fer, Ettahiri, Fassi, Felis, Fusco, Gay, Giordano, Glotin, Gr{\'e}goire, Ruiz, Graf,
  Hallmann, van Haren, Heijboer, Hello, {Hern{\'a}ndez-Rey}, H{\"o}ssl, Hofest{\"a}dt, Hugon, Illuminati, James, de~Jong, Jongen, Kadler, Kalekin, Katz, Kiessling, Kouchner, Kreter, Kreykenbohm, Kulikovskiy, Lachaud, Lahmann, Lef{\`e}vre, Leonora, Lotze, Loucatos, Marcelin, Margiotta, Marinelli, {Mart{\'i}nez-Mora}, Mele, Melis, Michael, Migliozzi, Moussa, Navas, Nezri, Organokov, P{\u a}v{\u a}la{\c s}, Pellegrino, Perrina, Piattelli, Popa, Pradier, Quinn, Racca, Riccobene, {S{\'a}nchez-Losa}, Salda{\~n}a, Salvadori, Samtleben, Sanguineti, Sapienza, Sieger, Spurio, Stolarczyk, Taiuti, Tayalati, Trovato, Turpin, T{\"o}nnis, Vallage, Elewyck, Versari, Vivolo, Vizzoca, Wilms, Zornoza, Z{\'u}{\~n}iga, Beardmore, Breeveld, Burrows, Cenko, Cusumano, D'A{\`i}, de~Pasquale, Emery, Evans, Giommi, Gronwall, Kennea, Krimm, Kuin, Lien, Marshall, Melandri, Nousek, Oates, Osborne, Pagani, Page, Palmer, Perri, Siegel, Sbarufatti, Tagliaferri, Tohuvavohu, Tavani, Verrecchia, Bulgarelli, Evangelista, Pacciani, Feroci,
  Pittori, Giuliani, Monte, Donnarumma, Argan, Trois, Ursi, Cardillo, Piano, Longo, Lucarelli, {Munar-Adrover}, Fuschino, Labanti, Marisaldi, Minervini, Fioretti, Parmiggiani, Gianotti, Trifoglio, Persio, Antonelli, Barbiellini, Caraveo, Cattaneo, Costa, Colafrancesco, D'Amico, Ferrari, Morselli, Paoletti, Picozza, Pilia, Rappoldi, Soffitta, Vercellone, Foley, Coulter, Kilpatrick, Drout, Piro, Shappee, Siebert, Simon, Ulloa, Kasen, Madore, {Murguia-Berthier}, Pan, Prochaska, {Ramirez-Ruiz}, Rest, {Rojas-Bravo}, Berger, {Soares-Santos}, Annis, Alexander, Allam, Balbinot, Blanchard, Brout, Butler, Chornock, Cook, Cowperthwaite, Diehl, {Drlica-Wagner}, Drout, Durret, Eftekhari, Finley, Fong, Frieman, Fryer, {Garc{\'i}a-Bellido}, Gruendl, Hartley, Herner, Kessler, Lin, Lopes, Louren{\c c}o, Margutti, Marshall, Matheson, Medina, Metzger, Mu{\~n}oz, Muir, Nicholl, Nugent, Palmese, {Paz-Chinch{\'o}n}, Quataert, Sako, Sauseda, Schlegel, Scolnic, Secco, Smith, Sobreira, Villar, Vivas, Wester, Williams, Yanny, Zenteno,
  Zhang, Abbott, Banerji, Bechtol, {Benoit-L{\'e}vy}, Bertin, Brooks, {Buckley-Geer}, Burke, Capozzi, Rosell, Kind, Castander, Crocce, Cunha, D'Andrea, da~Costa, Davis, DePoy, Desai, Dietrich, Eifler, Fernandez, Flaugher, Fosalba, Gaztanaga, Gerdes, Giannantonio, Goldstein, Gruen, Gschwend, Gutierrez, Honscheid, James, Jeltema, Johnson, Johnson, Kent, Krause, Kron, Kuehn, Lahav, Lima, Maia, March, Martini, McMahon, Menanteau, Miller, Miquel, Mohr, Nichol, Ogando, Plazas, Romer, Roodman, Rykoff, Sanchez, Scarpine, Schindler, Schubnell, {Sevilla-Noarbe}, Sheldon, Smith, Smith, Stebbins, Suchyta, Swanson, Tarle, Thomas, Troxel, Tucker, Vikram, Walker, Wechsler, Weller, Carlin, Gill, Li, Marriner, Neilsen, Haislip, Kouprianov, Reichart, Sand, Tartaglia, Valenti, Yang, Benetti, Brocato, Campana, Cappellaro, Covino, D'Avanzo, D'Elia, Getman, Ghirlanda, Ghisellini, Limatola, Nicastro, Palazzi, Pian, Piranomonte, Possenti, Rossi, Salafia, Tomasella, Amati, Antonelli, Bernardini, Bufano, Capaccioli, Casella, Dadina,
  Cesare, Paola, Giuffrida, Giunta, Israel, Lisi, Maiorano, Mapelli, Masetti, Pescalli, Pulone, Salvaterra, Schipani, Spera, Stamerra, Stella, Testa, Turatto, Vergani, Aresu, Bachetti, Buffa, Burgay, Buttu, Caria, Carretti, Casasola, Castangia, Carboni, Casu, Concu, Corongiu, Deiana, Egron, Fara, Gaudiomonte, Gusai, Ladu, Loru, Leurini, Marongiu, Melis, Melis, Migoni, Milia, Navarrini, Orlati, Ortu, Palmas, Pellizzoni, Perrodin, Pisanu, Poppi, Righini, Saba, Serra, Serrau, Stagni, Surcis, Vacca, Vargiu, Hunt, Jin, Klose, Kouveliotou, Mazzali, M{\o}ller, Nava, Piran, Selsing, Vergani, Wiersema, Toma, Higgins, Mundell, Alighieri, G{\'o}tz, Gao, Gomboc, Kaper, Kobayashi, Kopac, Mao, Starling, Steele, van~der Horst, Acero, Atwood, Baldini, Barbiellini, Bastieri, Berenji, Bellazzini, Bissaldi, Blandford, Bloom, Bonino, Bottacini, Bregeon, Buehler, Buson, Cameron, Caputo, Caraveo, Cavazzuti, Chekhtman, Cheung, Chiang, Ciprini, {Cohen-Tanugi}, Cominsky, Costantin, Cuoco, D'Ammando, de~Palma, Digel, Lalla, Mauro,
  Venere, Dubois, Fegan, Focke, Franckowiak, Fukazawa, Funk, Fusco, Gargano, Gasparrini, Giglietto, Giordano, Giroletti, Glanzman, Green, Grondin, Guillemot, Guiriec, Harding, Horan, J{\'o}hannesson, Kamae, Kensei, Kuss, Mura, Latronico, {Lemoine-Goumard}, Longo, Loparco, Lovellette, Lubrano, Magill, Maldera, Manfreda, Mazziotta, McEnery, Meyer, Michelson, Mirabal, Monzani, Moretti, Morselli, Moskalenko, Negro, Nuss, Ojha, Omodei, Orienti, Orlando, Palatiello, Paliya, Paneque, {Pesce-Rollins}, Piron, Porter, Principe, Rain{\`o}, Rando, Razzano, Razzaque, Reimer, Reimer, Reposeur, Rochester, Parkinson, Sgr{\`o}, Siskind, Spada, Spandre, Suson, Takahashi, Tanaka, Thayer, Thayer, Thompson, Tibaldo, Torres, Torresi, Troja, Venters, Vianello, Zaharijas, Allison, Bannister, Dobie, Kaplan, Lenc, Lynch, Murphy, Sadler, Hotan, James, Oslowski, Raja, Shannon, Whiting, Arcavi, Howell, McCully, Hosseinzadeh, Hiramatsu, Poznanski, Barnes, Zaltzman, Vasylyev, Maoz, Cooke, Bailes, Wolf, Deller, Lidman, Wang, Gendre,
  Andreoni, Ackley, Pritchard, Bessell, Chang, M{\"o}ller, Onken, Scalzo, {Ridden-Harper}, Sharp, Tucker, Farrell, Elmer, Johnston, Krishnan, Keane, Green, Jameson, Hu, Ma, Sun, Wu, Wang, Shang, Hu, Ashley, Yuan, Li, Tao, Zhu, Zhang, Suntzeff, Zhou, Yang, Orange, Morris, Cucchiara, Giblin, Klotz, Staff, Thierry, Schmidt, Tanvir, Levan, Cano, de~{Ugarte-Postigo}, {Gonz{\'a}lez-Fern{\'a}ndez}, Greiner, Hjorth, Irwin, Kr{\"u}hler, Mandel, {Milvang-Jensen}, O'Brien, Rol, Rosetti, Rosswog, Rowlinson, Steeghs, Th{\"o}ne, Ulaczyk, Watson, Bruun, Cutter, Jaimes, Fujii, Fruchter, Gompertz, Jakobsson, Hodosan, J{\`e}rgensen, Kangas, Kann, Rabus, Schr{\o}der, Stanway, Wijers, Lipunov, Gorbovskoy, Kornilov, Tyurina, Balanutsa, Kuznetsov, Vlasenko, Podesta, Lopez, Podesta, Levato, Saffe, Mallamaci, Budnev, Gress, Kuvshinov, Gorbunov, Vladimirov, Zimnukhov, Gabovich, Yurkov, Sergienko, Rebolo, {Serra-Ricart}, Tlatov, Ishmuhametova, Abe, Aoki, Aoki, Asakura, Baar, Barway, Bond, Doi, Finet, Fujiyoshi, Furusawa, Honda, Itoh,
  Kanda, Kawabata, Kawabata, Kim, Koshida, Kuroda, Lee, Liu, Matsubayashi, Miyazaki, Morihana, Morokuma, Motohara, Murata, Nagai, Nagashima, Nagayama, Nakaoka, Nakata, Ohsawa, Ohshima, Ohta, Okita, Saito, Saito, Sako, Sekiguchi, Sumi, Tajitsu, Takahashi, Takayama, Tamura, Tanaka, Tanaka, Terai, Tominaga, Tristram, Uemura, Utsumi, Yamaguchi, Yasuda, Yoshida, Zenko, Adams, Anupama, Bally, Barway, Bellm, Blagorodnova, Cannella, Chandra, Chatterjee, Clarke, Cobb, Cook, Copperwheat, De, Emery, Feindt, Foster, Fox, Frail, Fremling, Frohmaier, Garcia, Ghosh, Giacintucci, Goobar, Gottlieb, Grefenstette, Hallinan, Harrison, Heida, Helou, Ho, Horesh, Hotokezaka, Ip, Itoh, Jacobs, Jencson, Kasen, Kasliwal, Kassim, Kim, Kiran, Kuin, Kulkarni, Kupfer, Lau, Madsen, Mazzali, Miller, Miyasaka, Mooley, Myers, Nakar, Ngeow, Nugent, Ofek, Palliyaguru, Pavana, Perley, Peters, Pike, Piran, Qi, Quimby, Rana, Rosswog, Rusu, Sadler, Sistine, Sollerman, Xu, Yan, Yatsu, Yu, Zhang, Zhao, Chambers, Huber, Schultz, Bulger, Flewelling,
  Magnier, Lowe, Wainscoat, Waters, Willman, Ebisawa, Hanyu, Harita, Hashimoto, Hidaka, Hori, Ishikawa, Isobe, Iwakiri, Kawai, Kawai, Kawamuro, Kawase, Kitaoka, Makishima, Matsuoka, Mihara, Morita, Morita, Nakahira, Nakajima, Nakamura, Negoro, Oda, Sakamaki, Sasaki, Serino, Shidatsu, Shimomukai, Sugawara, Sugita, Sugizaki, Tachibana, Takao, Tanimoto, Tomida, Tsuboi, Tsunemi, Ueda, Ueno, Yamada, Yamaoka, Yamauchi, Yatabe, Yoneyama, Yoshii, Coward, Crisp, Macpherson, Andreoni, Laugier, Noysena, Klotz, Gendre, Thierry, Turpin, Im, Choi, Kim, Yoon, Lim, Lee, Lee, Kim, Ko, Joe, Kwon, Kim, Lim, Choi, Fynbo, Malesani, Xu, Smartt, Jerkstrand, Kankare, Sim, Fraser, Inserra, Maguire, Leloudas, Magee, Shingles, Smith, Young, Kotak, {Gal-Yam}, Lyman, Homan, Agliozzo, Anderson, Angus, Ashall, Barbarino, Bauer, Berton, Botticella, Bulla, Cannizzaro, Cartier, Cikota, Clark, Cia, Valle, Dennefeld, Dessart, Dimitriadis, {Elias-Rosa}, Firth, Fl{\"o}rs, Frohmaier, Galbany, {Gonz{\'a}lez-Gait{\'a}n}, Gromadzki, Guti{\'e}rrez,
  Hamanowicz, Harmanen, Heintz, Hernandez, Hodgkin, Hook, Izzo, James, Jonker, Kerzendorf, {Kostrzewa-Rutkowska}, Kromer, Kuncarayakti, Lawrence, Manulis, Mattila, McBrien, M{\"u}ller, Nordin, O'Neill, Onori, Palmerio, Pastorello, Patat, Pignata, Podsiadlowski, Razza, Reynolds, Roy, Ruiter, Rybicki, Salmon, Pumo, Prentice, Seitenzahl, Smith, Sollerman, Sullivan, Szegedi, Taddia, Taubenberger, Terreran, Soelen, Vos, Walton, Wright, Wyrzykowski, Yaron, Chen, Kr{\"u}hler, Schady, Wiseman, Greiner, Rau, Schweyer, Klose, Guelbenzu, Palliyaguru, Shara, Williams, Vaisanen, Potter, Colmenero, Crawford, Buckley, Mao, D{\'i}az, Macri, Lambas, de~Oliveira, Castell{\'o}n, Ribeiro, S{\'a}nchez, Schoenell, Abramo, Akras, Alcaniz, Artola, Beroiz, Bonoli, Cabral, Camuccio, Chavushyan, Coelho, Colazo, {Costa-Duarte}, Larenas, Romero, Dultzin, Fern{\'a}ndez, Garc{\'i}a, Girardini, Gon{\c c}alves, Gon{\c c}alves, Gurovich, {Jim{\'e}nez-Teja}, Kanaan, Lares, de~Oliveira, {L{\'o}pez-Cruz}, Melia, Molino, Padilla, Pe{\~n}uela,
  Placco, Qui{\~n}ones, Rivera, Renzi, Riguccini, {R{\'i}os-L{\'o}pez}, Rodriguez, Sampedro, Schneiter, Sodr{\'e}, Starck, {Torres-Flores}, Tornatore, Zadro{\textbackslash}.zny, Castillo, {Castro-Tirado}, Tello, Hu, Zhang, Cunniffe, Castell{\'o}n, Hiriart, {Caballero-Garc{\'i}a}, Jel{\'i}nek, Kub{\'a}nek, del Pulgar, Park, Jeong, Cer{\'o}n, Pandey, Yock, Querel, Fan, Wang, Beardsley, Brown, Crosse, Emrich, Franzen, Gaensler, Horsley, {Johnston-Hollitt}, Kenney, Morales, Pallot, Sokolowski, Steele, Tingay, Trott, Walker, Wayth, Williams, Wu, Yoshida, Sakamoto, Kawakubo, Yamaoka, Takahashi, Asaoka, Ozawa, Torii, Shimizu, Tamura, Ishizaki, Cherry, Ricciarini, Penacchioni, Marrocchesi, Pozanenko, Volnova, Mazaeva, Minaev, Krugov, Kusakin, Reva, Moskvitin, Rumyantsev, Inasaridze, Klunko, Tungalag, Schmalz, Burhonov, Abdalla, Abramowski, Aharonian, Benkhali, Ang{\"u}ner, Arakawa, Arrieta, Aubert, Backes, Balzer, Barnard, Becherini, Tjus, Berge, Bernhard, Bernl{\"o}hr, Blackwell, B{\"o}ttcher, Boisson, Bolmont,
  Bonnefoy, Bordas, Bregeon, Brun, Brun, Bryan, B{\"u}chele, Bulik, Capasso, Caroff, Carosi, Casanova, Cerruti, Chakraborty, Chaves, Chen, Chevalier, Colafrancesco, Condon, Conrad, Davids, Decock, Deil, Devin, {deWilt}, Dirson, {Djannati-Ata{\"i}}, Donath, Drury, Dutson, Dyks, Edwards, Egberts, Emery, Ernenwein, Eschbach, Farnier, Fegan, Fernandes, Fiasson, Fontaine, Funk, F{\"u}ssling, Gabici, Gallant, Garrigoux, Gat{\'e}, Giavitto, Giebels, Glawion, Glicenstein, Gottschall, Grondin, Hahn, Haupt, Hawkes, Heinzelmann, Henri, Hermann, Hinton, Hofmann, Hoischen, Holch, Holler, Horns, Ivascenko, Iwasaki, Jacholkowska, Jamrozy, Jankowsky, Jankowsky, Jingo, Jouvin, {Jung-Richardt}, Kastendieck, Katarzy{\'n}ski, Katsuragawa, Kerszberg, Khangulyan, Kh{\'e}lifi, King, Klepser, Klochkov, Klu{\'z}niak, Komin, Kosack, Krakau, Kraus, Kr{\"u}ger, Laffon, Lamanna, Lau, Lees, Lefaucheur, Lemi{\`e}re, {Lemoine-Goumard}, Lenain, Leser, Lohse, Lorentz, Liu, Lypova, Malyshev, Marandon, Marcowith, Mariaud, Marx, Maurin, Maxted,
  Mayer, Meintjes, Meyer, Mitchell, Moderski, Mohamed, Mohrmann, Mor{\aa}, Moulin, Murach, Nakashima, de~Naurois, Ndiyavala, Niederwanger, Niemiec, Oakes, O'Brien, Odaka, Ohm, Ostrowski, Oya, Padovani, Panter, Parsons, Pekeur, Pelletier, Perennes, Petrucci, Peyaud, Piel, Pita, Poireau, Poon, Prokhorov, Prokoph, P{\"u}hlhofer, Punch, Quirrenbach, Raab, Rauth, Reimer, Reimer, Renaud, de~los Reyes, Rieger, Rinchiuso, Romoli, Rowell, Rudak, Rulten, Sahakian, Saito, Sanchez, Santangelo, Sasaki, Schlickeiser, Sch{\"u}ssler, Schulz, Schwanke, Schwemmer, {Seglar-Arroyo}, Settimo, Seyffert, Shafi, Shilon, Shiningayamwe, Simoni, Sol, Spanier, {Spir-Jacob}, Stawarz, Steenkamp, Stegmann, Steppa, Sushch, Takahashi, Tavernet, Tavernier, Taylor, Terrier, Tibaldo, Tiziani, Tluczykont, Trichard, Tsirou, Tsuji, Tuffs, Uchiyama, van~der Walt, van Eldik, van Rensburg, van Soelen, Vasileiadis, Veh, Venter, Viana, Vincent, Vink, Voisin, V{\"o}lk, Vuillaume, Wadiasingh, Wagner, Wagner, Wagner, White, Wierzcholska, Willmann,
  W{\"o}rnlein, Wouters, Yang, Zaborov, Zacharias, Zanin, Zdziarski, Zech, Zefi, Ziegler, Zorn, {\textbackslash}.Zywucka, Fender, Broderick, Rowlinson, Wijers, Stewart, ter Veen, Shulevski, Kavic, Simonetti, League, Tsai, Obenberger, Nathaniel, Taylor, Dowell, Liebling, Estes, Lippert, Sharma, Vincent, Farella, Abeysekara, Albert, Alfaro, Alvarez, Arceo, {Arteaga-Vel{\'a}zquez}, Rojas, Solares, Barber, Gonzalez, Becerril, {Belmont-Moreno}, BenZvi, Berley, Bernal, Braun, Brisbois, {Caballero-Mora}, Capistr{\'a}n, Carrami{\~n}ana, Casanova, Castillo, Cotti, Cotzomi, de~Le{\'o}n, Le{\'o}n, la~Fuente, Hernandez, Dichiara, Dingus, DuVernois, {D{\'i}az-V{\'e}lez}, Ellsworth, Engel, {Enr{\'i}quez-Rivera}, Fiorino, Fleischhack, Fraija, {Garc{\'i}a-Gonz{\'a}lez}, Garfias, Gerhardt, Mu{\~n}oz, Gonz{\'a}lez, Goodman, {Hampel-Arias}, Harding, Hernandez, {Hernandez-Almada}, Hona, H{\"u}ntemeyer, Iriarte, {Jardin-Blicq}, Joshi, Kaufmann, Kieda, Lara, Lauer, Lennarz, Vargas, Linnemann, Longinotti, Raya, {Luna-Garc{\'i}a},
  {L{\'o}pez-Coto}, Malone, Marinelli, Martinez, {Martinez-Castellanos}, {Mart{\'i}nez-Castro}, {Mart{\'i}nez-Huerta}, Matthews, {Miranda-Romagnoli}, Moreno, Mostaf{\'a}, Nellen, Newbold, Nisa, {Noriega-Papaqui}, Pelayo, Pretz, {P{\'e}rez-P{\'e}rez}, Ren, Rho, Rivi{\`e}re, {Rosa-Gonz{\'a}lez}, Rosenberg, {Ruiz-Velasco}, Salazar, Greus, Sandoval, Schneider, Schoorlemmer, Sinnis, Smith, Springer, Surajbali, Tibolla, Tollefson, Torres, Ukwatta, Weisgarber, Westerhoff, Wisher, Wood, Yapici, Yodh, Younk, Zhou, {\'A}lvarez, Aab, Abreu, Aglietta, Albuquerque, Albury, Allekotte, Almela, Castillo, {Alvarez-Mu{\~n}iz}, Anastasi, Anchordoqui, Andrada, Andringa, Aramo, Arsene, Asorey, Assis, Avila, Badescu, Balaceanu, Barbato, Luz, Becker, Bellido, Berat, Bertaina, Bertou, Biermann, Biteau, Blaess, Blanco, Blazek, Bleve, Boh{\'a}{\v c}ov{\'a}, Bonifazi, Borodai, Botti, Brack, Brancus, Bretz, Bridgeman, Briechle, Buchholz, Bueno, Buitink, Buscemi, {Caballero-Mora}, Caccianiga, Cancio, Canfora, Caruso, Castellina,
  Catalani, Cataldi, Cazon, Chavez, Chinellato, Chudoba, Clay, Cerutti, Colalillo, Coleman, Collica, Coluccia, Concei{\c c}{\~a}o, Consolati, Contreras, Cooper, Coutu, Covault, Cronin, D'Amico, Daniel, Dasso, Daumiller, Dawson, Day, de~Almeida, de~Jong, Mauro, Neto, Mitri, de~Oliveira, de~Souza, Debatin, Deligny, Castro, Diogo, Dobrigkeit, D'Olivo, Dorosti, Anjos, Dova, Dundovic, Ebr, Engel, Erdmann, Erfani, Escobar, Espadanal, Etchegoyen, Falcke, Farmer, Farrar, Fauth, Fazzini, Feldbusch, Fenu, Fick, Figueira, Filip{\v c}i{\v c}, Freire, Fujii, Fuster, Ga{\"i}or, Garc{\'i}a, Gat{\'e}, Gemmeke, {Gherghel-Lascu}, Ghia, Giaccari, Giammarchi, Giller, G{\textbackslash}las, Glaser, Golup, Berisso, Vitale, Gonz{\'a}lez, Gorgi, Gottowik, Grillo, Grubb, Guarino, Guedes, Halliday, Hampel, Hansen, Harari, Harrison, Harvey, Haungs, Hebbeker, Heck, Heimann, Herve, Hill, Hojvat, Holt, Homola, H{\"o}randel, Horvath, Hrabovsk{\'y}, Huege, Hulsman, Insolia, Isar, Jandt, Johnsen, Josebachuili, Jurysek, K{\"a}{\"a}p{\"a},
  Kampert, Keilhauer, Kemmerich, Kemp, Kieckhafer, Klages, Kleifges, Kleinfeller, Krause, Krohm, Kuempel, Mezek, Kunka, Awad, Lago, LaHurd, Lang, Lauscher, Legumina, de~Oliveira, {Letessier-Selvon}, {Lhenry-Yvon}, Link, Presti, Lopes, L{\'o}pez, Casado, Lorek, Luce, Lucero, Malacari, Mallamaci, Mandat, Mantsch, Mariazzi, Maris, Marsella, Martello, Martinez, Bravo, Meza, Mathes, Mathys, Matthews, Matthiae, Mayotte, Mazur, Medina, {Medina-Tanco}, Melo, Menshikov, Merenda, Michal, Micheletti, Middendorf, Miramonti, Mitrica, Mockler, Mollerach, Montanet, Morello, Morlino, M{\"u}ller, M{\"u}ller, Muller, M{\"u}ller, Mussa, Naranjo, Nguyen, {Niculescu-Oglinzanu}, Niechciol, Niemietz, Niggemann, Nitz, Nosek, Novotny, No{\v z}ka, N{\'u}{\~n}ez, Oikonomou, Olinto, Palatka, Pallotta, Papenbreer, Parente, Parra, Paul, Pech, Pedreira, P{\c e}kala, {Pe{\~n}a-Rodriguez}, Pereira, Perlin, Perrone, Peters, Petrera, Phuntsok, Pierog, Pimenta, Pirronello, Platino, Plum, Poh, Porowski, Prado, Privitera, Prouza, Quel,
  Querchfeld, Quinn, {Ramos-Pollan}, Rautenberg, Ravignani, Ridky, Riehn, Risse, Ristori, Rizi, de~Carvalho, Fernandez, Rojo, Roncoroni, Roth, Roulet, Rovero, Ruehl, Saffi, Saftoiu, Salamida, Salazar, Saleh, Salina, S{\'a}nchez, {Sanchez-Lucas}, Santos, Santos, Sarazin, Sarmento, {Sarmiento-Cano}, Sato, Schauer, Scherini, Schieler, Schimp, Schmidt, Scholten, Schov{\'a}nek, Schr{\"o}der, Schr{\"o}der, Schulz, Schumacher, Sciutto, Segreto, Shadkam, Shellard, Sigl, Silli, {\v S}m{\'i}da, Snow, Sommers, Sonntag, Soriano, Squartini, Stanca, Stani{\v c}, Stasielak, Stassi, Stolpovskiy, Strafella, Streich, Suarez, {Suarez-Dur{\'a}n}, Sudholz, Suomij{\"a}rvi, Supanitsky, {\v S}up{\'i}k, Swain, Szadkowski, Taboada, Taborda, Timmermans, Peixoto, Tomankova, Tom{\'e}, Elipe, Travnicek, Trini, Tueros, Ulrich, Unger, Urban, Galicia, Vali{\~n}o, Valore, van Aar, van Bodegom, van~den Berg, van Vliet, Varela, C{\'a}rdenas, V{\'a}zquez, Veberi{\v c}, Ventura, Quispe, Verzi, Vicha, Villase{\~n}or, Vorobiov, Wahlberg, Wainberg,
  Walz, Watson, Weber, Weindl, Wiede{\'n}ski, Wiencke, Wilczy{\'n}ski, Wirtz, Wittkowski, Wundheiler, Yang, Yushkov, Zas, Zavrtanik, Zavrtanik, Zepeda, Zimmermann, Ziolkowski, Zong, Zuccarello, Kim, Schulze, Bauer, {Corral-Santana}, de~{Gregorio-Monsalvo}, {Gonz{\'a}lez-L{\'o}pez}, Hartmann, {Ishwara-Chandra}, Mart{\'i}n, Mehner, Misra, Micha{\textbackslash}lowski, Resmi, Paragi, Agudo, An, Beswick, Casadio, Frey, Jonker, Kettenis, Marcote, Moldon, Szomoru, van Langevelde, Yang, Cwiek, Cwiok, Czyrkowski, Dabrowski, Kasprowicz, Mankiewicz, Nawrocki, Opiela, Piotrowski, Wrochna, Zaremba, F.~{\textbackslash}.Zarnecki, Haggard, Nynka, Ruan, Bland, Booler, Devillepoix, de~Gois, Hancock, Howie, Paxman, Sansom, Towner, Tonry, Coughlin, Stubbs, Denneau, Heinze, Stalder, Weiland, Eatough, Kramer, Kraus, Troja, Piro, Gonz{\'a}lez, Butler, Fox, Khandrika, Kutyrev, Lee, Ricci, Jr, {S{\'a}nchez-Ram{\'i}rez}, Veilleux, Watson, Wieringa, Burgess, van Eerten, Fontes, Fryer, Korobkin, Wollaeger, Camilo, Foley, Goedhart,
  Makhathini, Oozeer, Smirnov, Fender, Woudt, {and}, {and}, {and}, {and}, {and}, {and}, {and}, {and}, {and}, {and}, {and}, {and}, {and}, {and}, {and}, {and}, {and}, {and}, {and}, {and}, {and}, {and}, {and}, {and}, \& {and}}]{abbott_multi-messenger_2017}
---. 2017{\natexlab{b}}, The Astrophysical Journal, 848, L12, \dodoi{10.3847/2041-8213/aa91c9}

\bibitem[{Abbott {et~al.}(2017{\natexlab{c}})Abbott, Abbott, Abbott, Acernese, Ackley, Adams, Adams, Addesso, Adhikari, Adya, Affeldt, Afrough, Agarwal, Agathos, Agatsuma, Aggarwal, Aguiar, Aiello, Ain, Ajith, Allen, Allen, Allocca, Altin, Amato, Ananyeva, Anderson, Anderson, Angelova, Antier, Appert, Arai, Araya, Areeda, Arnaud, Arun, Ascenzi, Ashton, Ast, Aston, Astone, Atallah, Aufmuth, Aulbert, AultONeal, Austin, {Avila-Alvarez}, Babak, Bacon, Bader, Bae, Baker, Baldaccini, Ballardin, Ballmer, Banagiri, Barayoga, Barclay, Barish, Barker, Barkett, Barone, Barr, Barsotti, Barsuglia, Barta, Bartlett, Bartos, Bassiri, Basti, Batch, Bawaj, Bayley, Bazzan, B{\'e}csy, Beer, Bejger, Belahcene, Bell, Berger, Bergmann, Bero, Berry, Bersanetti, Bertolini, Betzwieser, Bhagwat, Bhandare, Bilenko, Billingsley, Billman, Birch, Birney, Birnholtz, Biscans, Biscoveanu, Bisht, Bitossi, Biwer, Bizouard, Blackburn, Blackman, Blair, Blair, Blair, Bloemen, Bock, Bode, Boer, Bogaert, Bohe, Bondu, Bonilla, Bonnand, Boom, Bork,
  Boschi, Bose, Bossie, Bouffanais, Bozzi, Bradaschia, Brady, Branchesi, Brau, Briant, Brillet, Brinkmann, Brisson, Brockill, Broida, Brooks, Brown, Brown, Brunett, Buchanan, Buikema, Bulik, Bulten, Buonanno, Buskulic, Buy, Byer, Cabero, Cadonati, Cagnoli, Cahillane, Bustillo, Callister, Calloni, Camp, Canepa, Canizares, Cannon, Cao, Cao, Capano, Capocasa, Carbognani, Caride, Carney, Diaz, Casentini, Caudill, Cavagli{\`a}, Cavalier, Cavalieri, Cella, Cepeda, {Cerd{\'a}-Dur{\'a}n}, Cerretani, Cesarini, Chamberlin, Chan, Chao, Charlton, Chase, {Chassande-Mottin}, Chatterjee, Chatziioannou, Cheeseboro, Chen, Chen, Chen, Cheng, Chia, Chincarini, Chiummo, Chmiel, Cho, Cho, Chow, Christensen, Chu, Chua, Chua, Chung, Chung, Ciani, Ciolfi, Cirelli, Cirone, Clara, Clark, Clearwater, Cleva, Cocchieri, Coccia, Cohadon, Cohen, Colla, Collette, Cominsky, Constancio, Conti, Cooper, Corban, Corbitt, {Cordero-Carri{\'o}n}, Corley, Cornish, Corsi, Cortese, Costa, Coughlin, Coughlin, Coulon, Countryman, Couvares, Covas, Cowan,
  Coward, Cowart, Coyne, Coyne, Creighton, Creighton, Cripe, Crowder, Cullen, Cumming, Cunningham, Cuoco, Dal~Canton, D{\'a}lya, Danilishin, D'Antonio, Danzmann, Dasgupta, Da~Silva~Costa, Datrier, Dattilo, Dave, Davier, Davis, Daw, Day, De, DeBra, Degallaix, De~Laurentis, Del{\'e}glise, Del~Pozzo, Demos, Denker, Dent, De~Pietri, Dergachev, De~Rosa, DeRosa, De~Rossi, DeSalvo, {de Varona}, Devenson, Dhurandhar, D{\'i}az, Di~Fiore, Di~Giovanni, Di~Girolamo, Di~Lieto, Di~Pace, Di~Palma, Di~Renzo, Doctor, Dolique, Donovan, Dooley, Doravari, Dorrington, Douglas, {Dovale {\'a}lvarez}, Downes, Drago, Dreissigacker, Driggers, Du, Ducrot, Dupej, Dwyer, \& {The LIGO Scientific Collaboration and The Virgo Collaboration}}]{abbott_a_2017}
---. 2017{\natexlab{c}}, Nature, 551, 85, \dodoi{10.1038/nature24471}

\bibitem[{Abbott {et~al.}(2017{\natexlab{d}})Abbott, Abbott, Abbott, Acernese, Ackley, Adams, Adams, Addesso, Adhikari, Adya, Affeldt, Afrough, Agarwal, Agathos, Agatsuma, Aggarwal, Aguiar, Aiello, Ain, Ajith, Allen, Allen, Allocca, Altin, Amato, Ananyeva, Anderson, Anderson, Angelova, Antier, Appert, Arai, Araya, Areeda, Arnaud, Arun, Ascenzi, Ashton, Ast, Aston, Astone, Atallah, Aufmuth, Aulbert, AultONeal, Austin, {Avila-Alvarez}, Babak, Bacon, Bader, Bae, Baker, Baldaccini, Ballardin, Banagiri, Barayoga, Barclay, Barish, Barker, Barkett, Barone, Barr, Barsotti, Barsuglia, Barta, Bartlett, Bartos, Bassiri, Basti, Batch, Bawaj, Bayley, Bazzan, B{\'e}csy, Beer, Bejger, Belahcene, Bell, Bergmann, Bernuzzi, Bero, Berry, Bersanetti, Bertolini, Betzwieser, Bhagwat, Bhandare, Bilenko, Billingsley, Billman, Birch, Birney, Birnholtz, Biscans, Biscoveanu, Bisht, Bitossi, Biwer, Bizouard, Blackburn, Blackman, Blair, Blair, Blair, Bloemen, Bock, Bode, Boer, Bogaert, Bohe, Bondu, Bonilla, Bonnand, Boom, Bork, Boschi,
  Bose, Bossie, Bouffanais, Bozzi, Bradaschia, Brady, Branchesi, Brau, Briant, Brillet, Brinkmann, Brisson, Brockill, Broida, Brooks, Brown, Brunett, Buchanan, Buikema, Bulik, Bulten, Buonanno, Buskulic, Buy, Byer, Cabero, Cadonati, Cagnoli, Cahillane, Calder{\'o}n~Bustillo, Callister, Calloni, Camp, Canepa, Canizares, Cannon, Cao, Cao, Capano, Capocasa, Carbognani, Caride, Carney, Casanueva~Diaz, Casentini, Caudill, Cavagli{\`a}, Cavalier, Cavalieri, Cella, Cepeda, {Cerd{\'a}-Dur{\'a}n}, Cerretani, Cesarini, Chamberlin, Chan, Chao, Charlton, Chase, {Chassande-Mottin}, Chatterjee, Chatziioannou, Cheeseboro, Chen, Chen, Chen, Cheng, Chia, Chincarini, Chiummo, Chmiel, Cho, Cho, Chow, Christensen, Chu, Chua, Chua, Chung, Chung, Ciani, Ciolfi, Cirelli, Cirone, Clara, Clark, Clearwater, Cleva, Cocchieri, Coccia, Cohadon, Cohen, Colla, Collette, Cominsky, Constancio, Conti, Cooper, Corban, Corbitt, {Cordero-Carri{\'o}n}, Corley, Cornish, Corsi, Cortese, Costa, Coughlin, Coughlin, Coulon, Countryman, Couvares,
  Covas, Cowan, Coward, Cowart, Coyne, Coyne, Creighton, Creighton, Cripe, Crowder, Cullen, Cumming, Cunningham, Cuoco, Dal~Canton, D{\'a}lya, Danilishin, D'Antonio, Danzmann, Dasgupta, Da~Silva~Costa, Dattilo, Dave, Davier, Davis, Daw, Day, De, DeBra, Degallaix, De~Laurentis, Del{\'e}glise, Del~Pozzo, Demos, Denker, Dent, De~Pietri, Dergachev, De~Rosa, DeRosa, De~Rossi, DeSalvo, {de Varona}, Devenson, Dhurandhar, D{\'i}az, Dietrich, Di~Fiore, Di~Giovanni, Di~Girolamo, Di~Lieto, Di~Pace, Di~Palma, Di~Renzo, Doctor, Dolique, Donovan, Dooley, Doravari, Dorrington, Douglas, Dovale~{\'A}lvarez, Downes, Drago, Dreissigacker, Driggers, Du, Ducrot, Dupej, Dwyer, Edo, Edwards, Effler, Eggenstein, Ehrens, Eichholz, Eikenberry, Eisenstein, Essick, Estevez, Etienne, Etzel, Evans, Evans, Factourovich, Fafone, Fair, Fairhurst, Fan, Farinon, Farr, Farr, {Fauchon-Jones}, Favata, Fays, Fee, Fehrmann, Feicht, Fejer, {Fernandez-Galiana}, Ferrante, Ferreira, Ferrini, Fidecaro, Finstad, Fiori, Fiorucci, Fishbach, Fisher,
  {Fitz-Axen}, Flaminio, Fletcher, Fong, Font, Forsyth, Forsyth, Fournier, Frasca, Frasconi, Frei, Freise, Frey, Frey, Fries, Fritschel, Frolov, Fulda, Fyffe, Gabbard, Gadre, Gaebel, Gair, Gammaitoni, Ganija, Gaonkar, {Garcia-Quiros}, Garufi, Gateley, Gaudio, Gaur, Gayathri, Gehrels, Gemme, Genin, Gennai, George, George, Gergely, Germain, Ghonge, Ghosh, Ghosh, Ghosh, Giaime, Giardina, Giazotto, Gill, Glover, Goetz, Goetz, Gomes, Goncharov, Gonz{\'a}lez, Gonzalez~Castro, Gopakumar, Gorodetsky, Gossan, Gosselin, Gouaty, Grado, Graef, Granata, Grant, Gras, Gray, Greco, Green, Gretarsson, Groot, Grote, Grunewald, Gruning, Guidi, Guo, Gupta, Gupta, Gushwa, Gustafson, Gustafson, Halim, Hall, Hall, Hamilton, Hammond, Haney, Hanke, Hanks, Hanna, Hannam, Hannuksela, Hanson, Hardwick, Harms, Harry, Harry, Hart, Haster, Haughian, Healy, Heidmann, Heintze, Heitmann, Hello, Hemming, Hendry, Heng, Hennig, Heptonstall, Heurs, Hild, Hinderer, Hoak, Hofman, Holt, Holz, Hopkins, Horst, Hough, Houston, Howell, Hreibi, Hu,
  Huerta, Huet, Hughey, Husa, Huttner, {Huynh-Dinh}, Indik, Inta, Intini, Isa, Isac, Isi, Iyer, Izumi, Jacqmin, Jani, Jaranowski, Jawahar, {Jim{\'e}nez-Forteza}, Johnson, {Johnson-McDaniel}, Jones, Jones, Jonker, Ju, Junker, Kalaghatgi, Kalogera, Kamai, Kandhasamy, Kang, Kanner, Kapadia, Karki, Karvinen, Kasprzack, Kastaun, Katolik, Katsavounidis, Katzman, Kaufer, Kawabe, Kawaguchi, K{\'e}f{\'e}lian, Keitel, Kemball, Kennedy, Kent, Key, Khalili, Khan, Khan, Khan, Khazanov, Kijbunchoo, Kim, Kim, Kim, Kim, Kim, Kim, Kimbrell, King, King, {Kinley-Hanlon}, Kirchhoff, Kissel, Kleybolte, Klimenko, Knowles, Koch, Koehlenbeck, Koley, Kondrashov, Kontos, Korobko, Korth, Kowalska, Kozak, Kr{\"a}mer, Kringel, Kr{\'o}lak, Kuehn, Kumar, Kumar, Kumar, Kuo, Kutynia, Kwang, Lackey, Lai, Landry, Lang, Lange, Lantz, Lanza, Larson, {Lartaux-Vollard}, Lasky, Laxen, Lazzarini, Lazzaro, Leaci, Leavey, Lee, Lee, Lee, Lee, Lee, Lehmann, Lenon, Leonardi, Leroy, Letendre, Levin, Li, Linker, Littenberg, Liu, Liu, Lo, Lockerbie, London,
  Lord, Lorenzini, Loriette, Lormand, Losurdo, Lough, Lousto, Lovelace, L{\"u}ck, Lumaca, Lundgren, Lynch, Ma, Macas, Macfoy, Machenschalk, MacInnis, Macleod, Maga{\~n}a~Hernandez, {Maga{\~n}a-Sandoval}, Maga{\~n}a~Zertuche, Magee, Majorana, Maksimovic, Man, Mandic, Mangano, Mansell, Manske, Mantovani, Marchesoni, Marion, M{\'a}rka, M{\'a}rka, Markakis, Markosyan, Markowitz, Maros, Marquina, Martelli, Martellini, Martin, Martin, Martynov, Mason, Massera, Masserot, Massinger, {Masso-Reid}, Mastrogiovanni, Matas, Matichard, Matone, Mavalvala, Mazumder, McCarthy, McClelland, McCormick, McCuller, McGuire, McIntyre, McIver, McManus, McNeill, McRae, McWilliams, Meacher, Meadors, Mehmet, Meidam, {Mejuto-Villa}, Melatos, Mendell, Mercer, Merilh, Merzougui, Meshkov, Messenger, Messick, Metzdorff, Meyers, Miao, Michel, Middleton, Mikhailov, Milano, Miller, Miller, Miller, Millhouse, {Milovich-Goff}, Minazzoli, Minenkov, Ming, Mishra, Mitra, Mitrofanov, Mitselmakher, Mittleman, Moffa, Moggi, Mogushi, Mohan, Mohapatra,
  Montani, Moore, Moraru, Moreno, Morriss, Mours, {Mow-Lowry}, Mueller, Muir, Mukherjee, Mukherjee, Mukherjee, Mukund, Mullavey, Munch, Mu{\~n}iz, Muratore, Murray, Napier, Nardecchia, Naticchioni, Nayak, Neilson, Nelemans, Nelson, Nery, Neunzert, Nevin, Newport, Newton, Ng, Nguyen, Nichols, Nielsen, Nissanke, Nitz, Noack, Nocera, Nolting, North, Nuttall, Oberling, O'Dea, Ogin, Oh, Oh, Ohme, Okada, Oliver, Oppermann, Oram, O'Reilly, Ormiston, Ortega, O'Shaughnessy, Ossokine, Ottaway, Overmier, Owen, Pace, Page, Page, Pai, Pai, Palamos, Palashov, Palomba, {Pal-Singh}, Pan, Pan, Pang, Pang, Pankow, Pannarale, Pant, Paoletti, Paoli, Papa, Parida, Parker, Pascucci, Pasqualetti, Passaquieti, Passuello, Patil, Patricelli, Pearlstone, Pedraza, Pedurand, Pekowsky, Pele, Penn, Perez, Perreca, Perri, Pfeiffer, Phelps, Piccinni, Pichot, Piergiovanni, Pierro, Pillant, Pinard, Pinto, Pirello, Pitkin, Poe, Poggiani, Popolizio, Porter, Post, Powell, Prasad, Pratt, Pratten, Predoi, Prestegard, Prijatelj, Principe, Privitera,
  Prodi, Prokhorov, Puncken, Punturo, Puppo, P{\"u}rrer, Qi, Quetschke, Quintero, {Quitzow-James}, Rabeling, Radkins, Raffai, Raja, Rajan, Rajbhandari, Rakhmanov, Ramirez, {Ramos-Buades}, Rapagnani, Raymond, Razzano, Read, Regimbau, Rei, Reid, Reitze, Ren, Reyes, Ricci, Ricker, Rieger, Riles, Rizzo, Robertson, Robie, Robinet, Rocchi, Rolland, Rollins, Roma, Romano, Romel, Romie, Rosi{\'n}ska, Ross, Rowan, R{\"u}diger, Ruggi, Rutins, Ryan, Sachdev, Sadecki, Sadeghian, Sakellariadou, Salconi, Saleem, Salemi, Samajdar, Sammut, Sampson, Sanchez, Sanchez, {Sanchis-Gual}, Sandberg, Sanders, Sassolas, Sauter, Savage, Sawadsky, Schale, Scheel, Scheuer, Schmidt, Schmidt, Schnabel, Schofield, Sch{\"o}nbeck, Schreiber, Schuette, Schulte, Schutz, Schwalbe, Scott, Scott, Seidel, Sellers, Sengupta, Sentenac, Sequino, Sergeev, Shaddock, Shaffer, Shah, Shahriar, Shaner, Shao, Shapiro, Shawhan, Sheperd, Shoemaker, Shoemaker, Siellez, Siemens, Sieniawska, Sigg, Silva, Singer, Singh, Singhal, Sintes, Slagmolen, Smith, Smith,
  Smith, Somala, Son, Sonnenberg, Sorazu, Sorrentino, Souradeep, Spencer, Srivastava, Staats, Staley, Steinke, Steinlechner, Steinlechner, Steinmeyer, Stevenson, Stone, Stops, Strain, Stratta, Strigin, Strunk, Sturani, Stuver, Summerscales, Sun, Sunil, Suresh, Sutton, Swinkels, Szczepa{\'n}czyk, Tacca, Tait, Talbot, Talukder, Tanner, T{\'a}pai, Taracchini, Tasson, Taylor, Taylor, Tewari, Theeg, Thies, Thomas, Thomas, Thomas, Thorne, Thrane, Tiwari, Tiwari, Tokmakov, Toland, Tonelli, Tornasi, {Torres-Forn{\'e}}, Torrie, T{\"o}yr{\"a}, Travasso, Traylor, Trinastic, Tringali, Trozzo, Tsang, Tse, Tso, Tsukada, Tsuna, Tuyenbayev, Ueno, Ugolini, Unnikrishnan, Urban, Usman, Vahlbruch, Vajente, Valdes, {van Bakel}, {van Beuzekom}, {van den Brand}, Van Den~Broeck, {Vander-Hyde}, {van der Schaaf}, {van Heijningen}, {van Veggel}, Vardaro, Varma, Vass, Vas{\'u}th, Vecchio, Vedovato, Veitch, Veitch, Venkateswara, Venugopalan, Verkindt, Vetrano, Vicer{\'e}, Viets, Vinciguerra, Vine, Vinet, Vitale, Vo, Vocca, Vorvick,
  Vyatchanin, Wade, Wade, Wade, Walet, Walker, Wallace, Walsh, Wang, Wang, Wang, Wang, Wang, Ward, Warner, Was, Watchi, Weaver, Wei, Weinert, Weinstein, Weiss, Wen, Wessel, We{\ss}els, Westerweck, Westphal, Wette, Whelan, Whiting, Whittle, Wilken, Williams, Williams, Williamson, Willis, Willke, Wimmer, Winkler, Wipf, Wittel, Woan, Woehler, Wofford, Wong, Worden, Wright, Wu, Wysocki, Xiao, Yamamoto, Yancey, Yang, Yap, Yazback, Yu, Yu, Yvert, Zadro{\.z}ny, Zanolin, Zelenova, Zendri, Zevin, Zhang, Zhang, Zhang, Zhang, Zhao, Zhou, Zhou, Zhu, Zhu, Zimmerman, Zucker, Zweizig, {(LIGO Scientific Collaboration}, \& {Virgo Collaboration}}]{abbott_estimating_2017}
---. 2017{\natexlab{d}}, The Astrophysical Journal, 850, L39, \dodoi{10.3847/2041-8213/aa9478}

\bibitem[{Abbott {et~al.}(2017{\natexlab{e}})Abbott, Abbott, Abbott, Acernese, Ackley, Adams, Adams, Addesso, Adhikari, Adya, Affeldt, Afrough, Agarwal, Agathos, Agatsuma, Aggarwal, Aguiar, Aiello, Ain, Ajith, Allen, Allen, Allocca, Aloy, Altin, Amato, Ananyeva, Anderson, Anderson, Angelova, Antier, Appert, Arai, Araya, Areeda, Arnaud, Arun, Ascenzi, Ashton, Ast, Aston, Astone, Atallah, Aufmuth, Aulbert, AultONeal, Austin, {Avila-Alvarez}, Babak, Bacon, Bader, Bae, Baker, Baldaccini, Ballardin, Ballmer, Banagiri, Barayoga, Barclay, Barish, Barker, Barkett, Barone, Barr, Barsotti, Barsuglia, Barta, Bartlett, Bartos, Bassiri, Basti, Batch, Bawaj, Bayley, Bazzan, B{\'e}csy, Beer, Bejger, Belahcene, Bell, Berger, Bergmann, Bero, Berry, Bersanetti, Bertolini, Betzwieser, Bhagwat, Bhandare, Bilenko, Billingsley, Billman, Birch, Birney, Birnholtz, Biscans, Biscoveanu, Bisht, Bitossi, Biwer, Bizouard, Blackburn, Blackman, Blair, Blair, Blair, Bloemen, Bock, Bode, Boer, Bogaert, Bohe, Bondu, Bonilla, Bonnand, Boom,
  Bork, Boschi, Bose, Bossie, Bouffanais, Bozzi, Bradaschia, Brady, Branchesi, Brau, Briant, Brillet, Brinkmann, Brisson, Brockill, Broida, Brooks, Brown, Brown, Brunett, Buchanan, Buikema, Bulik, Bulten, Buonanno, Buskulic, Buy, Byer, Cabero, Cadonati, Cagnoli, Cahillane, Calder{\'o}n~Bustillo, Callister, Calloni, Camp, Canepa, Canizares, Cannon, Cao, Cao, Capano, Capocasa, Carbognani, Caride, Carney, Casanueva~Diaz, Casentini, Caudill, Cavagli{\`a}, Cavalier, Cavalieri, Cella, Cepeda, {Cerd{\'a}-Dur{\'a}n}, Cerretani, Cesarini, Chamberlin, Chan, Chao, Charlton, Chase, {Chassande-Mottin}, Chatterjee, Chatziioannou, Cheeseboro, Chen, Chen, Chen, Cheng, Chia, Chincarini, Chiummo, Chmiel, Cho, Cho, Chow, Christensen, Chu, Chua, Chua, Chung, Chung, Ciani, Ciolfi, Cirelli, Cirone, Clara, Clark, Clearwater, Cleva, Cocchieri, Coccia, Cohadon, Cohen, Colla, Collette, Cominsky, Constancio, Conti, Cooper, Corban, Corbitt, {Cordero-Carri{\'o}n}, Corley, Cornish, Corsi, Cortese, Costa, Coughlin, Coughlin, Coulon,
  Countryman, Couvares, Covas, Cowan, Coward, Cowart, Coyne, Coyne, Creighton, Creighton, Cripe, Crowder, Cullen, Cumming, Cunningham, Cuoco, Dal~Canton, D{\'a}lya, Danilishin, D'Antonio, Danzmann, Dasgupta, Da~Silva~Costa, Dattilo, Dave, Davier, Davis, Daw, Day, De, DeBra, Degallaix, De~Laurentis, Del{\'e}glise, Del~Pozzo, Demos, Denker, Dent, De~Pietri, Dergachev, De~Rosa, DeRosa, De~Rossi, DeSalvo, {de Varona}, Devenson, Dhurandhar, D{\'i}az, Di~Fiore, Di~Giovanni, Di~Girolamo, Di~Lieto, Di~Pace, Di~Palma, Di~Renzo, Doctor, Dolique, Donovan, Dooley, Doravari, Dorrington, Douglas, Dovale~{\'A}lvarez, Downes, Drago, Dreissigacker, Driggers, Du, Ducrot, Dupej, Dwyer, Edo, Edwards, Effler, Eggenstein, Ehrens, Eichholz, Eikenberry, Eisenstein, Essick, Estevez, Etienne, Etzel, Evans, Evans, Factourovich, Fafone, Fair, Fairhurst, Fan, Farinon, Farr, Farr, {Fauchon-Jones}, Favata, Fays, Fee, Fehrmann, Feicht, Fejer, {Fernandez-Galiana}, Ferrante, Ferreira, Ferrini, Fidecaro, Finstad, Fiori, Fiorucci, Fishbach,
  Fisher, {Fitz-Axen}, Flaminio, Fletcher, Fong, Font, Forsyth, Forsyth, Fournier, Frasca, Frasconi, Frei, Freise, Frey, Frey, Fries, Fritschel, Frolov, Fulda, Fyffe, Gabbard, Gadre, Gaebel, Gair, Gammaitoni, Ganija, Gaonkar, {Garcia-Quiros}, Garufi, Gateley, Gaudio, Gaur, Gayathri, Gehrels, Gemme, Genin, Gennai, George, George, Gergely, Germain, Ghonge, Ghosh, Ghosh, Ghosh, Giaime, Giardina, Giazotto, Gill, Glover, Goetz, Goetz, Gomes, Goncharov, Gonz{\'a}lez, Gonzalez~Castro, Gopakumar, Gorodetsky, Gossan, Gosselin, Gouaty, Grado, Graef, Granata, Grant, Gras, Gray, Greco, Green, Gretarsson, Groot, Grote, Grunewald, Gruning, Guidi, Guo, Gupta, Gupta, Gushwa, Gustafson, Gustafson, Halim, Hall, Hall, Hamilton, Hammond, Haney, Hanke, Hanks, Hanna, Hannam, Hannuksela, Hanson, Hardwick, Harms, Harry, Harry, Hart, Haster, Haughian, Healy, Heidmann, Heintze, Heitmann, Hello, Hemming, Hendry, Heng, Hennig, Heptonstall, Heurs, Hild, Hinderer, Hoak, Hofman, Holt, Holz, Hopkins, Horst, Hough, Houston, Howell, Hreibi,
  Hu, Huerta, Huet, Hughey, Husa, Huttner, {Huynh-Dinh}, Indik, Inta, Intini, Isa, Isac, Isi, Iyer, Izumi, Jacqmin, Jani, Jaranowski, Jawahar, {Jim{\'e}nez-Forteza}, Johnson, {Johnson-McDaniel}, Jones, Jones, Jonker, Ju, Junker, Kalaghatgi, Kalogera, Kamai, Kandhasamy, Kang, Kanner, Kapadia, Karki, Karvinen, Kasprzack, Kastaun, Katolik, Katsavounidis, Katzman, Kaufer, Kawabe, K{\'e}f{\'e}lian, Keitel, Kemball, Kennedy, Kent, Key, Khalili, Khan, Khan, Khan, Khazanov, Kijbunchoo, Kim, Kim, Kim, Kim, Kim, Kim, Kimbrell, King, King, {Kinley-Hanlon}, Kirchhoff, Kissel, Kleybolte, Klimenko, Knowles, Koch, Koehlenbeck, Koley, Kondrashov, Kontos, Korobko, Korth, Kowalska, Kozak, Kr{\"a}mer, Kringel, Krishnan, Kr{\'o}lak, Kuehn, Kumar, Kumar, Kumar, Kuo, Kutynia, Kwang, Lackey, Lai, Landry, Lang, Lange, Lantz, Lanza, {Lartaux-Vollard}, Lasky, Laxen, Lazzarini, Lazzaro, Leaci, Leavey, Lee, Lee, Lee, Lee, Lee, Lehmann, Lenon, Leonardi, Leroy, Letendre, Levin, Li, Linker, Littenberg, Liu, Lo, Lockerbie, London, Lord,
  Lorenzini, Loriette, Lormand, Losurdo, Lough, Lousto, Lovelace, L{\"u}ck, Lumaca, Lundgren, Lynch, Ma, Macas, Macfoy, Machenschalk, MacInnis, Macleod, Maga{\~n}a~Hernandez, {Maga{\~n}a-Sandoval}, Maga{\~n}a~Zertuche, Magee, Majorana, Maksimovic, Man, Mandic, Mangano, Mansell, Manske, Mantovani, Marchesoni, Marion, M{\'a}rka, M{\'a}rka, Markakis, Markosyan, Markowitz, Maros, Marquina, Martelli, Martellini, Martin, Martin, Martynov, Mason, Massera, Masserot, Massinger, {Masso-Reid}, Mastrogiovanni, Matas, Matichard, Matone, Mavalvala, Mazumder, McCarthy, McClelland, McCormick, McCuller, McGuire, McIntyre, McIver, McManus, McNeill, McRae, McWilliams, Meacher, Meadors, Mehmet, Meidam, {Mejuto-Villa}, Melatos, Mendell, Mercer, Merilh, Merzougui, Meshkov, Messenger, Messick, Metzdorff, Meyers, Miao, Michel, Middleton, Mikhailov, Milano, Miller, Miller, Miller, Millhouse, {Milovich-Goff}, Minazzoli, Minenkov, Ming, Mishra, Mitra, Mitrofanov, Mitselmakher, Mittleman, Moffa, Moggi, Mogushi, Mohan, Mohapatra,
  Montani, Moore, Moraru, Moreno, Morriss, Mours, {Mow-Lowry}, Mueller, Muir, Mukherjee, Mukherjee, Mukherjee, Mukund, Mullavey, Munch, Mu{\~n}iz, Muratore, Murray, Napier, Nardecchia, Naticchioni, Nayak, Neilson, Nelemans, Nelson, Nery, Neunzert, Nevin, Newport, Newton, Ng, Nguyen, Nichols, Nielsen, Nissanke, Nitz, Noack, Nocera, Nolting, North, Nuttall, Oberling, O'Dea, Ogin, Oh, Oh, Ohme, Okada, Oliver, Oppermann, Oram, O'Reilly, Ormiston, Ortega, O'Shaughnessy, Ossokine, Ottaway, Overmier, Owen, Pace, Page, Page, Pai, Pai, Palamos, Palashov, Palomba, {Pal-Singh}, Pan, Pan, Pang, Pang, Pankow, Pannarale, Pant, Paoletti, Paoli, Papa, Parida, Parker, Pascucci, Pasqualetti, Passaquieti, Passuello, Patil, Patricelli, Pearlstone, Pedraza, Pedurand, Pekowsky, Pele, Penn, Perez, Perreca, Perri, Pfeiffer, Phelps, Piccinni, Pichot, Piergiovanni, Pierro, Pillant, Pinard, Pinto, Pirello, Pitkin, Poe, Poggiani, Popolizio, Porter, Post, Powell, Prasad, Pratt, Pratten, Predoi, Prestegard, Prijatelj, Principe, Privitera,
  Prodi, Prokhorov, Puncken, Punturo, Puppo, P{\"u}rrer, Qi, Quetschke, Quintero, {Quitzow-James}, Raab, Rabeling, Radkins, Raffai, Raja, Rajan, Rajbhandari, Rakhmanov, Ramirez, {Ramos-Buades}, Rapagnani, Raymond, Razzano, Read, Regimbau, Rei, Reid, Reitze, Ren, Reyes, Ricci, Ricker, Rieger, Riles, Rizzo, Robertson, Robie, Robinet, Rocchi, Rolland, Rollins, Roma, Romano, Romel, Romie, Rosi{\'n}ska, Ross, Rowan, R{\"u}diger, Ruggi, Rutins, Ryan, Sachdev, Sadecki, Sadeghian, Sakellariadou, Salconi, Saleem, Salemi, Samajdar, Sammut, Sampson, Sanchez, Sanchez, {Sanchis-Gual}, Sandberg, Sanders, Sassolas, Sathyaprakash, Saulson, Sauter, Savage, Sawadsky, Schale, Scheel, Scheuer, Schmidt, Schmidt, Schnabel, Schofield, Sch{\"o}nbeck, Schreiber, Schuette, Schulte, Schutz, Schwalbe, Scott, Scott, Seidel, Sellers, Sengupta, Sentenac, Sequino, Sergeev, Shaddock, Shaffer, Shah, Shahriar, Shaner, Shao, Shapiro, Shawhan, Sheperd, Shoemaker, Shoemaker, Siellez, Siemens, Sieniawska, Sigg, Silva, Singer, Singh, Singhal,
  Sintes, Slagmolen, Smith, Smith, Smith, Somala, Son, Sonnenberg, Sorazu, Sorrentino, Souradeep, Spencer, Srivastava, Staats, Staley, Steinke, Steinlechner, Steinlechner, Steinmeyer, Stevenson, Stone, Stops, Strain, Stratta, Strigin, Strunk, Sturani, Stuver, Summerscales, Sun, Sunil, Suresh, Sutton, Swinkels, Szczepa{\'n}czyk, Tacca, Tait, Talbot, Talukder, Tanner, T{\'a}pai, Taracchini, Tasson, Taylor, Taylor, Tewari, Theeg, Thies, Thomas, Thomas, Thomas, Thorne, Thorne, Thrane, Tiwari, Tiwari, Tokmakov, Toland, Tonelli, Tornasi, {Torres-Forn{\'e}}, Torrie, T{\"o}yr{\"a}, Travasso, Traylor, Trinastic, Tringali, Trozzo, Tsang, Tse, Tso, Tsukada, Tsuna, Tuyenbayev, Ueno, Ugolini, Unnikrishnan, Urban, Usman, Vahlbruch, Vajente, Valdes, {van Bakel}, {van Beuzekom}, {van den Brand}, Van Den~Broeck, {Vander-Hyde}, {van der Schaaf}, {van Heijningen}, {van Veggel}, Vardaro, Varma, Vass, Vas{\'u}th, Vecchio, Vedovato, Veitch, Veitch, Venkateswara, Venugopalan, Verkindt, Vetrano, Vicer{\'e}, Viets, Vinciguerra, Vine,
  Vinet, Vitale, Vo, Vocca, Vorvick, Vyatchanin, Wade, Wade, Wade, Walet, Walker, Wallace, Walsh, Wang, Wang, Wang, Wang, Wang, Ward, Warner, Was, Watchi, Weaver, Wei, Weinert, Weinstein, Weiss, Wen, Wessel, We{\ss}els, Westerweck, Westphal, Wette, Whelan, Whitcomb, Whiting, Whittle, Wilken, Williams, Williams, Williamson, Willis, Willke, Wimmer, Winkler, Wipf, Wittel, Woan, Woehler, Wofford, Wong, Worden, Wright, Wu, Wysocki, Xiao, Yamamoto, Yancey, Yang, Yap, Yazback, Yu, Yu, Yvert, Zadro{\.z}ny, Zanolin, Zelenova, Zendri, Zevin, Zhang, Zhang, Zhang, Zhang, Zhao, Zhou, Zhou, Zhu, Zhu, Zimmerman, Zucker, Zweizig, {(LIGO Scientific Collaboration}, {Virgo Collaboration}, Burns, Veres, Kocevski, Racusin, Goldstein, Connaughton, Briggs, Blackburn, Hamburg, Hui, {von Kienlin}, McEnery, Preece, {Wilson-Hodge}, Bissaldi, Cleveland, Gibby, Giles, Kippen, McBreen, Meegan, Paciesas, Poolakkil, Roberts, Stanbro, {Gamma-ray Burst Monitor}, Savchenko, Ferrigno, Kuulkers, Bazzano, Bozzo, Brandt, Chenevez, Courvoisier,
  Diehl, Domingo, Hanlon, Jourdain, Laurent, Lebrun, Lutovinov, Mereghetti, Natalucci, Rodi, Roques, Sunyaev, Ubertini, \& {(INTEGRAL}}]{abbott_gravitational_2017}
---. 2017{\natexlab{e}}, The Astrophysical Journal, 848, L13, \dodoi{10.3847/2041-8213/aa920c}

\bibitem[{Abbott {et~al.}(2020{\natexlab{a}})Abbott, Abbott, Abbott, Abraham, Acernese, Ackley, Adams, Adhikari, Adya, Affeldt, Agathos, Agatsuma, Aggarwal, Aguiar, Aiello, Ain, Ajith, Allen, Allocca, Aloy, Altin, Amato, Anand, Ananyeva, Anderson, Anderson, Angelova, Antier, Appert, Arai, Araya, Areeda, Ar{\`e}ne, Arnaud, Aronson, Arun, Ascenzi, Ashton, Aston, Astone, Aubin, Aufmuth, AultONeal, Austin, Avendano, {Avila-Alvarez}, Babak, Bacon, Badaracco, Bader, Bae, Baird, Baker, Baldaccini, Ballardin, Ballmer, Bals, Banagiri, Barayoga, Barbieri, Barclay, Barish, Barker, Barkett, Barnum, Barone, Barr, Barsotti, Barsuglia, Barta, Bartlett, Bartos, Bassiri, Basti, Bawaj, Bayley, Baylor, Bazzan, B{\'e}csy, Bejger, Belahcene, Bell, Beniwal, Benjamin, Berger, Bergmann, Bernuzzi, Berry, Bersanetti, Bertolini, Betzwieser, Bhandare, Bidler, Biggs, Bilenko, Bilgili, Billingsley, Birney, Birnholtz, Biscans, Bischi, Biscoveanu, Bisht, Bitossi, Bizouard, Blackburn, Blackman, Blair, Blair, Blair, Bloemen, Bobba, Bode,
  Boer, Boetzel, Bogaert, Bondu, Bonnand, Booker, Boom, Bork, Boschi, Bose, Bossilkov, Bosveld, Bouffanais, Bozzi, Bradaschia, Brady, Bramley, Branchesi, Brau, Breschi, Briant, Briggs, Brighenti, Brillet, Brinkmann, Brockill, Brooks, Brooks, Brown, Brunett, Buikema, Bulik, Bulten, Buonanno, Buskulic, Buy, Byer, Cabero, Cadonati, Cagnoli, Cahillane, Bustillo, Callister, Calloni, Camp, Campbell, Canepa, Cannon, Cao, Cao, Carapella, Carbognani, Caride, Carney, Carullo, Diaz, Casentini, Caudill, Cavagli{\`a}, Cavalier, Cavalieri, Cella, {Cerd{\'a}-Dur{\'a}n}, Cesarini, Chaibi, Chakravarti, Chamberlin, Chan, Chao, Charlton, Chase, {Chassande-Mottin}, Chatterjee, Chaturvedi, Chatziioannou, Cheeseboro, Chen, Chen, Chen, Cheng, Cheong, Chia, Chiadini, Chincarini, Chiummo, Cho, Cho, Cho, Christensen, Chu, Chua, Chung, Chung, Ciani, Cie{\'s}lar, Ciobanu, Ciolfi, Cipriano, Cirone, Clara, Clark, Clearwater, Cleva, Coccia, Cohadon, Cohen, Colleoni, Collette, Collins, Colpi, Cominsky, Constancio, Conti, Cooper, Corban,
  Corbitt, {Cordero-Carri{\'o}n}, Corezzi, Corley, Cornish, Corre, Corsi, Cortese, Costa, Cotesta, Coughlin, Coughlin, Coulon, Countryman, Couvares, Covas, Cowan, Coward, Cowart, Coyne, Coyne, Creighton, Creighton, Cripe, Croquette, Crowder, Cullen, Cumming, Cunningham, Cuoco, Canton, D{\'a}lya, D'Angelo, Danilishin, D'Antonio, Danzmann, Dasgupta, Costa, Datrier, Dattilo, Dave, Davier, Davis, Daw, DeBra, Deenadayalan, Degallaix, Laurentis, Del{\'e}glise, Lillo, Pozzo, DeMarchi, Demos, Dent, Pietri, Rosa, Rossi, DeSalvo, de~Varona, Dhurandhar, D{\'i}az, Dietrich, Fiore, DiFronzo, Giorgio, Giovanni, Giovanni, Girolamo, Lieto, Ding, Pace, Palma, Renzo, Divakarla, Dmitriev, Doctor, Donovan, Dooley, Doravari, Dorrington, Downes, Drago, Driggers, Du, Ducoin, Dudi, Dupej, Durante, Dwyer, Easter, Eddolls, Edo, Effler, Ehrens, Eichholz, Eikenberry, Eisenmann, Eisenstein, Errico, Essick, Estelles, Estevez, Etienne, Etzel, Evans, Evans, Fafone, Fairhurst, Fan, Farinon, Farr, Farr, {Fauchon-Jones}, Favata, Fays, Fazio,
  Fee, Feicht, Fejer, Feng, {Fernandez-Galiana}, Ferrante, Ferreira, Ferreira, Fidecaro, Fiori, Fiorucci, Fishbach, Fisher, Fishner, Fittipaldi, {Fitz-Axen}, Fiumara, Flaminio, Fletcher, Floden, Flynn, Fong, Font, Forsyth, Fournier, Vivanco, Frasca, Frasconi, Frei, Freise, Frey, Frey, Fritschel, Frolov, Fronz{\`e}, Fulda, Fyffe, Gabbard, Gadre, Gaebel, Gair, Gamba, Gammaitoni, Gaonkar, {Garc{\'i}a-Quir{\'o}s}, Garufi, Gateley, Gaudio, Gaur, Gayathri, Gemme, Genin, Gennai, George, George, George, Gergely, Ghonge, Ghosh, Ghosh, Ghosh, Giacomazzo, Giaime, Giardina, Gibson, Gill, Glover, Gniesmer, Godwin, Goetz, Goetz, Goncharov, Gonz{\'a}lez, Castro, Gopakumar, Gossan, Gosselin, Gouaty, Grace, Grado, Granata, Grant, Gras, Grassia, Gray, Gray, Greco, Green, Green, Gretarsson, Grimaldi, Grimm, Groot, Grote, Grunewald, Gruning, Guidi, Gulati, Guo, Gupta, Gupta, Gupta, Gustafson, Gustafson, Haegel, Halim, Hall, Hall, Hamilton, Hammond, Haney, Hanke, Hanks, Hanna, Hannam, Hannuksela, Hansen, Hanson, Harder, Hardwick,
  Haris, Harms, Harry, Harry, Hasskew, Haster, Haughian, Hayes, Healy, Heidmann, Heintze, Heitmann, Hellman, Hello, Hemming, Hendry, Heng, Hennig, Heurs, Hild, Hinderer, Ho, Hochheim, Hofman, Holgado, Holland, Holt, Holz, Hopkins, Horst, Hough, Howell, Hoy, Huang, H{\"u}bner, Huerta, Huet, Hughey, Hui, Husa, Huttner, {Huynh-Dinh}, Idzkowski, Iess, Inchauspe, Ingram, Inta, Intini, Irwin, Isa, Isac, Isi, Iyer, Jacqmin, Jadhav, Jani, Janthalur, Jaranowski, Jariwala, Jenkins, Jiang, Johnson, {Johnson-McDaniel}, Jones, Jones, Jones, Jones, Jonker, Ju, Junker, Kalaghatgi, Kalogera, Kamai, Kandhasamy, Kang, Kanner, Kapadia, Karki, Kashyap, Kasprzack, Kastaun, Katsanevas, Katsavounidis, Katzman, Kaufer, Kawabe, Keerthana, K{\'e}f{\'e}lian, Keitel, Kennedy, Key, Khalili, Khan, Khan, Khazanov, Khetan, Khursheed, Kijbunchoo, Kim, Kim, Kim, Kim, Kim, Kim, Kimball, King, {Kinley-Hanlon}, Kirchhoff, Kissel, Kleybolte, Klika, Klimenko, Knowles, Koch, Koehlenbeck, Koekoek, Koley, Kondrashov, Kontos, Koper, Korobko, Korth,
  Kovalam, Kozak, Kr{\"a}mer, Kringel, Krishnendu, Kr{\'o}lak, Krupinski, Kuehn, Kumar, Kumar, Kumar, Kumar, Kuo, Kutynia, Kwang, Lackey, Laghi, Lai, Lam, Landry, Landry, Lane, Lang, Lange, Lantz, Lanza, {Lartaux-Vollard}, Lasky, Laxen, Lazzarini, Lazzaro, Leaci, Leavey, Lecoeuche, Lee, Lee, Lee, Lee, Lee, Lee, Lehmann, Lenon, Leroy, Letendre, Levin, Li, Li, Li, Li, Li, Lin, Linde, Linker, Littenberg, Liu, Liu, {Llorens-Monteagudo}, Lo, London, Longo, Lorenzini, Loriette, Lormand, Losurdo, Lough, Lousto, Lovelace, Lower, Lucaccioni, L{\"u}ck, Lumaca, Lundgren, Lynch, Ma, Macas, Macfoy, MacInnis, Macleod, Macquet, Hernandez, {Maga{\~n}a-Sandoval}, Magee, Majorana, Maksimovic, Malik, Man, Mandic, Mangano, Mansell, Manske, Mantovani, Mapelli, Marchesoni, Marion, M{\'a}rka, M{\'a}rka, Markakis, Markosyan, Markowitz, Maros, Marquina, Marsat, Martelli, Martin, Martin, Martinez, Martynov, Masalehdan, Mason, Massera, Masserot, Massinger, {Masso-Reid}, Mastrogiovanni, Matas, Matichard, Matone, Mavalvala, McCann,
  McCarthy, McClelland, McCormick, McCuller, McGuire, McIsaac, McIver, McManus, McRae, McWilliams, Meacher, Meadors, Mehmet, Mehta, Meidam, Villa, Melatos, Mendell, Mercer, Mereni, Merfeld, Merilh, Merzougui, Meshkov, Messenger, Messick, Messina, Metzdorff, Meyers, Meylahn, Miani, Miao, Michel, Middleton, Milano, Miller, Millhouse, Mills, {Milovich-Goff}, Minazzoli, Minenkov, Mishkin, Mishra, Mistry, Mitra, Mitrofanov, Mitselmakher, Mittleman, Mo, Moffa, Mogushi, Mohapatra, {Molina-Ruiz}, Mondin, Montani, Moore, Moraru, Morawski, Moreno, Morisaki, Mours, {Mow-Lowry}, Muciaccia, Mukherjee, Mukherjee, Mukherjee, Mukherjee, Mukund, Mullavey, Munch, Mu{\~n}iz, Muratore, Murray, Nagar, Nardecchia, Naticchioni, Nayak, Neil, Neilson, Nelemans, Nelson, Nery, Neunzert, Nevin, Ng, Ng, Nguyen, Nguyen, Nichols, Nichols, Nissanke, Nocera, North, Nuttall, Obergaulinger, Oberling, O'Brien, Oganesyan, Ogin, Oh, Oh, Ohme, Ohta, Okada, Oliver, Oppermann, Oram, O'Reilly, Ormiston, Ortega, O'Shaughnessy, Ossokine, Ottaway,
  Overmier, Owen, Pace, Pagano, Page, Pagliaroli, Pai, Pai, Palamos, Palashov, Palomba, Pan, Panda, Pang, Pankow, Pannarale, Pant, Paoletti, Paoli, Parida, Parker, Pascucci, Pasqualetti, Passaquieti, Passuello, Patil, Patricelli, Payne, Pearlstone, Pechsiri, Pedersen, Pedraza, Pedurand, Pele, Penn, Perego, Perez, P{\'e}rigois, Perreca, Petermann, Pfeiffer, Phelps, Phukon, Piccinni, Pichot, Piergiovanni, Pierro, Pillant, Pinard, Pinto, Pirello, Pitkin, Plastino, Poggiani, Pong, Ponrathnam, Popolizio, Porter, Powell, Prajapati, Prasad, Prasai, Prasanna, Pratten, Prestegard, Principe, Prodi, Prokhorov, Punturo, Puppo, P{\"u}rrer, Qi, Quetschke, Quinonez, Raab, Raaijmakers, Radkins, Radulesco, Raffai, Raja, Rajan, Rajbhandari, Rakhmanov, Ramirez, {Ramos-Buades}, Rana, Rao, Rapagnani, Raymond, Razzano, Read, Regimbau, Rei, Reid, Reitze, Rettegno, Ricci, Richardson, Richardson, Ricker, Riemenschneider, Riles, Rizzo, Robertson, Robinet, Rocchi, Rolland, Rollins, Roma, Romanelli, Romano, Romel, Romie, Rose, Rose,
  Rose, Rosell, Rosi{\'n}ska, Rosofsky, Ross, Rowan, Roy, R{\"u}diger, Ruggi, Rutins, Ryan, Sachdev, Sadecki, Sakellariadou, Salafia, Salconi, Saleem, Samajdar, Sammut, Sanchez, Sanchez, {Sanchis-Gual}, Sanders, Santiago, Santos, Sarin, Sassolas, Sathyaprakash, Sauter, Savage, Schale, Scheel, Scheuer, Schmidt, Schnabel, Schofield, Sch{\"o}nbeck, Schreiber, Schulte, Schutz, Scott, Scott, Seidel, Sellers, Sengupta, Sennett, Sentenac, Sequino, Sergeev, Setyawati, Shaddock, Shaffer, Shahriar, Shaner, Sharma, Sharma, Shawhan, Shen, Shink, Shoemaker, Shoemaker, Shukla, ShyamSundar, Siellez, Sieniawska, Sigg, Singer, Singh, Singh, Singhal, Sintes, Sitmukhambetov, Skliris, Slagmolen, {Slaven-Blair}, Smith, Smith, Somala, Son, Soni, Sorazu, Sorrentino, Souradeep, Sowell, Spencer, Spera, Srivastava, Srivastava, Staats, Stachie, Standke, Steer, Steinke, Steinlechner, Steinlechner, Steinmeyer, Stevenson, Stocks, Stone, Stops, Strain, Stratta, Strigin, Strunk, Sturani, Stuver, Sudhir, Summerscales, Sun, Sunil, Sur,
  Suresh, Sutton, Swinkels, Szczepa{\'n}czyk, Tacca, Tait, Talbot, Tanner, Tao, T{\'a}pai, Tapia, Tasson, Taylor, Tenorio, Terkowski, Thomas, Thomas, Thondapu, Thorne, Thrane, Tiwari, Tiwari, Tiwari, Toland, Tonelli, Tornasi, {Torres-Forn{\'e}}, Torrie, T{\"o}yr{\"a}, Travasso, Traylor, Tringali, Tripathee, Trovato, Trozzo, Tsang, Tse, Tso, Tsukada, Tsuna, Tsutsui, Tuyenbayev, Ueno, Ugolini, Unnikrishnan, Urban, Usman, Vahlbruch, Vajente, Valdes, Valentini, van Bakel, van Beuzekom, van~den Brand, Broeck, {Vander-Hyde}, van~der Schaaf, VanHeijningen, van Veggel, Vardaro, Varma, Vass, Vas{\'u}th, Vecchio, Vedovato, Veitch, Veitch, Venkateswara, Venugopalan, Verkindt, Vetrano, Vicer{\'e}, Viets, Vinciguerra, Vine, Vinet, Vitale, Vo, Vocca, Vorvick, Vyatchanin, Wade, Wade, Wade, Walet, Walker, Wallace, Walsh, Wang, Wang, Wang, Wang, Ward, Warden, Warner, Was, Watchi, Weaver, Wei, Weinert, Weinstein, Weiss, Wellmann, Wen, Wessel, We{\ss}els, Westhouse, Wette, Whelan, White, Whiting, Whittle, Wilken, Williams,
  Williamson, Willis, Willke, Winkler, Wipf, Wittel, Woan, Woehler, Wofford, Wright, Wu, Wysocki, Xiao, Xu, Yamamoto, Yancey, Yang, Yang, Yang, Yap, Yazback, Yeeles, Yu, Yu, Yuen, Zadro{\textbackslash}.zny, Zadro{\textbackslash}.zny, Zanolin, Zelenova, Zendri, Zevin, Zhang, Zhang, Zhang, Zhao, Zhao, Zhou, Zhou, Zhu, Zimmerman, Zucker, \& Zweizig}]{abbott_gw190425:_2020}
---. 2020{\natexlab{a}}, The Astrophysical Journal, 892, L3, \dodoi{10.3847/2041-8213/ab75f5}

\bibitem[{Abbott {et~al.}(2020{\natexlab{b}})Abbott, Abbott, Abbott, Abraham, Acernese, Ackley, Adams, Adya, Affeldt, Agathos, Agatsuma, Aggarwal, Aguiar, Aiello, Ain, Ajith, Akutsu, Allen, Allocca, Aloy, Altin, Amato, Ananyeva, Anderson, Anderson, Ando, Angelova, Antier, Appert, Arai, Arai, Arai, Araki, Araya, Araya, Areeda, Ar{\`e}ne, Aritomi, Arnaud, Arun, Ascenzi, Ashton, Aso, Aston, Astone, Aubin, Aufmuth, AultONeal, Austin, Avendano, {Avila-Alvarez}, Babak, Bacon, Badaracco, Bader, Bae, Bae, Baiotti, Bajpai, Baker, Baldaccini, Ballardin, Ballmer, Banagiri, Barayoga, Barclay, Barish, Barker, Barkett, Barnum, Barone, Barr, Barsotti, Barsuglia, Barta, Bartlett, Barton, Bartos, Bassiri, Basti, Bawaj, Bayley, Bazzan, B{\'e}csy, Bejger, Belahcene, Bell, Beniwal, Berger, Bergmann, Bernuzzi, Bero, Berry, Bersanetti, Bertolini, Betzwieser, Bhandare, Bidler, Bilenko, Bilgili, Billingsley, Birch, Birney, Birnholtz, Biscans, Biscoveanu, Bisht, Bitossi, Bizouard, Blackburn, Blair, Blair, Blair, Bloemen, Bode, Boer,
  Boetzel, Bogaert, Bondu, Bonilla, Bonnand, Booker, Boom, Booth, Bork, Boschi, Bose, Bossie, Bossilkov, Bosveld, Bouffanais, Bozzi, Bradaschia, Brady, Bramley, Branchesi, Brau, Briant, Briggs, Brighenti, Brillet, Brinkmann, Brisson, Brockill, Brooks, Brown, Brown, Brunett, Buikema, Bulik, Bulten, Buonanno, Buskulic, Buy, Byer, Cabero, Cadonati, Cagnoli, Cahillane, Bustillo, Callister, Calloni, Camp, Campbell, Canepa, Cannon, Cannon, Cao, Cao, Capocasa, Carbognani, Caride, Carney, Carullo, Diaz, Casentini, Caudill, Cavagli{\`a}, Cavalier, Cavalieri, Cella, {Cerd{\'a}-Dur{\'a}n}, Cerretani, Cesarini, Chaibi, Chakravarti, Chamberlin, Chan, Chan, Chao, Charlton, Chase, {Chassande-Mottin}, Chatterjee, Chaturvedi, Chatziioannou, Cheeseboro, Chen, Chen, Chen, Chen, Chen, Chen, Cheng, Cheong, Chia, Chincarini, Chiummo, Cho, Cho, Cho, Christensen, Chu, Chu, Chu, Chua, Chung, Chung, Ciani, Ciobanu, Ciolfi, Cipriano, Cirone, Clara, Clark, Clearwater, Cleva, Cocchieri, Coccia, Cohadon, Cohen, Colgan, Colleoni, Collette,
  Collins, Cominsky, Constancio, Conti, Cooper, Corban, Corbitt, {Cordero-Carri{\'o}n}, Corley, Cornish, Corsi, Cortese, Costa, Cotesta, Coughlin, Coughlin, Coulon, Countryman, Couvares, Covas, Cowan, Coward, Cowart, Coyne, Coyne, Creighton, Creighton, Cripe, Croquette, Crowder, Cullen, Cumming, Cunningham, Cuoco, Canton, D{\'a}lya, Danilishin, D'Antonio, Danzmann, Dasgupta, Da~Silva~Costa, Datrier, Dattilo, Dave, Davier, Davis, Daw, DeBra, Deenadayalan, Degallaix, De~Laurentis, Del{\'e}glise, Pozzo, DeMarchi, Demos, Dent, De~Pietri, Derby, De~Rosa, De~Rossi, DeSalvo, {de Varona}, Dhurandhar, D{\'i}az, Dietrich, \& Fiore}]{abbott_prospects_2020a}
---. 2020{\natexlab{b}}, Living Reviews in Relativity, 23, 3, \dodoi{10.1007/s41114-020-00026-9}

\bibitem[{Abbott {et~al.}(2021{\natexlab{a}})Abbott, Abbott, Abraham, Acernese, Ackley, Adams, Adams, Adhikari, Adya, Affeldt, Agarwal, Agathos, Agatsuma, Aggarwal, Aguiar, Aiello, Ain, Ajith, Akutsu, Aleman, Allen, Allocca, Altin, Amato, Anand, Ananyeva, Anderson, Anderson, Ando, Angelova, Ansoldi, Antelis, Antier, Appert, Arai, Arai, Arai, Araki, Araya, Araya, Areeda, Ar{\`e}ne, Aritomi, Arnaud, Aronson, Arun, Asada, Asali, Ashton, Aso, Aston, Astone, Aubin, Aufmuth, AultONeal, Austin, Babak, Badaracco, Bader, Bae, Bae, Baer, Bagnasco, Bai, Baiotti, Baird, Bajpai, Ball, Ballardin, Ballmer, Bals, Balsamo, Baltus, Banagiri, Bankar, Bankar, Barayoga, Barbieri, Barish, Barker, Barneo, Barone, Barr, Barsotti, Barsuglia, Barta, Bartlett, Barton, Bartos, Bassiri, Basti, Bawaj, Bayley, Baylor, Bazzan, B{\'e}csy, Bedakihale, Bejger, Belahcene, Benedetto, Beniwal, Benjamin, Benkel, Bennett, Bentley, BenYaala, Bergamin, Berger, Bernuzzi, Berry, Bersanetti, Bertolini, Betzwieser, Bhandare, Bhandari, Bhattacharjee,
  Bhaumik, Bidler, Bilenko, Billingsley, Birney, Birnholtz, Biscans, Bischi, Biscoveanu, Bisht, Biswas, Bitossi, Bizouard, Blackburn, Blackman, Blair, Blair, Blair, Bobba, Bode, Boer, Bogaert, Boldrini, Bondu, Bonilla, Bonnand, Booker, Boom, Bork, Boschi, Bose, Bose, Bossilkov, Boudart, Bouffanais, Bozzi, Bradaschia, Brady, Bramley, Branch, Branchesi, Brau, Breschi, Briant, Briggs, Brillet, Brinkmann, Brockill, Brooks, Brooks, Brown, Brunett, Bruno, Bruntz, Bryant, Buikema, Bulik, Bulten, Buonanno, Buscicchio, Buskulic, Byer, Cadonati, Caesar, Cagnoli, Cahillane, III, Bustillo, Callaghan, Callister, Calloni, Camp, Canepa, Cannavacciuolo, Cannon, Cao, Cao, Cao, Capocasa, Capote, Carapella, Carbognani, Carlin, Carney, Carpinelli, Carullo, Carver, Diaz, Casentini, Castaldi, Caudill, Cavagli{\`a}, Cavalier, Cavalieri, Cella, {Cerd{\'a}-Dur{\'a}n}, Cesarini, Chaibi, Chakravarti, Champion, Chan, Chan, Chan, Chan, Chandra, Chanial, Chao, Charlton, Chase, {Chassande-Mottin}, Chatterjee, Chaturvedi, Chatziioannou,
  Chen, Chen, Chen, Chen, Chen, Chen, Chen, Chen, Chen, Cheng, Cheong, Cheung, Chia, Chiadini, Chiang, Chierici, Chincarini, Chiofalo, Chiummo, Cho, Cho, Choate, Choudhary, Choudhary, Christensen, Chu, Chu, Chu, Chua, Chung, Ciani, Ciecielag, Cie{\'s}lar, Cifaldi, Ciobanu, Ciolfi, Cipriano, Cirone, Clara, Clark, Clark, Clarke, Clearwater, Clesse, Cleva, Coccia, Cohadon, Cohen, Cohen, Colleoni, Collette, Colpi, Compton, Jr, Conti, Cooper, Corban, Corbitt, {Cordero-Carri{\'o}n}, Corezzi, Corley, Cornish, Corre, Corsi, Cortese, Costa, Cotesta, Coughlin, Coughlin, Coulon, Countryman, Cousins, Couvares, Covas, Coward, Cowart, Coyne, Coyne, Creighton, Creighton, Criswell, Croquette, Crowder, Cudell, Cullen, Cumming, Cummings, Cuoco, Cury{\l}o, Canton, D{\'a}lya, Dana, DaneshgaranBajastani, D'Angelo, Danilishin, D'Antonio, Danzmann, {Darsow-Fromm}, Dasgupta, Datrier, Dattilo, Dave, Davier, Davies, Davis, Daw, Dean, DeBra, Deenadayalan, Degallaix, Laurentis, Del{\'e}glise, Favero, Lillo, Lillo, Pozzo, DeMarchi,
  Matteis, D'Emilio, Demos, Dent, Depasse, Pietri, Rosa, Rossi, DeSalvo, Simone, Dhurandhar, D{\'i}az, Jr, Didio, Dietrich, Fiore, Fronzo, Giorgio, Giovanni, Girolamo, Lieto, Ding, Pace, Palma, Renzo, Divakarla, Dmitriev, Doctor, D'Onofrio, Donovan, Dooley, Doravari, Dorrington, Drago, Driggers, Drori, Du, Ducoin, Dupej, Durante, D'Urso, Duverne, Dwyer, Easter, Ebersold, Eddolls, Edelman, Edo, Edy, Effler, Eguchi, Eichholz, Eikenberry, Eisenmann, Eisenstein, Ejlli, Enomoto, Errico, Essick, Estell{\'e}s, Estevez, Etienne, Etzel, Evans, Evans, Ewing, Fafone, Fair, Fairhurst, Fan, Farah, Farinon, Farr, Farr, Farrow, {Fauchon-Jones}, Favata, Fays, Fazio, Feicht, Fejer, Feng, Fenyvesi, Ferguson, {Fernandez-Galiana}, Ferrante, Ferreira, Fidecaro, Figura, Fiori, Fishbach, Fisher, Fittipaldi, Fiumara, Flaminio, Floden, Flynn, Fong, Font, Fornal, Forsyth, Franke, Frasca, Frasconi, Frederick, Frei, Freise, Frey, Fritschel, Frolov, Fronz{\'e}, Fujii, Fujikawa, Fukunaga, Fukushima, Fulda, Fyffe, Gabbard, Gadre, Gaebel,
  Gair, Gais, Galaudage, Gamba, Ganapathy, Ganguly, Gao, Gaonkar, Garaventa, {Garc{\'i}a-N{\'u}{\~n}ez}, {Garc{\'i}a-Quir{\'o}s}, Garufi, Gateley, Gaudio, Gayathri, Ge, Gemme, Gennai, George, Gergely, Gewecke, Ghonge, Ghosh, Ghosh, Ghosh, Ghosh, Ghosh, Giacomazzo, Giacoppo, Giaime, Giardina, Gibson, Gier, Giesler, Giri, Gissi, Glanzer, Gleckl, Godwin, Goetz, Goetz, Gohlke, Goncharov, Gonz{\'a}lez, Gopakumar, Gosselin, Gouaty, Grace, Grado, Granata, Granata, Grant, Gras, Grassia, Gray, Gray, Greco, Green, Green, Gretarsson, Gretarsson, Griffith, Griffiths, Griggs, Grignani, Grimaldi, Grimes, Grimm, Grote, Grunewald, Gruning, Guerrero, Guidi, Guimaraes, Guix{\'e}, Gulati, Guo, Guo, Gupta, Gupta, Gupta, Gustafson, Gustafson, Guzman, Ha, Haegel, Hagiwara, Haino, Halim, Hall, Hamilton, Hammond, Han, Haney, Hanks, Hanna, Hannam, Hannuksela, Hansen, Hansen, Hanson, Harder, Hardwick, Haris, Harms, Harry, Harry, Hartwig, Hasegawa, Haskell, Hasskew, Haster, Hattori, Haughian, Hayakawa, Hayama, Hayes, Healy, Heidmann,
  Heintze, Heinze, Heinzel, Heitmann, Hellman, Hello, {Helmling-Cornell}, Hemming, Hendry, Heng, Hennes, Hennig, Hennig, Vivanco, Heurs, Hild, Hill, Himemoto, Hinderer, Hines, Hiranuma, Hirata, Hirose, Ho, Hochheim, Hofman, Hohmann, Holgado, Holland, Hollows, Holmes, Holt, Holz, Hong, Hopkins, Hough, Howell, Hoy, Hoyland, Hreibi, Hsieh, Hsu, Huang, Huang, Huang, Huang, Huang, Huang, H{\"u}bner, Huddart, Huerta, Hughey, Hui, Hui, Husa, Huttner, Huxford, {Huynh-Dinh}, Ide, Idzkowski, Iess, Ikenoue, Imam, Inayoshi, Inchauspe, Ingram, Inoue, Intini, Ioka, Isi, Isleif, Ito, Itoh, Iyer, Izumi, JaberianHamedan, Jacqmin, Jadhav, Jadhav, James, Jan, Jani, Janssens, Janthalur, Jaranowski, Jariwala, Jaume, Jenkins, Jeon, Jeunon, Jia, Jiang, Jin, Johns, Jones, Jones, Jones, Jones, Jones, Jonker, Ju, Jung, Jung, Junker, Kaihotsu, Kajita, Kakizaki, Kalaghatgi, Kalogera, Kamai, Kamiizumi, Kanda, Kandhasamy, Kang, Kanner, Kao, Kapadia, Kapasi, Karat, Karathanasis, Karki, Kashyap, Kasprzack, Kastaun, Katsanevas,
  Katsavounidis, Katzman, Kaur, Kawabe, Kawaguchi, Kawai, Kawasaki, K{\'e}f{\'e}lian, Keitel, Key, Khadka, Khalili, Khan, Khan, Khazanov, Khetan, Khursheed, Kijbunchoo, Kim, Kim, Kim, Kim, Kim, Kim, Kimball, Kimura, King, {Kinley-Hanlon}, Kirchhoff, Kissel, Kita, Kitazawa, Kleybolte, Klimenko, Knee, Knowles, Knyazev, Koch, Koekoek, Kojima, Kokeyama, Koley, Kolitsidou, Kolstein, Komori, Kondrashov, Kong, Kontos, Koper, Korobko, Kotake, Kovalam, Kozak, Kozakai, Kozu, Kringel, Krishnendu, Kr{\'o}lak, Kuehn, Kuei, Kumar, Kumar, Kumar, Kumar, Kume, Kuns, Kuo, Kuo, Kuromiya, Kuroyanagi, Kusayanagi, Kwak, Kwang, Laghi, Lalande, Lam, Lamberts, Landry, Landry, Lane, Lang, Lange, Lantz, Rosa, {Lartaux-Vollard}, Lasky, Laxen, Lazzarini, Lazzaro, Leaci, Leavey, Lecoeuche, Lee, Lee, Lee, Lee, Lee, Lee, Lehmann, Lema{\^i}tre, Leon, Leonardi, Leroy, Letendre, Levin, Leviton, Li, Li, Li, Li, Li, Li, Lin, Lin, Lin, Lin, Lin, Linde, Linker, Linley, Littenberg, Liu, Liu, Liu, Liu, {Llorens-Monteagudo}, Lo, Lockwood, Lollie,
  London, Longo, Lopez, Lorenzini, Loriette, Lormand, Losurdo, Lough, Lousto, Lovelace, L{\"u}ck, Lumaca, Lundgren, Luo, Macas, MacInnis, Macleod, MacMillan, Macquet, Hernandez, {Maga{\~n}a-Sandoval}, Magazz{\`u}, Magee, Maggiore, Majorana, Makarem, Maksimovic, Maliakal, Malik, Man, Mandic, Mangano, Mango, Mansell, Manske, Mantovani, Mapelli, Marchesoni, Marchio, Marion, Mark, M{\'a}rka, M{\'a}rka, Markakis, Markosyan, Markowitz, Maros, Marquina, Marsat, Martelli, Martin, Martin, Martinez, Martinez, Martinovic, Martynov, Marx, Masalehdan, Mason, Massera, Masserot, Massinger, {Masso-Reid}, Mastrogiovanni, Matas, {Mateu-Lucena}, Matichard, Matiushechkina, Mavalvala, McCann, McCarthy, McClelland, McClincy, McCormick, McCuller, McGhee, McGuire, McIsaac, McIver, McManus, McRae, McWilliams, Meacher, Mehmet, Mehta, Melatos, Melchor, Mendell, {Menendez-Vazquez}, Menoni, Mercer, Mereni, Merfeld, Merilh, Merritt, Merzougui, Meshkov, Messenger, Messick, Meyers, Meylahn, Mhaske, Miani, Miao, Michaloliakos, Michel,
  Michimura, Middleton, Milano, Miller, Millhouse, Mills, Milotti, {Milovich-Goff}, Minazzoli, Minenkov, Mio, Mir, Mishkin, Mishra, Mishra, Mistry, Mitra, Mitrofanov, Mitselmakher, Mittleman, Miyakawa, Miyamoto, Miyazaki, Miyo, Miyoki, Mo, Mogushi, Mohapatra, Mohite, Molina, {Molina-Ruiz}, Mondin, Montani, Moore, Moraru, Morawski, More, Moreno, Moreno, Mori, Morisaki, Moriwaki, Mours, {Mow-Lowry}, Mozzon, Muciaccia, Mukherjee, Mukherjee, Mukherjee, Mukherjee, Mukund, Mullavey, Munch, Mu{\~n}iz, Murray, Musenich, Nadji, Nagano, Nagano, Nagar, Nakamura, Nakano, Nakano, Nakashima, Nakayama, Nardecchia, Narikawa, Naticchioni, Nayak, Nayak, Negishi, Neil, Neilson, Nelemans, Nelson, Nery, Neunzert, Ng, Ng, Nguyen, Nguyen, Nguyen, Quynh, Ni, Nichols, Nishizawa, Nissanke, Nocera, Noh, Norman, North, Nozaki, Nuttall, Oberling, O'Brien, Obuchi, O'Dell, Ogaki, Oganesyan, Oh, Oh, Oh, Ohashi, Ohishi, Ohkawa, Ohme, Ohta, Okada, Okutani, Okutomi, Olivetto, Oohara, Ooi, Oram, O'Reilly, Ormiston, Ormsby, Ortega,
  O'Shaughnessy, O'Shea, Oshino, Ossokine, Osthelder, Otabe, Ottaway, Overmier, Pace, Pagano, Page, Pagliaroli, Pai, Pai, Palamos, Palashov, Palomba, Pan, Panda, Pang, Pang, Pankow, Pannarale, Pant, Paoletti, Paoli, Paolone, Parisi, Park, Parker, Pascucci, Pasqualetti, Passaquieti, Passuello, Patel, Patricelli, Payne, Pechsiri, Pedraza, Pegoraro, Pele, Arellano, Penn, Perego, Pereira, Pereira, Perez, P{\'e}rigois, Perreca, Perri{\`e}s, Petermann, Petterson, Pfeiffer, Pham, Phukon, Piccinni, Pichot, Piendibene, Piergiovanni, Pierini, Pierro, Pillant, Pilo, Pinard, Pinto, Piotrzkowski, Piotrzkowski, Pirello, Pitkin, Placidi, Plastino, Pluchar, Poggiani, Polini, Pong, Ponrathnam, Popolizio, Porter, Powell, Pracchia, Pradier, Prajapati, Prasai, Prasanna, Pratten, Prestegard, Principe, Prodi, Prokhorov, Prosposito, Prudenzi, Puecher, Punturo, Puosi, Puppo, P{\"u}rrer, Qi, Quetschke, Quinonez, {Quitzow-James}, Raab, Raaijmakers, Radkins, Radulesco, Raffai, Rail, Raja, Rajan, Ramirez, Ramirez, {Ramos-Buades}, Rana,
  Rapagnani, Rapol, Ratto, Ray, Raymond, Raza, Razzano, Read, Rees, Regimbau, Rei, Reid, Reitze, Relton, Rettegno, Ricci, Richardson, Richardson, Richardson, Ricker, Riemenschneider, Riles, Rizzo, Robertson, Robie, Robinet, Rocchi, Rocha, Rodriguez, {Rodriguez-Soto}, Rolland, Rollins, Roma, Romanelli, Romano, Romel, Romero, {Romero-Shaw}, Romie, Rose, Rosi{\'n}ska, Rosofsky, Ross, Rowan, Rowlinson, Roy, Roy, Rozza, Ruggi, Ryan, Sachdev, Sadecki, Sadiq, Sago, Saito, Saito, Sakai, Sakai, Sakellariadou, Sakuno, Salafia, Salconi, Saleem, Salemi, Samajdar, Sanchez, Sanchez, Sanchez, {Sanchis-Gual}, Sanders, Sanuy, Saravanan, Sarin, Sassolas, Satari, Sathyaprakash, Sato, Sato, Sauter, Savage, Savant, Sawada, Sawant, Sawant, Sayah, Schaetzl, Scheel, Scheuer, {Schindler-Tyka}, Schmidt, Schnabel, Schneewind, Schofield, Sch{\"o}nbeck, Schulte, Schutz, Schwartz, Scott, Scott, {Seglar-Arroyo}, Seidel, Sekiguchi, Sekiguchi, Sellers, Sengupta, Sennett, Sentenac, Seo, Sequino, Sergeev, Setyawati, Shaffer, Shahriar, Shams,
  Shao, Sharifi, Sharma, Sharma, Shawhan, Shcheblanov, Shen, Shibagaki, Shikauchi, Shimizu, Shimoda, Shimode, Shink, Shinkai, Shishido, Shoda, Shoemaker, Shoemaker, Shukla, ShyamSundar, Sieniawska, Sigg, Singer, Singh, Singh, Singha, Sintes, Sipala, Skliris, Slagmolen, {Slaven-Blair}, Smetana, Smith, Smith, Somala, Somiya, Son, Soni, Soni, Sorazu, Sordini, Sorrentino, Sorrentino, Sotani, Soulard, Souradeep, Sowell, Spagnuolo, Spencer, Spera, Srivastava, Srivastava, Staats, Stachie, Steer, Steinlechner, Steinlechner, Stops, Stevenson, Stover, Strain, Strang, Stratta, Strunk, Sturani, Stuver, S{\"u}dbeck, Sudhagar, Sudhir, Sugimoto, Suh, Summerscales, Sun, Sun, Sunil, Sur, Suresh, Sutton, Suzuki, Suzuki, Swinkels, Szczepa{\'n}czyk, Szewczyk, Tacca, Tagoshi, Tait, Takahashi, Takahashi, Takamori, Takano, Takeda, Takeda, Talbot, Tanaka, Tanaka, Tanaka, Tanaka, Tanaka, Tanasijczuk, Tanioka, Tanner, Tao, Tapia, Martin, Tasson, Telada, Tenorio, Terkowski, Test, Thirugnanasambandam, Thomas, Thomas, Thompson, Thondapu,
  Thorne, Thrane, Tiwari, Tiwari, Tiwari, Toland, Tolley, Tomaru, Tomigami, Tomura, Tonelli, {Torres-Forn{\'e}}, Torrie, e~Melo, T{\"o}yr{\"a}, Trapananti, Travasso, Traylor, Tringali, Tripathee, Troiano, Trovato, Trozzo, Trudeau, Tsai, Tsai, Tsang, Tsang, Tsao, Tse, Tso, Tsubono, Tsuchida, Tsukada, Tsuna, Tsutsui, Tsuzuki, Turconi, Tuyenbayev, Ubhi, Uchikata, Uchiyama, Udall, Ueda, Uehara, Ueno, Ueshima, Ugolini, Unnikrishnan, Uraguchi, Urban, Ushiba, Usman, Utina, Vahlbruch, Vajente, Vajpeyi, Valdes, Valentini, Valsan, van Bakel, van Beuzekom, van~den Brand, Broeck, {Vander-Hyde}, van~der Schaaf, van Heijningen, Vanosky, van Putten, Vardaro, Vargas, Varma, Vas{\'u}th, Vecchio, Vedovato, Veitch, Veitch, Venkateswara, Venneberg, Venugopalan, Verkindt, Verma, Veske, Vetrano, Vicer{\'e}, Viets, {Villa-Ortega}, Vinet, Vitale, Vo, Vocca, von Reis, von Wrangel, Vorvick, Vyatchanin, Wade, Wade, Wagner, Walet, Walker, Wallace, Wallace, Walsh, Wang, Wang, Wang, Ward, Warner, Was, Washimi, Washington, Watchi, Weaver,
  Wei, Weinert, Weinstein, Weiss, Weller, Wellmann, Wen, We{\ss}els, Westhouse, Wette, Whelan, White, Whiting, Whittle, Wilken, Williams, Williams, Williamson, Willis, Willke, Wilson, Winkler, Wipf, Wlodarczyk, Woan, Woehler, Wofford, Wong, Wu, Wu, Wu, Wu, Wysocki, Xiao, Xu, Yamada, Yamamoto, Yamamoto, Yamamoto, Yamamoto, Yamashita, Yamazaki, Yang, Yang, Yang, Yang, Yang, Yap, Yeeles, Yelikar, Ying, Yokogawa, Yokoyama, Yokozawa, Yoon, Yoshioka, Yu, Yu, Yuzurihara, Zadro{\.z}ny, Zanolin, Zappa, Zeidler, Zelenova, Zendri, Zevin, Zhan, Zhang, Zhang, Zhang, Zhang, Zhang, Zhao, Zhao, Zhao, Zhao, Zhou, Zhu, Zhu, Zimmerman, Zlochower, Zucker, Zweizig, {the LIGO Scientific Collaboration}, \& Collaboration}]{abbott_observation_2021}
Abbott, R., Abbott, T.~D., Abraham, S., {et~al.} 2021{\natexlab{a}}, The Astrophysical Journal Letters, 915, L5, \dodoi{10.3847/2041-8213/ac082e}

\bibitem[{Abbott {et~al.}(2021{\natexlab{b}})Abbott, Abbott, Abraham, Acernese, Ackley, Adams, Adams, Adhikari, Adya, Affeldt, Agathos, Agatsuma, Aggarwal, Aguiar, Aiello, Ain, Ajith, Akcay, Allen, Allocca, Altin, Amato, Anand, Ananyeva, Anderson, Anderson, Angelova, Ansoldi, Antelis, Antier, Appert, Arai, Araya, Areeda, Ar{\`e}ne, Arnaud, Aronson, Arun, Asali, Ascenzi, Ashton, Aston, Astone, Aubin, Aufmuth, AultONeal, Austin, Avendano, Babak, Badaracco, Bader, Bae, Baer, Bagnasco, Baird, Ball, Ballardin, Ballmer, Bals, Balsamo, Baltus, Banagiri, Bankar, Bankar, Barayoga, Barbieri, Barish, Barker, Barneo, Barnum, Barone, Barr, Barsotti, Barsuglia, Barta, Bartlett, Bartos, Bassiri, Basti, Bawaj, Bayley, Bazzan, Becher, B{\'e}csy, Bedakihale, Bejger, Belahcene, Beniwal, Benjamin, Bennett, Bentley, Bergamin, Berger, Bergmann, Bernuzzi, Berry, Bersanetti, Bertolini, Betzwieser, Bhandare, Bhandari, Bhattacharjee, Bidler, Bilenko, Billingsley, Birney, Birnholtz, Biscans, Bischi, Biscoveanu, Bisht, Bitossi,
  Bizouard, Blackburn, Blackman, Blair, Blair, Blair, Blanch, Bobba, Bode, Boer, Boetzel, Bogaert, Boldrini, Bondu, Bonilla, Bonnand, Booker, Boom, Bork, Boschi, Bose, Bossilkov, Boudart, Bouffanais, Bozzi, Bradaschia, Brady, Bramley, Branchesi, Brau, Breschi, Briant, Briggs, Brighenti, Brillet, Brinkmann, Brockill, Brooks, Brooks, Brown, Brunett, Bruno, Bruntz, Buikema, Bulik, Bulten, Buonanno, Buscicchio, Buskulic, Byer, Cabero, Cadonati, Caesar, Cagnoli, Cahillane, Calder{\'o}n~Bustillo, Callaghan, Callister, Calloni, Camp, Canepa, Cannon, Cao, Cao, Carapella, Carbognani, Carney, Carpinelli, Carullo, Carver, Casanueva~Diaz, Casentini, Caudill, Cavagli{\`a}, Cavalier, Cavalieri, Cella, {Cerd{\'a}-Dur{\'a}n}, Cesarini, Chaibi, Chakravarti, Chan, Chan, Chandra, Chanial, Chao, Charlton, Chase, {Chassande-Mottin}, Chatterjee, Chattopadhyay, Chaturvedi, Chatziioannou, Chen, Chen, Chen, Chen, Cheng, Cheong, Chia, Chiadini, Chierici, Chincarini, Chiummo, Cho, Cho, Cho, Choate, Christensen, Chu, Chua, Chung, Chung,
  Ciani, Ciecielag, Cie{\'s}lar, Cifaldi, Ciobanu, Ciolfi, Cipriano, Cirone, Clara, Clark, Clark, Clarke, Clearwater, Clesse, Cleva, Coccia, Cohadon, Cohen, Colleoni, Collette, Collins, Colpi, Constancio, Conti, Cooper, Corban, Corbitt, {Cordero-Carri{\'o}n}, Corezzi, Corley, Cornish, Corre, Corsi, Cortese, Costa, Cotesta, Coughlin, Coughlin, Coulon, Countryman, Cousins, Couvares, Covas, Coward, Cowart, Coyne, Coyne, Creighton, Creighton, Croquette, Crowder, Cudell, Cullen, Cumming, Cummings, Cunningham, Cuoco, Cury{\l}o, Canton, D{\'a}lya, Dana, DaneshgaranBajastani, D'Angelo, Danila, Danilishin, D'Antonio, Danzmann, {Darsow-Fromm}, Dasgupta, Datrier, Dattilo, Dave, Davier, Davies, Davis, Daw, Dean, DeBra, Deenadayalan, Degallaix, De~Laurentis, Del{\'e}glise, Del~Favero, De~Lillo, De~Lillo, Del~Pozzo, DeMarchi, De~Matteis, D'Emilio, Demos, Denker, Dent, Depasse, De~Pietri, De~Rosa, De~Rossi, DeSalvo, {de Varona}, Dhurandhar, D{\'i}az, {Diaz-Ortiz}, Didio, Dietrich, Di~Fiore, DiFronzo, Di~Giorgio,
  Di~Giovanni, Di~Giovanni, Di~Girolamo, Di~Lieto, Ding, Di~Pace, Di~Palma, Di~Renzo, Divakarla, Dmitriev, Doctor, D'Onofrio, Donovan, Dooley, Doravari, Dorrington, Downes, Drago, Driggers, Du, Ducoin, Dupej, Durante, D'Urso, Duverne, Dwyer, Easter, Eddolls, Edelman, Edo, Edy, Effler, Eichholz, Eikenberry, Eisenmann, Eisenstein, Ejlli, Errico, Essick, Estell{\'e}s, Estevez, Etienne, Etzel, Evans, Evans, Ewing, Fafone, Fair, Fairhurst, Fan, Farah, Farinon, Farr, Farr, {Fauchon-Jones}, Favata, Fays, Fazio, Feicht, Fejer, Feng, Fenyvesi, Ferguson, {Fernandez-Galiana}, Ferrante, Ferreira, Fidecaro, Figura, Fiori, Fiorucci, Fishbach, Fisher, Fishner, Fittipaldi, {Fitz-Axen}, Fiumara, Flaminio, Floden, Flynn, Fong, Font, Forsyth, Fournier, Frasca, Frasconi, Frei, Freise, Frey, Frey, Fritschel, Frolov, Fronz{\'e}, Fulda, Fyffe, Gabbard, Gadre, Gaebel, Gair, Gais, Galaudage, Gamba, Ganapathy, Ganguly, Gaonkar, Garaventa, {Garc{\'i}a-Quir{\'o}s}, Garufi, Gateley, Gaudio, Gayathri, Gemme, Gennai, George, George,
  George, Gergely, Ghonge, Ghosh, Ghosh, Ghosh, Giacomazzo, Giacoppo, Giaime, Giardina, Gibson, Gier, Gill, Giri, Glanzer, Gleckl, Godwin, Goetz, Goetz, Gohlke, Goncharov, Gonz{\'a}lez, Gopakumar, Gossan, Gosselin, Gouaty, Grace, Grado, Granata, Granata, Grant, Gras, Grassia, Gray, Gray, Greco, Green, Green, Gretarsson, Griggs, Grignani, Grimaldi, Grimes, Grimm, Grote, Grunewald, Gruning, Guerrero, Guidi, Guimaraes, Guix{\'e}, Gulati, Guo, Gupta, Gupta, Gupta, Gustafson, Gustafson, Guzman, Haegel, Halim, Hall, Hamilton, Hammond, Haney, Hanke, Hanks, Hanna, Hannam, Hannuksela, Hannuksela, Hansen, Hansen, Hanson, Harder, Hardwick, Haris, Harms, Harry, Harry, Hartwig, Hasskew, Haster, Haughian, Hayes, Healy, Heidmann, Heintze, Heinze, Heinzel, Heitmann, Hellman, Hello, {Helmling-Cornell}, Hemming, Hendry, Heng, Hennes, Hennig, Hennig, Hernandez~Vivanco, Heurs, Hild, Hill, Hines, Hochheim, Hofgard, Hofman, Hohmann, Holgado, Holland, Hollows, Holmes, Holt, Holz, Hopkins, Horst, Hough, Howell, Hoy, Hoyland, Huang,
  H{\"u}bner, Huddart, Huerta, Hughey, Hui, Husa, Huttner, Hutzler, Huxford, {Huynh-Dinh}, Idzkowski, Iess, Imperato, Inchauspe, Ingram, Intini, Isi, Iyer, JaberianHamedan, Jacqmin, Jadhav, Jadhav, James, Jani, Janssens, Janthalur, Jaranowski, Jariwala, Jaume, Jenkins, Jeunon, Jiang, Johns, {Johnson-McDaniel}, Jones, Jones, Jones, Jones, Jones, Jonker, Ju, Junker, Kalaghatgi, Kalogera, Kamai, Kandhasamy, Kang, Kanner, Kapadia, Kapasi, Karathanasis, Karki, Kashyap, Kasprzack, Kastaun, Katsanevas, Katsavounidis, Katzman, Kawabe, K{\'e}f{\'e}lian, Keitel, Key, Khadka, Khalili, Khan, Khan, Khazanov, Khetan, Khursheed, Kijbunchoo, Kim, Kim, Kim, Kim, Kim, Kim, Kimball, King, {Kinley-Hanlon}, Kirchhoff, Kissel, Kleybolte, Klimenko, Knowles, Knyazev, Koch, Koehlenbeck, Koekoek, Koley, Kolstein, Komori, Kondrashov, Kontos, Koper, Korobko, Korth, Kovalam, Kozak, Kr{\"a}mer, Kringel, Krishnendu, Kr{\'o}lak, Kuehn, Kumar, Kumar, Kumar, Kumar, Kuns, Kwang, Lackey, Laghi, Lalande, Lam, Lamberts, Landry, Lane, Lang, Lange,
  Lantz, Lanza, La~Rosa, {Lartaux-Vollard}, Lasky, Laxen, Lazzarini, Lazzaro, Leaci, Leavey, Lecoeuche, Lee, Lee, Lee, Lee, Lehmann, Leon, Leroy, Letendre, Levin, Li, Li, Li, Li, Li, Linde, Linker, Linley, Littenberg, Liu, Liu, {Llorens-Monteagudo}, Lo, Lockwood, London, Longo, Lorenzini, Loriette, Lormand, Losurdo, Lough, Lousto, Lovelace, L{\"u}ck, Lumaca, Lundgren, Ma, Macas, MacInnis, Macleod, MacMillan, Macquet, Maga{\~n}a~Hernandez, {Maga{\~n}a-Sandoval}, Magazz{\`u}, Magee, Majorana, Maksimovic, Maliakal, Malik, Man, Mandic, Mangano, Mansell, Manske, Mantovani, Mapelli, Marchesoni, Marion, M{\'a}rka, M{\'a}rka, Markakis, Markosyan, Markowitz, Maros, Marquina, Marsat, Martelli, Martin, Martin, Martinez, Martinez, Martynov, Masalehdan, Mason, Massera, Masserot, Massinger, {Masso-Reid}, Mastrogiovanni, Matas, {Mateu-Lucena}, Matichard, Matiushechkina, Mavalvala, Maynard, McCann, McCarthy, McClelland, McCormick, McCuller, McGuire, McIsaac, McIver, McManus, McRae, McWilliams, Meacher, Meadors, Mehmet,
  Mehta, Melatos, Melchor, Mendell, {Menendez-Vazquez}, Mercer, Mereni, Merfeld, Merilh, Merritt, Merzougui, Meshkov, Messenger, Messick, Metzdorff, Meyers, Meylahn, Mhaske, Miani, Miao, Michaloliakos, Michel, Middleton, Milano, Miller, Millhouse, Mills, Milotti, {Milovich-Goff}, Minazzoli, Minenkov, Mir, Mishkin, Mishra, Mistry, Mitra, Mitrofanov, Mitselmakher, Mittleman, Mo, Mogushi, Mohapatra, Mohite, Molina, {Molina-Ruiz}, Mondin, Montani, Moore, Moraru, Morawski, Moreno, Morisaki, Mours, {Mow-Lowry}, Mozzon, Muciaccia, Mukherjee, Mukherjee, Mukherjee, Mukherjee, Mukund, Mullavey, Munch, Mu{\~n}iz, Murray, Nadji, Nagar, Nardecchia, Naticchioni, Nayak, Neil, Neilson, Nelemans, Nelson, Nery, Neunzert, Nitz, Ng, Ng, Nguyen, Nguyen, Nguyen, Nichols, Nissanke, Nocera, Noh, North, Nothard, Nuttall, Oberling, O'Brien, O'Dell, Oganesyan, Ogin, Oh, Oh, Ohme, Ohta, Okada, Olivetto, Oppermann, Oram, O'Reilly, Ormiston, Ortega, O'Shaughnessy, Ossokine, Osthelder, Ottaway, Overmier, Owen, Pace, Pagano, Page,
  Pagliaroli, Pai, Pai, Palamos, Palashov, Palomba, Pan, Panda, Pang, Pankow, Pannarale, Pant, Paoletti, Paoli, Paolone, Parker, Pascucci, Pasqualetti, Passaquieti, Passuello, Patel, Patricelli, Payne, Pechsiri, Pedraza, Pegoraro, Pele, Penn, Perego, Perez, P{\'e}rigois, Perreca, Perri{\`e}s, Petermann, Petterson, Pfeiffer, Pham, Phukon, Piccinni, Pichot, Piendibene, Piergiovanni, Pierini, Pierro, Pillant, Pilo, Pinard, Pinto, Piotrzkowski, Pirello, Pitkin, Placidi, Plastino, Pluchar, Poggiani, Polini, Pong, Ponrathnam, Popolizio, Porter, Poverman, Powell, Pracchia, Prajapati, Prasai, Prasanna, Pratten, Prestegard, Principe, Prodi, Prokhorov, Prosposito, Prudenzi, Puecher, Punturo, Puosi, Puppo, P{\"u}rrer, Qi, Quetschke, Quinonez, {Quitzow-James}, Raab, Raaijmakers, Radkins, Radulesco, Raffai, Rafferty, Rail, Raja, Rajan, Rajbhandari, Rakhmanov, Ramirez, Ramirez, {Ramos-Buades}, Rana, Rao, Rapagnani, Rapol, Ratto, Raymond, Razzano, Read, Regimbau, Rei, Reid, Reitze, Rettegno, Ricci, Richardson, Richardson,
  Richardson, Ricker, Riemenschneider, Riles, Rizzo, Robertson, Robinet, Rocchi, Rocha, Rodriguez, {Rodriguez-Soto}, Rolland, Rollins, Roma, Romanelli, Romano, Romel, Romero, {Romero-Shaw}, Romie, Ronchini, Rose, Rose, Rose, Rosell, Rosi{\'n}ska, Rosofsky, Ross, Rowan, Rowlinson, Roy, Roy, Ruggi, Ryan, Sachdev, Sadecki, Sadiq, Sakellariadou, Salafia, Salconi, Saleem, Samajdar, Sanchez, Sanchez, Sanchez, {Sanchis-Gual}, Sanders, Sandles, Santiago, Santos, Saravanan, Sarin, Sassolas, Sathyaprakash, Sauter, Savage, Savant, Sawant, Sayah, Schaetzl, Schale, Scheel, Scheuer, {Schindler-Tyka}, Schmidt, Schnabel, Schofield, Sch{\"o}nbeck, Schreiber, Schulte, Schutz, Schwarm, Schwartz, Scott, Scott, {Seglar-Arroyo}, Seidel, Sellers, Sengupta, Sennett, Sentenac, Sequino, Sergeev, Setyawati, Shaffer, Shahriar, Sharifi, Sharma, Sharma, Shawhan, Shen, Shikauchi, Shink, Shoemaker, Shoemaker, Shukla, ShyamSundar, Sieniawska, Sigg, Singer, Singh, Singh, Singha, Singhal, Sintes, Sipala, Skliris, Slagmolen, {Slaven-Blair},
  Smetana, Smith, Smith, Somala, Son, Soni, Soni, Sorazu, Sordini, Sorrentino, Sorrentino, Soulard, Souradeep, Sowell, Spencer, Spera, Srivastava, Srivastava, Staats, Stachie, Steer, Steinhoff, Steinke, Steinlechner, Steinlechner, Steinmeyer, Stevenson, {Stolle-McAllister}, Stops, Stover, Strain, Stratta, Strunk, Sturani, Stuver, S{\"u}dbeck, Sudhagar, Sudhir, Suh, Summerscales, Sun, Sun, Sunil, Sur, Suresh, Sutton, Swinkels, Szczepa{\'n}czyk, Tacca, Tait, Talbot, Tanasijczuk, Tanner, Tao, Tapia, Tapia San~Martin, Tasson, Taylor, Tenorio, Terkowski, Thirugnanasambandam, Thomas, Thomas, Thomas, Thompson, Thondapu, Thorne, Thrane, Tiwari, Tiwari, Tiwari, Toland, Tolley, Tonelli, Tornasi, {Torres-Forn{\'e}}, Torrie, {e Melo}, T{\"o}yr{\"a}, Tran, Trapananti, Travasso, Traylor, Tringali, Tripathee, Trovato, Trudeau, Tsai, Tsang, Tse, Tso, Tsukada, Tsuna, Tsutsui, Turconi, Ubhi, Udall, Ueno, Ugolini, Unnikrishnan, Urban, Usman, Utina, Vahlbruch, Vajente, Vajpeyi, Valdes, Valentini, Valsan, {van Bakel}, {van
  Beuzekom}, {van den Brand}, Van Den~Broeck, {Vander-Hyde}, {van der Schaaf}, {van Heijningen}, Vardaro, Vargas, Varma, Vass, Vas{\'u}th, Vecchio, Vedovato, Veitch, Veitch, Venkateswara, Venneberg, Venugopalan, Verkindt, Verma, Veske, Vetrano, Vicer{\'e}, Viets, Vijaykumar, {Villa-Ortega}, Vinet, Vitale, Vo, Vocca, Vorvick, Vyatchanin, Wade, Wade, Wade, Walet, Walker, Wallace, Wallace, Walsh, Wang, Wang, Wang, Wang, Ward, Warner, Was, Washington, Watchi, Weaver, Wei, Weinert, Weinstein, Weiss, Wellmann, Wen, We{\ss}els, Westhouse, Wette, Whelan, White, White, Whiting, Whittle, Wilken, Williams, Williams, Williamson, Willis, Willke, Wilson, Wimmer, Winkler, Wipf, Woan, Woehler, Wofford, Wong, Wrangel, Wright, Wu, Wysocki, Xiao, Yamamoto, Yang, Yang, Yang, Yap, Yeeles, Yoon, Yu, Yu, Yuen, Zadro{\.z}ny, Zanolin, Zelenova, Zendri, Zevin, Zhang, Zhang, Zhang, Zhang, Zhao, Zhao, Zheng, Zhou, Zhou, Zhu, Zimmerman, Zlochower, Zucker, \& Zweizig}]{abbott_gwtc-2:_2021}
---. 2021{\natexlab{b}}, Physical Review X, 11, 021053, \dodoi{10.1103/PhysRevX.11.021053}

\bibitem[{Abbott {et~al.}(2023)Abbott, Abbott, Acernese, Ackley, Adams, Adhikari, Adhikari, Adya, Affeldt, Agarwal, Agathos, Agatsuma, Aggarwal, Aguiar, Aiello, Ain, Ajith, Akcay, Akutsu, Albanesi, Allocca, Altin, Amato, Anand, Anand, Ananyeva, Anderson, Anderson, Ando, Andrade, Andres, Andri{\'c}, Angelova, Ansoldi, Antelis, Antier, Appert, Arai, Arai, Arai, Araki, Araya, Araya, Areeda, Ar{\`e}ne, Aritomi, Arnaud, Arogeti, Aronson, Arun, Asada, Asali, Ashton, Aso, Assiduo, Aston, Astone, Aubin, Austin, Babak, Badaracco, Bader, Badger, Bae, Bae, Baer, Bagnasco, Bai, Baiotti, Baird, Bajpai, Ball, Ballardin, Ballmer, Balsamo, Baltus, Banagiri, Bankar, Barayoga, Barbieri, Barish, Barker, Barneo, Barone, Barr, Barsotti, Barsuglia, Barta, Bartlett, Barton, Bartos, Bassiri, Basti, Bawaj, Bayley, Baylor, Bazzan, B{\'e}csy, Bedakihale, Bejger, Belahcene, Benedetto, Beniwal, Bennett, Bentley, BenYaala, Bergamin, Berger, Bernuzzi, Berry, Bersanetti, Bertolini, Betzwieser, Beveridge, Bhandare, Bhardwaj, Bhattacharjee,
  Bhaumik, Bilenko, Billingsley, Bini, Birney, Birnholtz, Biscans, Bischi, Biscoveanu, Bisht, Biswas, Bitossi, Bizouard, Blackburn, Blair, Blair, Blair, Bobba, Bode, Boer, Bogaert, Boldrini, Bonavena, Bondu, Bonilla, Bonnand, Booker, Boom, Bork, Boschi, Bose, Bose, Bossilkov, Boudart, Bouffanais, Bozzi, Bradaschia, Brady, Bramley, Branch, Branchesi, Brandt, Brau, Breschi, Briant, Briggs, Brillet, Brinkmann, Brockill, Brooks, Brooks, Brown, Brunett, Bruno, Bruntz, Bryant, Bulik, Bulten, Buonanno, Buscicchio, Buskulic, Buy, Byer, Davies, Cadonati, Cagnoli, Cahillane, Bustillo, Callaghan, Callister, Calloni, Cameron, Camp, Canepa, Canevarolo, Cannavacciuolo, Cannon, Cao, Cao, Capocasa, Capote, Carapella, Carbognani, Carlin, Carney, Carpinelli, Carrillo, Carullo, Carver, Diaz, Casentini, Castaldi, Caudill, Cavagli{\`a}, Cavalier, Cavalieri, Ceasar, Cella, {Cerd{\'a}-Dur{\'a}n}, Cesarini, Chaibi, Chakravarti, Subrahmanya, Champion, Chan, Chan, Chan, Chan, Chan, Chandra, Chanial, Chao, {Chapman-Bird}, Charlton,
  Chase, {Chassande-Mottin}, Chatterjee, Chatterjee, Chatterjee, Chaturvedi, Chaty, Chatziioannou, Chen, Chen, Chen, Chen, Chen, Chen, Chen, Chen, Cheng, Cheong, Cheung, Chia, Chiadini, Chiang, Chiarini, Chierici, Chincarini, Chiofalo, Chiummo, Cho, Cho, Choudhary, Choudhary, Christensen, Chu, Chu, Chu, Chua, Chung, Ciani, Ciecielag, Cie{\'s}lar, Cifaldi, Ciobanu, Ciolfi, Cipriano, Cirone, Clara, Clark, Clark, Clarke, Clearwater, Clesse, Cleva, Coccia, Codazzo, Cohadon, Cohen, Cohen, Colleoni, Collette, Colombo, Colpi, Compton, Constancio, Conti, Cooper, Corban, Corbitt, {Cordero-Carri{\'o}n}, Corezzi, Corley, Cornish, Corre, Corsi, Cortese, Costa, Cotesta, Coughlin, Coulon, Countryman, Cousins, Couvares, Coward, Cowart, Coyne, Coyne, Creighton, Creighton, Criswell, Croquette, Crowder, Cudell, Cullen, Cumming, Cummings, Cunningham, Cuoco, Cury{\l}o, Dabadie, Canton, Dall'Osso, D{\'a}lya, Dana, DaneshgaranBajastani, D'Angelo, Danila, Danilishin, D'Antonio, Danzmann, {Darsow-Fromm}, Dasgupta, Datrier, Dattilo,
  Dave, Davier, Davis, Davis, Daw, {de Alarc{\'o}n}, Dean, DeBra, Deenadayalan, Degallaix, De~Laurentis, Del{\'e}glise, Del~Favero, De~Lillo, De~Lillo, Del~Pozzo, DeMarchi, De~Matteis, D'Emilio, Demos, Dent, Depasse, De~Pietri, De~Rosa, De~Rossi, DeSalvo, De~Simone, Dhurandhar, D{\'i}az, {Diaz-Ortiz}, Didio, Dietrich, Di~Fiore, Di~Fronzo, Di~Giorgio, Di~Giovanni, Di~Giovanni, Di~Girolamo, Di~Lieto, Ding, Di~Pace, Di~Palma, Di~Renzo, Divakarla, Dmitriev, Doctor, D'Onofrio, Donovan, Dooley, Doravari, Dorrington, Drago, Driggers, Drori, Ducoin, Dupej, Durante, D'Urso, Duverne, Dwyer, Eassa, Easter, Ebersold, Eckhardt, Eddolls, Edelman, Edo, Edy, Effler, Eguchi, Eichholz, Eikenberry, Eisenmann, Eisenstein, Ejlli, Engelby, Enomoto, Errico, Essick, Estell{\'e}s, Estevez, Etienne, Etzel, Evans, Evans, Ewing, Fafone, Fair, Fairhurst, Farah, Farinon, Farr, Farr, Farrow, {Fauchon-Jones}, Favaro, Favata, Fays, Fazio, Feicht, Fejer, Fenyvesi, Ferguson, {Fernandez-Galiana}, Ferrante, Ferreira, Fidecaro, Figura, Fiori,
  Fishbach, Fisher, Fittipaldi, Fiumara, Flaminio, Floden, Fong, Font, Fornal, Forsyth, Franke, Frasca, Frasconi, Frederick, Freed, Frei, Freise, Frey, Fritschel, Frolov, Fronz{\'e}, Fujii, Fujikawa, Fukunaga, Fukushima, Fulda, Fyffe, Gabbard, Gabella, Gadre, Gair, Gais, Galaudage, Gamba, Ganapathy, Ganguly, Gao, Gaonkar, Garaventa, Garc{\'i}a, {Garc{\'i}a-N{\'u}{\~n}ez}, {Garc{\'i}a-Quir{\'o}s}, Garufi, Gateley, Gaudio, Gayathri, Ge, Gemme, Gennai, George, George, Gerberding, Gergely, Gewecke, Ghonge, Ghosh, Ghosh, Ghosh, Ghosh, Giacomazzo, Giacoppo, Giaime, Giardina, Gibson, Gier, Giesler, Giri, Gissi, Glanzer, Gleckl, Godwin, Goetz, Goetz, Gohlke, Golomb, Goncharov, Gonz{\'a}lez, Gopakumar, Gosselin, Gouaty, Gould, Grace, Grado, Granata, Granata, Grant, Gras, Grassia, Gray, Gray, Greco, Green, Green, Gretarsson, Gretarsson, Griffith, Griffiths, Griggs, Grignani, Grimaldi, Grimm, Grote, Grunewald, Gruning, Guerra, Guidi, Guimaraes, Guix{\'e}, Gulati, Guo, Guo, Gupta, Gupta, Gupta, Gustafson, Gustafson,
  Guzman, Ha, Haegel, Hagiwara, Haino, Halim, Hall, Hamilton, Hammond, Han, Haney, Hanks, Hanna, Hannam, Hannuksela, Hansen, Hansen, Hanson, Harder, Hardwick, Haris, Harms, Harry, Harry, Hartwig, Hasegawa, Haskell, Hasskew, Haster, Hattori, Haughian, Hayakawa, Hayama, Hayes, Healy, Heidmann, Heidt, Heintze, Heinze, Heinzel, Heitmann, Hellman, Hello, {Helmling-Cornell}, Hemming, Hendry, Heng, Hennes, Hennig, Hennig, Hernandez, Hernandez~Vivanco, Heurs, Hild, Hill, Himemoto, Hines, Hiranuma, Hirata, Hirose, Hochheim, Hofman, Hohmann, Holcomb, Holland, {Holley-Bockelmann}, Hollows, Holmes, Holt, Holz, Hong, Hopkins, Hough, Hourihane, Howell, Hoy, Hoyland, Hreibi, Hsieh, Hsu, Huang, Huang, Huang, Huang, Huang, Huang, H{\"u}bner, Huddart, Hughey, Hui, Hui, Husa, Huttner, Huxford, {Huynh-Dinh}, Ide, Idzkowski, Iess, Ikenoue, Imam, Inayoshi, Ingram, Inoue, Ioka, Isi, Isleif, Ito, Itoh, Iyer, Izumi, JaberianHamedan, Jacqmin, Jadhav, Jadhav, James, Jan, Jani, Janquart, Janssens, Janthalur, Jaranowski, Jariwala, Jaume,
  Jenkins, Jenner, Jeon, Jeunon, Jia, Jin, Johns, {Johnson-McDaniel}, Jones, Jones, Jones, Jones, Jones, Jonker, Ju, Jung, Jung, Junker, Juste, Kaihotsu, Kajita, Kakizaki, Kalaghatgi, Kalogera, Kamai, Kamiizumi, Kanda, Kandhasamy, Kang, Kanner, Kao, Kapadia, Kapasi, Karat, Karathanasis, Karki, Kashyap, Kasprzack, Kastaun, Katsanevas, Katsavounidis, Katzman, Kaur, Kawabe, Kawaguchi, Kawai, Kawasaki, K{\'e}f{\'e}lian, Keitel, Key, Khadka, Khalili, Khan, Khazanov, Khetan, Khursheed, Kijbunchoo, Kim, Kim, Kim, Kim, Kim, Kim, Kimball, Kimura, {Kinley-Hanlon}, Kirchhoff, Kissel, Kita, Kitazawa, Kleybolte, Klimenko, Knee, Knowles, Knyazev, Koch, Koekoek, Kojima, Kokeyama, Koley, Kolitsidou, Kolstein, Komori, Kondrashov, Kong, Kontos, Koper, Korobko, Kotake, Kovalam, Kozak, Kozakai, Kozu, Kringel, Krishnendu, Kr{\'o}lak, Kuehn, Kuei, Kuijer, Kulkarni, Kumar, Kumar, Kumar, Kumar, Kume, Kuns, Kuo, Kuo, Kuromiya, Kuroyanagi, Kusayanagi, Kuwahara, Kwak, Lagabbe, Laghi, Lalande, Lam, Lamberts, Landry, Lane, Lang, Lange,
  Lantz, La~Rosa, {Lartaux-Vollard}, Lasky, Laxen, Lazzarini, Lazzaro, Leaci, Leavey, Lecoeuche, Lee, Lee, Lee, Lee, Lee, Lee, Lehmann, Lema{\^i}tre, Leonardi, Leroy, Letendre, Levesque, Levin, Leviton, Leyde, Li, Li, Li, Li, Li, Li, Lin, Lin, Lin, Lin, Lin, Linde, Linker, Linley, Littenberg, Liu, Liu, Liu, Liu, Llamas, {Llorens-Monteagudo}, Lo, Lockwood, Loh, London, Longo, Lopez, Portilla, Lorenzini, Loriette, Lormand, Losurdo, Lott, Lough, Lousto, Lovelace, Lucaccioni, L{\"u}ck, Lumaca, Lundgren, Luo, Lynam, Macas, MacInnis, Macleod, MacMillan, Macquet, Hernandez, Magazz{\`u}, Magee, Maggiore, Magnozzi, Mahesh, Majorana, Makarem, Maksimovic, Maliakal, Malik, Man, Mandic, Mangano, Mango, Mansell, Manske, Mantovani, Mapelli, Marchesoni, Marchio, Marion, Mark, M{\'a}rka, M{\'a}rka, Markakis, Markosyan, Markowitz, Maros, Marquina, Marsat, Martelli, Martin, Martin, Martinez, Martinez, Martinez, Martinovic, Martynov, Marx, Masalehdan, Mason, Massera, Masserot, Massinger, {Masso-Reid}, Mastrogiovanni, Matas,
  {Mateu-Lucena}, Matichard, Matiushechkina, Mavalvala, McCann, McCarthy, McClelland, McClincy, McCormick, McCuller, McGhee, McGuire, McIsaac, McIver, McRae, McWilliams, Meacher, Mehmet, Mehta, Meijer, Melatos, Melchor, Mendell, {Menendez-Vazquez}, Menoni, Mercer, Mereni, Merfeld, Merilh, Merritt, Merzougui, Meshkov, Messenger, Messick, Meyers, Meylahn, Mhaske, Miani, Miao, Michaloliakos, Michel, Michimura, Middleton, Milano, Miller, Miller, Miller, Millhouse, Mills, Milotti, Minazzoli, Minenkov, Mio, Mir, {Miravet-Ten{\'e}s}, Mishra, Mishra, Mistry, Mitra, Mitrofanov, Mitselmakher, Mittleman, Miyakawa, Miyamoto, Miyazaki, Miyo, Miyoki, Mo, Modafferi, Moguel, Mogushi, Mohapatra, Mohite, Molina, {Molina-Ruiz}, Mondin, Montani, Moore, Moraru, Morawski, More, Moreno, Moreno, Mori, Morisaki, Moriwaki, Morr{\'a}s, Mours, {Mow-Lowry}, Mozzon, Muciaccia, Mukherjee, Mukherjee, Mukherjee, Mukherjee, Mukherjee, Mukund, Mullavey, Munch, Mu{\~n}iz, Murray, Musenich, Muusse, Nadji, Nagano, Nagano, Nagar, Nakamura, Nakano,
  Nakano, Nakashima, Nakayama, Napolano, Nardecchia, Narikawa, Naticchioni, Nayak, Nayak, Negishi, Neil, Neilson, Nelemans, Nelson, Nery, Neubauer, Neunzert, Ng, Ng, Nguyen, Nguyen, Nguyen, Quynh, Ni, Nichols, Nishizawa, Nissanke, Nitoglia, Nocera, Norman, North, Nozaki, Siles, Nuttall, Oberling, O'Brien, Obuchi, O'Dell, Oelker, Ogaki, Oganesyan, Oh, Oh, Oh, Ohashi, Ohishi, Ohkawa, Ohme, Ohta, Okada, Okutani, Okutomi, Olivetto, Oohara, Ooi, Oram, O'Reilly, Ormiston, Ormsby, Ortega, O'Shaughnessy, O'Shea, Oshino, Ossokine, Osthelder, Otabe, Ottaway, Overmier, Pace, Pagano, Page, Pagliaroli, Pai, Pai, Palamos, Palashov, Palomba, Pan, Pan, Panda, Pang, Pang, Pankow, Pannarale, Pant, Panther, Paoletti, Paoli, Paolone, Parisi, Park, Park, Parker, Pascucci, Pasqualetti, Passaquieti, Passuello, Patel, Pathak, Patricelli, Patron, Paul, Payne, Pedraza, Pegoraro, Pele, Arellano, Penn, Perego, Pereira, Pereira, Perez, P{\'e}rigois, Perkins, Perreca, Perri{\`e}s, Petermann, Petterson, Pfeiffer, Pham, Phukon, Piccinni,
  Pichot, Piendibene, Piergiovanni, Pierini, Pierro, Pillant, Pillas, Pilo, Pinard, Pinto, Pinto, Piotrzkowski, Piotrzkowski, Pirello, Pitkin, Placidi, Planas, Plastino, Pluchar, Poggiani, Polini, Pong, Ponrathnam, Popolizio, Porter, Poulton, Powell, Pracchia, Pradier, Prajapati, Prasai, Prasanna, Pratten, Principe, Prodi, Prokhorov, Prosposito, Prudenzi, Puecher, Punturo, Puosi, Puppo, P{\"u}rrer, Qi, Quetschke, {Quitzow-James}, Qutob, Raab, Raaijmakers, Radkins, Radulesco, Raffai, Rail, Raja, Rajan, Ramirez, Ramirez, {Ramos-Buades}, Rana, Rapagnani, Rapol, Ray, Raymond, Raza, Razzano, Read, Rees, Regimbau, Rei, Reid, Reid, Reitze, Relton, Renzini, Rettegno, Reza, Rezac, Ricci, Richards, Richardson, Richardson, Riemenschneider, Riles, Rinaldi, Rink, Rizzo, Robertson, Robie, Robinet, Rocchi, Rodriguez, Rolland, Rollins, Romanelli, Romano, Romel, {Romero-Rodr{\'i}guez}, {Romero-Shaw}, Romie, Ronchini, Rosa, Rose, Rosi{\'n}ska, Ross, Rowan, Rowlinson, Roy, Roy, Roy, Rozza, Ruggi, {Ruiz-Rocha}, Ryan, Sachdev,
  Sadecki, Sadiq, Sago, Saito, Saito, Sakai, Sakai, Sakellariadou, Sakuno, Salafia, Salconi, Saleem, Salemi, Samajdar, Sanchez, Sanchez, Sanchez, {Sanchis-Gual}, Sanders, Sanuy, Saravanan, Sarin, Sassolas, Satari, Sathyaprakash, Sato, Sato, Sauter, Savage, Sawada, Sawant, Sawant, Sayah, Schaetzl, Scheel, Scheuer, Schiworski, Schmidt, Schmidt, Schnabel, Schneewind, Schofield, Sch{\"o}nbeck, Schulte, Schutz, Schwartz, Scott, Scott, {Seglar-Arroyo}, Sekiguchi, Sekiguchi, Sellers, Sengupta, Sentenac, Seo, Sequino, Sergeev, Setyawati, Shaffer, Shahriar, Shams, Shao, Sharma, Sharma, Shawhan, Shcheblanov, Shibagaki, Shikauchi, Shimizu, Shimoda, Shimode, Shinkai, Shishido, Shoda, Shoemaker, Shoemaker, ShyamSundar, Sieniawska, Sigg, Singer, Singh, Singh, Singha, Sintes, Sipala, Skliris, Slagmolen, {Slaven-Blair}, Smetana, Smith, Smith, Soldateschi, Somala, Somiya, Son, Soni, Soni, Sordini, Sorrentino, Sorrentino, Sotani, Soulard, Souradeep, Sowell, Spagnuolo, Spencer, Spera, Srinivasan, Srivastava, Srivastava, Staats,
  Stachie, Steer, Steinhoff, Steinlechner, Steinlechner, Stevenson, Stops, Stover, Strain, Strang, Stratta, Strunk, Sturani, Stuver, Sudhagar, Sudhir, Sugimoto, Suh, Sullivan, Sullivan, Summerscales, Sun, Sun, Sunil, Sur, Suresh, Sutton, Suzuki, Suzuki, Swinkels, Szczepa{\'n}czyk, Szewczyk, Tacca, Tagoshi, Tait, Takahashi, Takahashi, Takamori, Takano, Takeda, Takeda, Talbot, Talbot, Tanaka, Tanaka, Tanaka, Tanaka, Tanaka, Tanasijczuk, Tanioka, Tanner, Tao, Tao, Mart{\'i}n, Taranto, Tasson, Telada, Tenorio, Terhune, Terkowski, Thirugnanasambandam, Thomas, Thomas, Thomas, Thompson, Thondapu, Thorne, Thrane, Tiwari, Tiwari, Tiwari, Toivonen, Toland, Tolley, Tomaru, Tomigami, Tomura, Tonelli, {Torres-Forn{\'e}}, Torrie, {e Melo}, T{\"o}yr{\"a}, Trapananti, Travasso, Traylor, Trevor, Tringali, Tripathee, Troiano, Trovato, Trozzo, Trudeau, Tsai, Tsai, Tsang, Tsang, Tsao, Tse, Tso, Tsubono, Tsuchida, Tsukada, Tsuna, Tsutsui, Tsuzuki, Turbang, Turconi, Tuyenbayev, Ubhi, Uchikata, Uchiyama, Udall, Ueda, Uehara, Ueno,
  Ueshima, Unnikrishnan, Uraguchi, Urban, Ushiba, Utina, Vahlbruch, Vajente, Vajpeyi, Valdes, Valentini, Valsan, {van Bakel}, {van Beuzekom}, {van den Brand}, Van Den~Broeck, {Vander-Hyde}, {van der Schaaf}, {van Heijningen}, Vanosky, {van Putten}, {van Remortel}, Vardaro, Vargas, Varma, Vas{\'u}th, Vecchio, Vedovato, Veitch, Veitch, Venneberg, Venugopalan, Verkindt, Verma, Verma, Veske, Vetrano, Vicer{\'e}, Vidyant, Viets, Vijaykumar, {Villa-Ortega}, Vinet, Virtuoso, Vitale, Vo, Vocca, {von Reis}, {von Wrangel}, Vorvick, Vyatchanin, Wade, Wade, Wagner, Walet, Walker, Wallace, Wallace, Walsh, Wang, Wang, Wang, Ward, Warner, Was, Washimi, Washington, Watchi, Weaver, Webster, Weinert, Weinstein, Weiss, Weller, Weller, Wellmann, Wen, We{\ss}els, Wette, Whelan, White, Whiting, Whittle, Wilken, Williams, Williams, Williams, Williamson, Willis, Willke, Wilson, Winkler, Wipf, Wlodarczyk, Woan, Woehler, Wofford, Wong, Wu, Wu, Wu, Wu, Wysocki, Xiao, Xu, Yamada, Yamamoto, Yamamoto, Yamamoto, Yamamoto, Yamashita,
  Yamazaki, Yang, Yang, Yang, Yang, Yang, Yap, Yeeles, Yelikar, Ying, Yokogawa, Yokoyama, Yokozawa, Yoo, Yoshioka, Yu, Yu, Yuzurihara, Zadro{\.z}ny, Zanolin, Zeidler, Zelenova, Zendri, Zevin, Zhan, Zhang, Zhang, Zhang, Zhang, Zhang, Zhao, Zhao, Zhao, Zhao, Zheng, Zhou, Zhou, Zhu, Zhu, Zimmerman, Zlochower, Zucker, \& Zweizig}]{abbott_gwtc-3:_2023}
Abbott, R., Abbott, T.~D., Acernese, F., {et~al.} 2023, Physical Review X, 13, 041039, \dodoi{10.1103/PhysRevX.13.041039}

\bibitem[{Ahumada {et~al.}(2024)Ahumada, Anand, Coughlin, Gupta, Kasliwal, Karambelkar, Stein, Waratkar, Swain, {Jegou du Laz}, Anumarlapudi, Andreoni, Bulla, Srinivasaragavan, Toivonen, Wold, Bellm, Cenko, Kaplan, Sollerman, Bhalerao, Perley, Salgundi, Suresh, Hinds, Reusch, Necker, Cook, Pletskova, Singer, Banerjee, Barna, Copperwheat, Healy, Kiendrebeogo, Kumar, Kumar, Pezzella, {Sagu{\'e}s-Carracedo}, Sravan, Bloom, Chen, Graham, Helou, Laher, Mahabal, Purdum, Anupama, Barway, Basu, Raman, \& Roychowdhury}]{ahumada_searching_2024}
Ahumada, T., Anand, S., Coughlin, M.~W., {et~al.} 2024, Publications of the Astronomical Society of the Pacific, 136, 114201, \dodoi{10.1088/1538-3873/ad8265}

\bibitem[{Akeson {et~al.}(2019)Akeson, Armus, Bachelet, Bailey, Bartusek, Bellini, Benford, Bennett, Bhattacharya, Bohlin, Boyer, Bozza, Bryden, Novati, Carpenter, Casertano, Choi, Content, Dayal, Dressler, Dor{\'e}, Fall, Fan, Fang, Filippenko, Finkelstein, Foley, Furlanetto, Kalirai, Gaudi, Gilbert, Girard, Grady, Greene, Guhathakurta, Heinrich, Hemmati, Hendel, Henderson, Henning, Hirata, Ho, Huff, Hutter, Jansen, Jha, Johnson, Jones, Kasdin, Kelly, Kirshner, Koekemoer, Kruk, Lewis, Macintosh, Madau, Malhotra, Mandel, Massara, Masters, McEnery, McQuinn, Melchior, Melton, Mennesson, Peeples, Penny, Perlmutter, Pisani, Plazas, Poleski, Postman, Ranc, Rauscher, Rest, Roberge, Robertson, Rodney, Rhoads, Rhodes, Ryan~Jr., Sahu, Sand, Scolnic, Seth, Shvartzvald, Siellez, Smith, Spergel, Stassun, Street, Strolger, Szalay, Trauger, Troxel, Turnbull, {van der Marel}, {von der Linden}, Wang, Weinberg, Williams, Windhorst, Wollack, Wu, Yee, \& Zimmerman}]{akeson_the_2019}
Akeson, R., Armus, L., Bachelet, E., {et~al.} 2019, The {{Wide Field Infrared Survey Telescope}}: 100 {{Hubbles}} for the 2020s,  arXiv, \dodoi{10.48550/arXiv.1902.05569}

\bibitem[{{Al-Mamun} {et~al.}(2021){Al-Mamun}, Steiner, N{\"a}ttil{\"a}, Lange, O'Shaughnessy, Tews, Gandolfi, Heinke, \& Han}]{al-mamun_combining_2021}
{Al-Mamun}, M., Steiner, A.~W., N{\"a}ttil{\"a}, J., {et~al.} 2021, Physical Review Letters, 126, 061101, \dodoi{10.1103/PhysRevLett.126.061101}

\bibitem[{Alexander {et~al.}(2017)Alexander, Berger, Fong, Williams, Guidorzi, Margutti, Metzger, Annis, Blanchard, Brout, Brown, Chen, Chornock, Cowperthwaite, Drout, Eftekhari, Frieman, Holz, Nicholl, Rest, Sako, {Soares-Santos}, \& Villar}]{alexander_the_2017}
Alexander, K.~D., Berger, E., Fong, W., {et~al.} 2017, The Astrophysical Journal, 848, L21, \dodoi{10.3847/2041-8213/aa905d}

\bibitem[{Anand {et~al.}(2021)Anand, Coughlin, Kasliwal, Bulla, Ahumada, Sagu{\'e}s~Carracedo, Almualla, Andreoni, Stein, Foucart, Singer, Sollerman, Bellm, Bolin, {Caballero-Garc{\'i}a}, {Castro-Tirado}, Cenko, De, Dekany, Duev, Feeney, Fremling, Goldstein, Golkhou, Graham, Guessoum, Hankins, Hu, Kong, Kool, Kulkarni, Kumar, Laher, Masci, Mr{\'o}z, Nissanke, Porter, Reusch, Riddle, Rosnet, Rusholme, Serabyn, {S{\'a}nchez-Ram{\'i}rez}, Rigault, Shupe, Smith, Soumagnac, Walters, \& Valeev}]{anand_optical_2021}
Anand, S., Coughlin, M.~W., Kasliwal, M.~M., {et~al.} 2021, Nature Astronomy, 5, 46, \dodoi{10.1038/s41550-020-1183-3}

\bibitem[{Antier {et~al.}(2020{\natexlab{a}})Antier, Agayeva, Aivazyan, Alishov, Arbouch, Baransky, Barynova, Bai, Basa, Beradze, Bertin, Berthier, Bla{\v z}ek, Bo{\"e}r, Burkhonov, Burrell, Cailleau, Chabert, Chen, Christensen, Coleiro, Cordier, Corre, Coughlin, Coward, Crisp, Delattre, Dietrich, Ducoin, Duverne, {Marchal-Duval}, Gendre, Eymar, {Fock-Hang}, Han, Hello, Howell, Inasaridze, Ismailov, Kann, Kapanadze, Klotz, Kochiashvili, Lachaud, Leroy, Le~Van~Su, Lin, Li, Lognone, Marron, Mo, Moore, Natsvlishvili, Noysena, Perrigault, Peyrot, Samadov, Sadibekova, Simon, Stachie, Teng, Thierry, Th{\"o}ne, Tillayev, Turpin, {de~Ugarte~Postigo}, Vachier, Vardosanidze, Vasylenko, Vidadi, Wang, Wang, Wei, Yan, Zhang, Zhang, \& Zhang}]{antier_the_2020}
Antier, S., Agayeva, S., Aivazyan, V., {et~al.} 2020{\natexlab{a}}, Monthly Notices of the Royal Astronomical Society, 492, 3904, \dodoi{10.1093/mnras/stz3142}

\bibitem[{Antier {et~al.}(2020{\natexlab{b}})Antier, Agayeva, Almualla, Awiphan, Baransky, Barynova, Beradze, Bla{\v z}ek, Bo{\"e}r, Burkhonov, Christensen, Coleiro, Corre, Coughlin, Crisp, Dietrich, Ducoin, Duverne, {Marchal-Duval}, Gendre, Gokuldass, Eggenstein, Eymar, Hello, Howell, Ismailov, Kann, Karpov, Klotz, Kochiashvili, Lachaud, Leroy, Lin, Li, Ma{\v s}ek, Mo, Menard, Morris, Noysena, Orange, Prouza, Rattanamala, Sadibekova, {Saint-Gelais}, Serrau, Simon, Stachie, Th{\"o}ne, Tillayev, Turpin, Postigo, Vasylenko, Vidadi, Was, Wang, Zhang, Zhang, \& Zhang}]{antier_grandma_2020}
Antier, S., Agayeva, S., Almualla, M., {et~al.} 2020{\natexlab{b}}, Monthly Notices of the Royal Astronomical Society, 497, 5518, \dodoi{10.1093/mnras/staa1846}

\bibitem[{{Astropy Collaboration} {et~al.}(2022){Astropy Collaboration}, {Price-Whelan}, Lim, Earl, Starkman, Bradley, Shupe, Patil, Corrales, Brasseur, N{\"o}the, Donath, Tollerud, Morris, Ginsburg, Vaher, Weaver, Tocknell, Jamieson, {van Kerkwijk}, Robitaille, Merry, Bachetti, G{\"u}nther, Aldcroft, {Alvarado-Montes}, Archibald, B{\'o}di, Bapat, Barentsen, Baz{\'a}n, Biswas, Boquien, Burke, Cara, Cara, Conroy, Conseil, Craig, Cross, Cruz, D'Eugenio, Dencheva, Devillepoix, Dietrich, Eigenbrot, Erben, Ferreira, {Foreman-Mackey}, Fox, Freij, Garg, Geda, Glattly, Gondhalekar, Gordon, Grant, Greenfield, Groener, Guest, Gurovich, Handberg, Hart, {Hatfield-Dodds}, Homeier, Hosseinzadeh, Jenness, Jones, Joseph, Kalmbach, Karamehmetoglu, Ka{\l}uszy{\'n}ski, Kelley, Kern, Kerzendorf, Koch, Kulumani, Lee, Ly, Ma, MacBride, Maljaars, Muna, Murphy, Norman, O'Steen, Oman, Pacifici, Pascual, {Pascual-Granado}, Patil, Perren, Pickering, Rastogi, Roulston, Ryan, Rykoff, Sabater, Sakurikar, Salgado, Sanghi, Saunders,
  Savchenko, Schwardt, {Seifert-Eckert}, Shih, Jain, Shukla, Sick, Simpson, Singanamalla, Singer, Singhal, Sinha, Sip{\H o}cz, Spitler, Stansby, Streicher, {\v S}umak, Swinbank, Taranu, Tewary, Tremblay, {de Val-Borro}, Van~Kooten, Vasovi{\'c}, Verma, {de Miranda Cardoso}, Williams, Wilson, Winkel, {Wood-Vasey}, Xue, Yoachim, Zhang, Zonca, \& {Astropy Project Contributors}}]{astropycollaboration_the_2022a}
{Astropy Collaboration}, {Price-Whelan}, A.~M., Lim, P.~L., {et~al.} 2022, The Astrophysical Journal, 935, 167, \dodoi{10.3847/1538-4357/ac7c74}

\bibitem[{Banerjee {et~al.}(2024)Banerjee, Tanaka, Kato, \& Gaigalas}]{banerjee_diversity_2024}
Banerjee, S., Tanaka, M., Kato, D., \& Gaigalas, G. 2024, The Astrophysical Journal, 968, 64, \dodoi{10.3847/1538-4357/ad4029}

\bibitem[{Banerjee {et~al.}(2020)Banerjee, Tanaka, Kawaguchi, Kato, \& Gaigalas}]{banerjee_simulations_2020}
Banerjee, S., Tanaka, M., Kawaguchi, K., Kato, D., \& Gaigalas, G. 2020, The Astrophysical Journal, 901, 29, \dodoi{10.3847/1538-4357/abae61}

\bibitem[{Biscoveanu {et~al.}(2023)Biscoveanu, Landry, \& Vitale}]{biscoveanu_population_2023}
Biscoveanu, S., Landry, P., \& Vitale, S. 2023, Monthly Notices of the Royal Astronomical Society, 518, 5298, \dodoi{10.1093/mnras/stac3052}

\bibitem[{Biswas(2021)}]{biswas_impact_2021}
Biswas, B. 2021, The Astrophysical Journal, 921, 63, \dodoi{10.3847/1538-4357/ac1c72}

\bibitem[{Breschi {et~al.}(2024)Breschi, Gamba, Carullo, Godzieba, Bernuzzi, Perego, \& Radice}]{breschi_bayesian_2024}
Breschi, M., Gamba, R., Carullo, G., {et~al.} 2024, Bayesian Inference of Multimessenger Astrophysical Data: {{Joint}} and Coherent Inference of Gravitational Waves and Kilonovae,  arXiv, \dodoi{10.48550/arXiv.2401.03750}

\bibitem[{Breschi {et~al.}(2021)Breschi, Perego, Bernuzzi, Del~Pozzo, Nedora, Radice, \& Vescovi}]{breschi_at2017gfo:_2021}
Breschi, M., Perego, A., Bernuzzi, S., {et~al.} 2021, Monthly Notices of the Royal Astronomical Society, 505, 1661, \dodoi{10.1093/mnras/stab1287}

\bibitem[{Bulla {et~al.}(2022)Bulla, Coughlin, Dhawan, \& Dietrich}]{bulla_multi-messenger_2022}
Bulla, M., Coughlin, M.~W., Dhawan, S., \& Dietrich, T. 2022, Universe, 8, 289, \dodoi{10.3390/universe8050289}

\bibitem[{Capano {et~al.}(2020)Capano, Tews, Brown, Margalit, De, Kumar, Brown, Krishnan, \& Reddy}]{capano_stringent_2020}
Capano, C.~D., Tews, I., Brown, S.~M., {et~al.} 2020, Nature Astronomy, 4, 625, \dodoi{10.1038/s41550-020-1014-6}

\bibitem[{Chaudhary {et~al.}(2024)Chaudhary, Toivonen, Waratkar, Mo, Chatterjee, Antier, Brockill, Coughlin, Essick, Ghosh, Morisaki, Baral, Baylor, Adhikari, Brady, Cabourn~Davies, Dal~Canton, Cavaglia, Creighton, Choudhary, Chu, Clearwater, Davis, Dent, Drago, Ewing, Godwin, Guo, Hanna, Huxford, Harry, Katsavounidis, Kovalam, Li, Magee, Marx, Meacher, Messick, {Morice-Atkinson}, Pace, De~Pietri, Piotrzkowski, Roy, Sachdev, Singer, Singh, Szczepanczyk, Tang, Trevor, Tsukada, {Villa-Ortega}, Wen, \& Wysocki}]{chaudhary_low-latency_2024}
Chaudhary, S.~S., Toivonen, A., Waratkar, G., {et~al.} 2024, Proceedings of the National Academy of Sciences, 121, e2316474121, \dodoi{10.1073/pnas.2316474121}

\bibitem[{Chornock {et~al.}(2017)Chornock, Berger, Kasen, Cowperthwaite, Nicholl, Villar, Alexander, Blanchard, Eftekhari, Fong, Margutti, Williams, Annis, Brout, Brown, Chen, Drout, Farr, Foley, Frieman, Fryer, Herner, Holz, Kessler, Matheson, Metzger, Quataert, Rest, Sako, Scolnic, Smith, \& {Soares-Santos}}]{chornock_the_2017}
Chornock, R., Berger, E., Kasen, D., {et~al.} 2017, The Astrophysical Journal Letters, 848, L19, \dodoi{10.3847/2041-8213/aa905c}

\bibitem[{Coughlin {et~al.}(2019{\natexlab{a}})Coughlin, Dietrich, Margalit, \& Metzger}]{coughlin_multi-messenger_2019}
Coughlin, M.~W., Dietrich, T., Margalit, B., \& Metzger, B.~D. 2019{\natexlab{a}}, Monthly Notices of the Royal Astronomical Society: Letters, 489, L91, \dodoi{10.1093/mnrasl/slz133}

\bibitem[{Coughlin {et~al.}(2019{\natexlab{b}})Coughlin, Ahumada, Anand, De, Hankins, Kasliwal, Singer, Bellm, Andreoni, Cenko, Cooke, Copperwheat, Dugas, Jencson, Perley, Yu, Bhalerao, Kumar, Bloom, Anupama, Ashley, Bagdasaryan, Biswas, Buckley, Burdge, Cook, Cromer, Cunningham, D'A{\`i}, Dekany, Delacroix, Dichiara, Duev, Dutta, Feeney, Frederick, Gatkine, Ghosh, Goldstein, Golkhou, Goobar, Graham, Hanayama, Horiuchi, Hung, Jha, Kong, Giomi, Kaplan, Karambelkar, Kowalski, Kulkarni, Kupfer, Masci, Mazzali, Moore, Mogotsi, Neill, Ngeow, {Mart{\'i}nez-Palomera}, Parola, Pavana, Ofek, Patil, Riddle, Rigault, Rusholme, Serabyn, Shupe, Sharma, Singh, Sollerman, Soon, Staats, Taggart, Tan, Travouillon, Troja, Waratkar, \& Yatsu}]{coughlin_growth_2019}
Coughlin, M.~W., Ahumada, T., Anand, S., {et~al.} 2019{\natexlab{b}}, The Astrophysical Journal Letters, 885, L19, \dodoi{10.3847/2041-8213/ab4ad8}

\bibitem[{Coughlin {et~al.}(2020{\natexlab{a}})Coughlin, Antier, Dietrich, Foley, Heinzel, Bulla, Christensen, Coulter, Issa, \& Khetan}]{coughlin_measuring_2020}
Coughlin, M.~W., Antier, S., Dietrich, T., {et~al.} 2020{\natexlab{a}}, Nature Communications, 11, 4129, \dodoi{10.1038/s41467-020-17998-5}

\bibitem[{Coughlin {et~al.}(2020{\natexlab{b}})Coughlin, Dietrich, Heinzel, Khetan, Antier, Bulla, Christensen, Coulter, \& Foley}]{coughlin_standardizing_2020}
Coughlin, M.~W., Dietrich, T., Heinzel, J., {et~al.} 2020{\natexlab{b}}, Physical Review Research, 2, 022006, \dodoi{10.1103/PhysRevResearch.2.022006}

\bibitem[{Coughlin {et~al.}(2020{\natexlab{c}})Coughlin, Dietrich, Antier, Bulla, Foucart, Hotokezaka, Raaijmakers, Hinderer, \& Nissanke}]{coughlin_implications_2020a}
Coughlin, M.~W., Dietrich, T., Antier, S., {et~al.} 2020{\natexlab{c}}, Monthly Notices of the Royal Astronomical Society, 492, 863, \dodoi{10.1093/mnras/stz3457}

\bibitem[{Coulter {et~al.}(2017)Coulter, Foley, Kilpatrick, Drout, Piro, Shappee, Siebert, Simon, Ulloa, Kasen, Madore, {Murguia-Berthier}, Pan, Prochaska, {Ramirez-Ruiz}, Rest, \& {Rojas-Bravo}}]{coulter_swope_2017}
Coulter, D.~A., Foley, R.~J., Kilpatrick, C.~D., {et~al.} 2017, Science, 358, 1556, \dodoi{10.1126/science.aap9811}

\bibitem[{Cowperthwaite {et~al.}(2017)Cowperthwaite, Berger, Villar, Metzger, Nicholl, Chornock, Blanchard, Fong, Margutti, {Soares-Santos}, Alexander, Allam, Annis, Brout, Brown, Butler, Chen, Diehl, Doctor, Drout, Eftekhari, Farr, Finley, Foley, Frieman, Fryer, {Garc{\'i}a-Bellido}, Gill, Guillochon, Herner, Holz, Kasen, Kessler, Marriner, Matheson, Neilsen, Quataert, Palmese, Rest, Sako, Scolnic, Smith, Tucker, Williams, Balbinot, Carlin, Cook, Durret, Li, Lopes, Louren{\c c}o, Marshall, Medina, Muir, Mu{\~n}oz, Sauseda, Schlegel, Secco, Vivas, Wester, Zenteno, Zhang, Abbott, Banerji, Bechtol, {Benoit-L{\'e}vy}, Bertin, {Buckley-Geer}, Burke, Capozzi, Rosell, Kind, Castander, Crocce, Cunha, D'Andrea, da~Costa, Davis, DePoy, Desai, Dietrich, {Drlica-Wagner}, Eifler, Evrard, Fernandez, Flaugher, Fosalba, Gaztanaga, Gerdes, Giannantonio, Goldstein, Gruen, Gruendl, Gutierrez, Honscheid, Jain, James, Jeltema, Johnson, Johnson, Kent, Krause, Kron, Kuehn, Nuropatkin, Lahav, Lima, Lin, Maia, March, Martini,
  McMahon, Menanteau, Miller, Miquel, Mohr, Neilsen, Nichol, Ogando, Plazas, Roe, Romer, Roodman, Rykoff, Sanchez, Scarpine, Schindler, Schubnell, {Sevilla-Noarbe}, Smith, Smith, Sobreira, Suchyta, Swanson, Tarle, Thomas, Thomas, Troxel, Vikram, Walker, Wechsler, Weller, Yanny, \& Zuntz}]{cowperthwaite_the_2017}
Cowperthwaite, P.~S., Berger, E., Villar, V.~A., {et~al.} 2017, The Astrophysical Journal Letters, 848, L17, \dodoi{10.3847/2041-8213/aa8fc7}

\bibitem[{Criswell(2024)}]{criswell_criswellalexander/uvex-followup:_2024}
Criswell, A. 2024, Criswellalexander/Uvex-Followup: {{Code}} Release for {{Electromagnetic Follow-up}} to {{Gravitational Wave Events}} with the {{UltraViolet EXplorer}} ({{UVEX}}), Zenodo, \dodoi{10.5281/zenodo.14014700}

\bibitem[{Dichiara {et~al.}(2021)Dichiara, Becerra, Chase, Troja, Lee, Watson, Butler, O'Connor, Pereyra, L{\'o}pez, Lien, Gottlieb, \& Kutyrev}]{dichiara_constraints_2021}
Dichiara, S., Becerra, R.~L., Chase, E.~A., {et~al.} 2021, The Astrophysical Journal Letters, 923, L32, \dodoi{10.3847/2041-8213/ac4259}

\bibitem[{Dietrich {et~al.}(2020{\natexlab{a}})Dietrich, Coughlin, Pang, Bulla, Heinzel, Issa, Tews, \& Antier}]{dietrich_multimessenger_2020a}
Dietrich, T., Coughlin, M.~W., Pang, P. T.~H., {et~al.} 2020{\natexlab{a}}, Science, 370, 1450, \dodoi{10.1126/science.abb4317}

\bibitem[{Dietrich {et~al.}(2020{\natexlab{b}})Dietrich, Coughlin, Pang, Bulla, Heinzel, Issa, Tews, \& Antier}]{dietrich_multimessenger_2020}
---. 2020{\natexlab{b}}, Science, 370, 1450, \dodoi{10.1126/science.abb4317}

\bibitem[{Essick {et~al.}(2020)Essick, Landry, \& Holz}]{essick_nonparametric_2020}
Essick, R., Landry, P., \& Holz, D.~E. 2020, Physical Review D, 101, 063007, \dodoi{10.1103/PhysRevD.101.063007}

\bibitem[{Evans {et~al.}(2017)Evans, Cenko, Kennea, Emery, Kuin, Korobkin, Wollaeger, Fryer, Madsen, Harrison, Xu, Nakar, Hotokezaka, Lien, Campana, Oates, Troja, Breeveld, Marshall, Barthelmy, Beardmore, Burrows, Cusumano, D'A{\`i}, D'Avanzo, D'Elia, {de Pasquale}, Even, Fontes, Forster, Garcia, Giommi, Grefenstette, Gronwall, Hartmann, Heida, Hungerford, Kasliwal, Krimm, Levan, Malesani, Melandri, Miyasaka, Nousek, O'Brien, Osborne, Pagani, Page, Palmer, Perri, Pike, Racusin, Rosswog, Siegel, Sakamoto, Sbarufatti, Tagliaferri, Tanvir, \& Tohuvavohu}]{evans_swift_2017}
Evans, P.~A., Cenko, S.~B., Kennea, J.~A., {et~al.} 2017, Science, 358, 1565, \dodoi{10.1126/science.aap9580}

\bibitem[{Farah {et~al.}(2022)Farah, Fishbach, Essick, Holz, \& Galaudage}]{farah_bridging_2022}
Farah, A., Fishbach, M., Essick, R., Holz, D.~E., \& Galaudage, S. 2022, The Astrophysical Journal, 931, 108, \dodoi{10.3847/1538-4357/ac5f03}

\bibitem[{Fisher(2024)}]{fisher_new_2024}
Fisher, A. 2024, New {{NASA Mission}} Will {{Study Ultraviolet Sky}}, {{Stars}}, {{Stellar Explosions}} - {{NASA}}

\bibitem[{Gehrels {et~al.}(2004)Gehrels, Chincarini, Giommi, Mason, Nousek, Wells, White, Barthelmy, Burrows, Cominsky, Hurley, Marshall, M{\'e}sz{\'a}ros, Roming, Angelini, Barbier, Belloni, Campana, Caraveo, Chester, Citterio, Cline, Cropper, Cummings, Dean, Feigelson, Fenimore, Frail, Fruchter, Garmire, Gendreau, Ghisellini, Greiner, Hill, Hunsberger, Krimm, Kulkarni, Kumar, Lebrun, {Lloyd-Ronning}, Markwardt, Mattson, Mushotzky, Norris, Osborne, Paczynski, Palmer, Park, Parsons, Paul, Rees, Reynolds, Rhoads, Sasseen, Schaefer, Short, Smale, Smith, Stella, Tagliaferri, Takahashi, Tashiro, Townsley, Tueller, Turner, Vietri, Voges, Ward, Willingale, Zerbi, \& Zhang}]{gehrels_the_2004}
Gehrels, N., Chincarini, G., Giommi, P., {et~al.} 2004, The Astrophysical Journal, 611, 1005, \dodoi{10.1086/422091}

\bibitem[{Gianfagna {et~al.}(2023)Gianfagna, Piro, Pannarale, Van~Eerten, Ricci, Ryan, \& Troja}]{gianfagna_joint_2023}
Gianfagna, G., Piro, L., Pannarale, F., {et~al.} 2023, Monthly Notices of the Royal Astronomical Society, 523, 4771, \dodoi{10.1093/mnras/stad1728}

\bibitem[{Gill {et~al.}(2019)Gill, Nathanail, \& Rezzolla}]{gill_when_2019a}
Gill, R., Nathanail, A., \& Rezzolla, L. 2019, The Astrophysical Journal, 876, 139, \dodoi{10.3847/1538-4357/ab16da}

\bibitem[{Gompertz {et~al.}(2020)Gompertz, Cutter, Steeghs, Galloway, Lyman, Ulaczyk, Dyer, Ackley, Dhillon, O'Brien, Ramsay, Poshyachinda, Kotak, Nuttall, Breton, Pall{\'e}, Pollacco, Thrane, Aukkaravittayapun, Awiphan, Brown, Burhanudin, Chote, Chrimes, Daw, Duffy, {Eyles-Ferris}, Heikkil{\"a}, Irawati, Kennedy, Killestein, Levan, Littlefair, Makrygianni, Marsh, Mata~S{\'a}nchez, Mattila, Maund, McCormac, Mkrtichian, Mong, Mullaney, M{\"u}ller, Obradovic, Rol, Sawangwit, Stanway, Starling, Str{\o}m, Tooke, West, \& Wiersema}]{gompertz_searching_2020}
Gompertz, B.~P., Cutter, R., Steeghs, D., {et~al.} 2020, Monthly Notices of the Royal Astronomical Society, 497, 726, \dodoi{10.1093/mnras/staa1845}

\bibitem[{Gottlieb \& Loeb(2020)}]{gottlieb_electromagnetic_2020}
Gottlieb, O., \& Loeb, A. 2020, Monthly Notices of the Royal Astronomical Society, 493, 1753, \dodoi{10.1093/mnras/staa363}

\bibitem[{Guidorzi {et~al.}(2017)Guidorzi, Margutti, Brout, Scolnic, Fong, Alexander, Cowperthwaite, Annis, Berger, Blanchard, Chornock, Coppejans, Eftekhari, Frieman, Huterer, Nicholl, {Soares-Santos}, Terreran, Villar, \& Williams}]{guidorzi_improved_2017}
Guidorzi, C., Margutti, R., Brout, D., {et~al.} 2017, The Astrophysical Journal Letters, 851, L36, \dodoi{10.3847/2041-8213/aaa009}

\bibitem[{Harris {et~al.}(2020)Harris, Millman, {van der Walt}, Gommers, Virtanen, Cournapeau, Wieser, Taylor, Berg, Smith, Kern, Picus, Hoyer, {van Kerkwijk}, Brett, Haldane, {del R{\'i}o}, Wiebe, Peterson, {G{\'e}rard-Marchant}, Sheppard, Reddy, Weckesser, Abbasi, Gohlke, \& Oliphant}]{harris_array_2020a}
Harris, C.~R., Millman, K.~J., {van der Walt}, S.~J., {et~al.} 2020, Nature, 585, 357, \dodoi{10.1038/s41586-020-2649-2}

\bibitem[{Hjorth {et~al.}(2017)Hjorth, Levan, Tanvir, Lyman, Wojtak, Schr{\o}der, Mandel, Gall, \& Bruun}]{hjorth_the_2017}
Hjorth, J., Levan, A.~J., Tanvir, N.~R., {et~al.} 2017, The Astrophysical Journal Letters, 848, L31, \dodoi{10.3847/2041-8213/aa9110}

\bibitem[{Hosseinzadeh {et~al.}(2019)Hosseinzadeh, Cowperthwaite, Gomez, Villar, Nicholl, Margutti, Berger, Chornock, Paterson, Fong, Savchenko, Short, Alexander, Blanchard, Braga, Calkins, Cartier, Coppejans, Eftekhari, Laskar, Ly, Patton, Pelisoli, Reichart, Terreran, \& Williams}]{hosseinzadeh_follow-up_2019}
Hosseinzadeh, G., Cowperthwaite, P.~S., Gomez, S., {et~al.} 2019, The Astrophysical Journal Letters, 880, L4, \dodoi{10.3847/2041-8213/ab271c}

\bibitem[{Hotokezaka \& Nakar(2020)}]{hotokezaka_radioactive_2020}
Hotokezaka, K., \& Nakar, E. 2020, The Astrophysical Journal, 891, 152, \dodoi{10.3847/1538-4357/ab6a98}

\bibitem[{Hotokezaka {et~al.}(2019)Hotokezaka, Nakar, Gottlieb, Nissanke, Masuda, Hallinan, Mooley, \& Deller}]{hotokezaka_a_2019}
Hotokezaka, K., Nakar, E., Gottlieb, O., {et~al.} 2019, Nature Astronomy, 3, 940, \dodoi{10.1038/s41550-019-0820-1}

\bibitem[{Hunter(2007)}]{hunter_matplotlib:_2007}
Hunter, J.~D. 2007, Computing in Science \& Engineering, 9, 90, \dodoi{10.1109/MCSE.2007.55}

\bibitem[{Huth {et~al.}(2022)Huth, Pang, Tews, Dietrich, F{\`e}vre, Schwenk, Trautmann, Agarwal, Bulla, Coughlin, \& Broeck}]{huth_constraining_2022}
Huth, S., Pang, P. T.~H., Tews, I., {et~al.} 2022, Nature, 606, 276, \dodoi{10.1038/s41586-022-04750-w}

\bibitem[{Ivezi{\'c} {et~al.}(2019)Ivezi{\'c}, Kahn, Tyson, Abel, Acosta, Allsman, Alonso, AlSayyad, Anderson, Andrew, Angel, Angeli, Ansari, Antilogus, Araujo, Armstrong, Arndt, Astier, Aubourg, Auza, Axelrod, Bard, Barr, Barrau, Bartlett, Bauer, Bauman, Baumont, Bechtol, Bechtol, Becker, Becla, Beldica, Bellavia, Bianco, Biswas, Blanc, Blazek, Blandford, Bloom, Bogart, Bond, Booth, Borgland, Borne, Bosch, Boutigny, Brackett, Bradshaw, Brandt, Brown, Bullock, Burchat, Burke, Cagnoli, Calabrese, Callahan, Callen, Carlin, Carlson, Chandrasekharan, {Charles-Emerson}, Chesley, Cheu, Chiang, Chiang, Chirino, Chow, Ciardi, Claver, {Cohen-Tanugi}, Cockrum, Coles, Connolly, Cook, Cooray, Covey, Cribbs, Cui, Cutri, Daly, Daniel, Daruich, Daubard, Daues, Dawson, Delgado, Dellapenna, {de Peyster}, {de Val-Borro}, Digel, Doherty, Dubois, {Dubois-Felsmann}, Durech, Economou, Eifler, Eracleous, Emmons, Fausti~Neto, Ferguson, Figueroa, {Fisher-Levine}, Focke, Foss, Frank, Freemon, Gangler, Gawiser, Geary, Gee, Geha,
  Gessner, Gibson, Gilmore, Glanzman, Glick, Goldina, Goldstein, Goodenow, Graham, Gressler, Gris, Guy, Guyonnet, Haller, Harris, Hascall, Haupt, Hernandez, Herrmann, Hileman, Hoblitt, Hodgson, Hogan, Howard, Huang, Huffer, Ingraham, Innes, Jacoby, Jain, Jammes, Jee, Jenness, Jernigan, Jevremovi{\'c}, Johns, Johnson, Johnson, Jones, {Juramy-Gilles}, Juri{\'c}, Kalirai, Kallivayalil, Kalmbach, Kantor, Karst, Kasliwal, Kelly, Kessler, Kinnison, Kirkby, Knox, Kotov, Krabbendam, Krughoff, Kub{\'a}nek, Kuczewski, Kulkarni, Ku, Kurita, Lage, Lambert, Lange, Langton, Le~Guillou, Levine, Liang, Lim, Lintott, Long, Lopez, Lotz, Lupton, Lust, MacArthur, Mahabal, Mandelbaum, Markiewicz, Marsh, Marshall, Marshall, May, McKercher, McQueen, Meyers, Migliore, Miller, Mills, Miraval, Moeyens, Moolekamp, Monet, Moniez, Monkewitz, Montgomery, Morrison, Mueller, Muller, Mu{\~n}oz~Arancibia, Neill, Newbry, Nief, Nomerotski, Nordby, O'Connor, Oliver, Olivier, Olsen, O'Mullane, Ortiz, Osier, Owen, Pain, Palecek, Parejko, Parsons,
  Pease, Peterson, Peterson, Petravick, Libby~Petrick, Petry, Pierfederici, Pietrowicz, Pike, Pinto, Plante, Plate, Plutchak, Price, Prouza, Radeka, Rajagopal, Rasmussen, Regnault, Reil, Reiss, Reuter, Ridgway, Riot, Ritz, Robinson, Roby, Roodman, Rosing, Roucelle, Rumore, Russo, Saha, Sassolas, Schalk, Schellart, Schindler, Schmidt, Schneider, Schneider, Schoening, Schumacher, Schwamb, Sebag, Selvy, Sembroski, Seppala, Serio, Serrano, Shaw, Shipsey, Sick, Silvestri, Slater, Smith, Smith, Sobhani, Soldahl, {Storrie-Lombardi}, Stover, Strauss, Street, Stubbs, Sullivan, Sweeney, Swinbank, Szalay, Takacs, Tether, Thaler, Thayer, Thomas, Thornton, Thukral, Tice, Trilling, Turri, Van~Berg, Vanden~Berk, Vetter, Virieux, Vucina, Wahl, Walkowicz, Walsh, Walter, Wang, Wang, Warner, Wiecha, Willman, Winters, Wittman, Wolff, {Wood-Vasey}, Wu, Xin, Yoachim, \& Zhan}]{ivezic_lsst:_2019}
Ivezi{\'c}, {\v Z}., Kahn, S.~M., Tyson, J.~A., {et~al.} 2019, The Astrophysical Journal, 873, 111, \dodoi{10.3847/1538-4357/ab042c}

\bibitem[{Kasen {et~al.}(2017)Kasen, Metzger, Barnes, Quataert, \& {Ramirez-Ruiz}}]{kasen_origin_2017}
Kasen, D., Metzger, B., Barnes, J., Quataert, E., \& {Ramirez-Ruiz}, E. 2017, Nature, 551, 80, \dodoi{10.1038/nature24453}

\bibitem[{Kasliwal {et~al.}(2017)Kasliwal, Nakar, Singer, Kaplan, Cook, Van~Sistine, Lau, Fremling, Gottlieb, Jencson, Adams, Feindt, Hotokezaka, Ghosh, Perley, Yu, Piran, Allison, Anupama, Balasubramanian, Bannister, Bally, Barnes, Barway, Bellm, Bhalerao, Bhattacharya, Blagorodnova, Bloom, Brady, Cannella, Chatterjee, Cenko, Cobb, Copperwheat, Corsi, De, Dobie, Emery, Evans, Fox, Frail, Frohmaier, Goobar, Hallinan, Harrison, Helou, Hinderer, Ho, Horesh, Ip, Itoh, Kasen, Kim, Kuin, Kupfer, Lynch, Madsen, Mazzali, Miller, Mooley, Murphy, Ngeow, Nichols, Nissanke, Nugent, Ofek, Qi, Quimby, Rosswog, Rusu, Sadler, Schmidt, Sollerman, Steele, Williamson, Xu, Yan, Yatsu, Zhang, \& Zhao}]{kasliwal_illuminating_2017}
Kasliwal, M.~M., Nakar, E., Singer, L.~P., {et~al.} 2017, Science, 358, 1559, \dodoi{10.1126/science.aap9455}

\bibitem[{Kasliwal {et~al.}(2020)Kasliwal, Anand, Ahumada, Stein, Carracedo, Andreoni, Coughlin, Singer, Kool, De, Kumar, AlMualla, Yao, Bulla, Dobie, Reusch, Perley, Cenko, Bhalerao, Kaplan, Sollerman, Goobar, Copperwheat, Bellm, Anupama, Corsi, Nissanke, Agudo, Bagdasaryan, Barway, Belicki, Bloom, Bolin, Buckley, Burdge, Burruss, {Caballero-Garc{\'i}a}, Cannella, {Castro-Tirado}, Cook, Cooke, Cunningham, Dahiwale, Deshmukh, Dichiara, Duev, Dutta, Feeney, Franckowiak, Frederick, Fremling, {Gal-Yam}, Gatkine, Ghosh, Goldstein, Golkhou, Graham, Graham, Hankins, Helou, Hu, Ip, Jaodand, Karambelkar, Kong, Kowalski, Khandagale, Kulkarni, Kumar, Laher, Li, Mahabal, Masci, Miller, Mogotsi, Mohite, Mooley, Mroz, Newman, Ngeow, Oates, Patil, Pandey, Pavana, Pian, Riddle, {S{\'a}nchez-Ram{\'i}rez}, Sharma, Singh, Smith, Soumagnac, Taggart, Tan, Tzanidakis, Troja, Valeev, Walters, Waratkar, Webb, Yu, Zhang, Zhou, \& Zolkower}]{kasliwal_kilonova_2020}
Kasliwal, M.~M., Anand, S., Ahumada, T., {et~al.} 2020, The Astrophysical Journal, 905, 145, \dodoi{10.3847/1538-4357/abc335}

\bibitem[{Kasliwal {et~al.}(2022)Kasliwal, Kasen, Lau, Perley, Rosswog, Ofek, Hotokezaka, Chary, Sollerman, Goobar, \& Kaplan}]{kasliwal_spitzer_2022}
Kasliwal, M.~M., Kasen, D., Lau, R.~M., {et~al.} 2022, Monthly Notices of the Royal Astronomical Society: Letters, 510, L7, \dodoi{10.1093/mnrasl/slz007}

\bibitem[{Kiendrebeogo {et~al.}(2023)Kiendrebeogo, Farah, Foley, Gray, Kunert, Puecher, Toivonen, VandenBerg, Anand, Ahumada, Karambelkar, Coughlin, Dietrich, Kam, Pang, Singer, \& Sravan}]{kiendrebeogo_updated_2023a}
Kiendrebeogo, R.~W., Farah, A.~M., Foley, E.~M., {et~al.} 2023, The Astrophysical Journal, 958, 158, \dodoi{10.3847/1538-4357/acfcb1}

\bibitem[{Kilpatrick {et~al.}(2017)Kilpatrick, Foley, Kasen, {Murguia-Berthier}, {Ramirez-Ruiz}, Coulter, Drout, Piro, Shappee, Boutsia, Contreras, Di~Mille, Madore, Morrell, Pan, Prochaska, Rest, {Rojas-Bravo}, Siebert, Simon, \& Ulloa}]{kilpatrick_electromagnetic_2017}
Kilpatrick, C.~D., Foley, R.~J., Kasen, D., {et~al.} 2017, Science, 358, 1583, \dodoi{10.1126/science.aaq0073}

\bibitem[{Klion {et~al.}(2021)Klion, Duffell, Kasen, \& Quataert}]{klion_the_2021}
Klion, H., Duffell, P.~C., Kasen, D., \& Quataert, E. 2021, Monthly Notices of the Royal Astronomical Society, 502, 865, \dodoi{10.1093/mnras/stab042}

\bibitem[{Kulkarni {et~al.}(2023)Kulkarni, Harrison, Grefenstette, Earnshaw, Andreoni, Berg, Bloom, Cenko, Chornock, Christiansen, Coughlin, Criswell, Darvish, Das, De, Dessart, Dixon, Dorsman, {El-Badry}, Evans, Ford, Fremling, Gansicke, Gezari, Goetberg, Green, Graham, Heida, Ho, Jaodand, {Johns-Krull}, Kasliwal, Lazzarini, Lu, Margutti, Martin, Masters, McKernan, Naze, Nissanke, Parazin, Perley, Phinney, Piro, Raaijmakers, Rauw, Rodriguez, Sana, Senchyna, Singer, Spake, Stassun, Stern, Teplitz, Weisz, \& Yao}]{kulkarni_science_2023}
Kulkarni, S.~R., Harrison, F.~A., Grefenstette, B.~W., {et~al.} 2023, Science with the {{Ultraviolet Explorer}} ({{UVEX}}),  arXiv, \dodoi{10.48550/arXiv.2111.15608}

\bibitem[{Laureijs {et~al.}(2011)Laureijs, Amiaux, Arduini, Augu{\`e}res, Brinchmann, Cole, Cropper, Dabin, Duvet, Ealet, Garilli, Gondoin, Guzzo, Hoar, Hoekstra, Holmes, Kitching, Maciaszek, Mellier, Pasian, Percival, Rhodes, Saavedra~Criado, Sauvage, Scaramella, Valenziano, Warren, Bender, Castander, Cimatti, Le~F{\`e}vre, {Kurki-Suonio}, Levi, Lilje, Meylan, Nichol, Pedersen, Popa, Rebolo~Lopez, Rix, Rottgering, Zeilinger, Grupp, Hudelot, Massey, Meneghetti, Miller, Paltani, {Paulin-Henriksson}, Pires, Saxton, Schrabback, Seidel, Walsh, Aghanim, Amendola, Bartlett, Baccigalupi, Beaulieu, Benabed, Cuby, Elbaz, Fosalba, Gavazzi, Helmi, Hook, Irwin, Kneib, Kunz, Mannucci, Moscardini, Tao, Teyssier, Weller, Zamorani, Zapatero~Osorio, Boulade, Foumond, Di~Giorgio, Guttridge, James, Kemp, Martignac, Spencer, Walton, Bl{\"u}mchen, Bonoli, Bortoletto, Cerna, Corcione, Fabron, Jahnke, Ligori, Madrid, Martin, Morgante, Pamplona, Prieto, Riva, Toledo, Trifoglio, Zerbi, Abdalla, Douspis, Grenet, Borgani, Bouwens,
  Courbin, Delouis, Dubath, Fontana, Frailis, Grazian, Koppenh{\"o}fer, Mansutti, Melchior, Mignoli, Mohr, Neissner, Noddle, Poncet, Scodeggio, Serrano, Shane, Starck, Surace, Taylor, {Verdoes-Kleijn}, Vuerli, Williams, Zacchei, Altieri, Escudero~Sanz, Kohley, Oosterbroek, Astier, Bacon, Bardelli, Baugh, Bellagamba, Benoist, Bianchi, Biviano, Branchini, Carbone, Cardone, Clements, Colombi, Conselice, Cresci, Deacon, Dunlop, Fedeli, Fontanot, Franzetti, Giocoli, {Garcia-Bellido}, Gow, Heavens, Hewett, Heymans, Holland, Huang, Ilbert, Joachimi, Jennins, Kerins, Kiessling, Kirk, Kotak, Krause, Lahav, {van Leeuwen}, Lesgourgues, Lombardi, Magliocchetti, Maguire, Majerotto, Maoli, Marulli, Maurogordato, McCracken, McLure, Melchiorri, Merson, Moresco, Nonino, Norberg, Peacock, Pello, Penny, Pettorino, Di~Porto, Pozzetti, Quercellini, Radovich, Rassat, Roche, Ronayette, Rossetti, Sartoris, Schneider, Semboloni, Serjeant, Simpson, Skordis, Smadja, Smartt, Spano, Spiro, Sullivan, Tilquin, Trotta, Verde, Wang,
  Williger, Zhao, Zoubian, \& Zucca}]{laureijs_euclid_2011}
Laureijs, R., Amiaux, J., Arduini, S., {et~al.} 2011, Euclid {{Definition Study Report}}, \dodoi{10.48550/arXiv.1110.3193}

\bibitem[{Leggio \& Criswell(2024)}]{leggio_data_2024}
Leggio, S., \& Criswell, A. 2024, Data Products for "{{Electromagnetic Follow-up}} to {{Gravitational Wave Events}} with the {{UltraViolet EXplorer}} ({{UVEX}})",  Zenodo, \dodoi{10.5281/zenodo.14014902}

\bibitem[{Legred {et~al.}(2021)Legred, Chatziioannou, Essick, Han, \& Landry}]{legred_impact_2021}
Legred, I., Chatziioannou, K., Essick, R., Han, S., \& Landry, P. 2021, Physical Review D, 104, 063003, \dodoi{10.1103/PhysRevD.104.063003}

\bibitem[{Li \& Paczy{\'n}ski(1998)}]{li_transient_1998}
Li, L.-X., \& Paczy{\'n}ski, B. 1998, The Astrophysical Journal, 507, L59, \dodoi{10.1086/311680}

\bibitem[{Margalit \& Metzger(2019)}]{margalit_the_2019}
Margalit, B., \& Metzger, B.~D. 2019, The Astrophysical Journal, 880, L15, \dodoi{10.3847/2041-8213/ab2ae2}

\bibitem[{Margutti \& Chornock(2021)}]{margutti_first_2021}
Margutti, R., \& Chornock, R. 2021, Annual Review of Astronomy and Astrophysics, 59, 155, \dodoi{10.1146/annurev-astro-112420-030742}

\bibitem[{Margutti {et~al.}(2017)Margutti, Berger, Fong, Guidorzi, Alexander, Metzger, Blanchard, Cowperthwaite, Chornock, Eftekhari, Nicholl, Villar, Williams, Annis, Brown, Chen, Doctor, Frieman, Holz, Sako, \& {Soares-Santos}}]{margutti_the_2017}
Margutti, R., Berger, E., Fong, W., {et~al.} 2017, The Astrophysical Journal, 848, L20, \dodoi{10.3847/2041-8213/aa9057}

\bibitem[{McKinney(2010)}]{mckinney_data_2010a}
McKinney, W. 2010, in Proceedings of the 9th {{Python}} in {{Science Conference}}, ed. S.~{van der Walt} \& J.~Millman, 56--61, \dodoi{10.25080/Majora-92bf1922-00a}

\bibitem[{Metzger(2019)}]{metzger_kilonovae_2019}
Metzger, B.~D. 2019, Living Reviews in Relativity, 23, 1, \dodoi{10.1007/s41114-019-0024-0}

\bibitem[{Metzger {et~al.}(2015)Metzger, Bauswein, Goriely, \& Kasen}]{metzger_neutron-powered_2015}
Metzger, B.~D., Bauswein, A., Goriely, S., \& Kasen, D. 2015, Monthly Notices of the Royal Astronomical Society, 446, 1115, \dodoi{10.1093/mnras/stu2225}

\bibitem[{Metzger {et~al.}(2010)Metzger, {Mart{\'i}nez-Pinedo}, Darbha, Quataert, Arcones, Kasen, Thomas, Nugent, Panov, \& Zinner}]{metzger_electromagnetic_2010a}
Metzger, B.~D., {Mart{\'i}nez-Pinedo}, G., Darbha, S., {et~al.} 2010, Monthly Notices of the Royal Astronomical Society, 406, 2650, \dodoi{10.1111/j.1365-2966.2010.16864.x}

\bibitem[{Miller {et~al.}(2021)Miller, Lamb, Dittmann, Bogdanov, Arzoumanian, Gendreau, Guillot, Ho, Lattimer, Loewenstein, Morsink, Ray, Wolff, Baker, Cazeau, Manthripragada, Markwardt, Okajima, Pollard, Cognard, Cromartie, Fonseca, Guillemot, Kerr, Parthasarathy, Pennucci, Ransom, \& Stairs}]{miller_the_2021}
Miller, M.~C., Lamb, F.~K., Dittmann, A.~J., {et~al.} 2021, The Astrophysical Journal Letters, 918, L28, \dodoi{10.3847/2041-8213/ac089b}

\bibitem[{{Murguia-Berthier} {et~al.}(2021){Murguia-Berthier}, {Ramirez-Ruiz}, Colle, Janiuk, Rosswog, \& Lee}]{murguia-berthier_the_2021}
{Murguia-Berthier}, A., {Ramirez-Ruiz}, E., Colle, F.~D., {et~al.} 2021, The Astrophysical Journal, 908, 152, \dodoi{10.3847/1538-4357/abd08e}

\bibitem[{Nakar(2020)}]{nakar_the_2020}
Nakar, E. 2020, Physics Reports, 886, 1, \dodoi{10.1016/j.physrep.2020.08.008}

\bibitem[{Nicholl {et~al.}(2021)Nicholl, Margalit, Schmidt, Smith, Ridley, \& Nuttall}]{nicholl_tight_2021}
Nicholl, M., Margalit, B., Schmidt, P., {et~al.} 2021, Monthly Notices of the Royal Astronomical Society, 505, 3016, \dodoi{10.1093/mnras/stab1523}

\bibitem[{Nicholl {et~al.}(2017)Nicholl, Berger, Kasen, Metzger, Elias, Brice{\~n}o, Alexander, Blanchard, Chornock, Cowperthwaite, Eftekhari, Fong, Margutti, Villar, Williams, Brown, Annis, Bahramian, Brout, Brown, Chen, Clemens, Dennihy, Dunlap, Holz, Marchesini, Massaro, Moskowitz, Pelisoli, Rest, Ricci, Sako, {Soares-Santos}, \& Strader}]{nicholl_the_2017}
Nicholl, M., Berger, E., Kasen, D., {et~al.} 2017, The Astrophysical Journal, 848, L18, \dodoi{10.3847/2041-8213/aa9029}

\bibitem[{Paek {et~al.}(2024)Paek, Im, Kim, Lim, Park, Choi, Kim, Barbieri, Salafia, Paek, Shin, Seo, Lee, Lee, Kim, \& Sung}]{paek_gravitational-wave_2024}
Paek, G. S.~H., Im, M., Kim, J., {et~al.} 2024, The Astrophysical Journal, 960, 113, \dodoi{10.3847/1538-4357/ad0238}

\bibitem[{Page {et~al.}(2020)Page, Evans, Tohuvavohu, Kennea, Klingler, Cenko, Oates, Ambrosi, Barthelmy, Beardmore, Bernardini, Breeveld, Brown, Burrows, Campana, Caputo, Cusumano, D'A{\`i}, D'Avanzo, D'Elia, De~Pasquale, Emery, Giommi, Gronwall, Hartmann, Krimm, Kuin, Malesani, Marshall, Melandri, Nousek, O'Brien, Osborne, Pagani, Page, Palmer, Perri, Racusin, Sakamoto, Sbarufatti, Schlieder, Siegel, Tagliaferri, \& Troja}]{page_swift-xrt_2020}
Page, K.~L., Evans, P.~A., Tohuvavohu, A., {et~al.} 2020, Monthly Notices of the Royal Astronomical Society, 499, 3459, \dodoi{10.1093/mnras/staa3032}

\bibitem[{Palmese {et~al.}(2024)Palmese, Kaur, Hajela, Margutti, McDowell, \& MacFadyen}]{palmese_standard_2024}
Palmese, A., Kaur, R., Hajela, A., {et~al.} 2024, Physical Review D, 109, 063508, \dodoi{10.1103/PhysRevD.109.063508}

\bibitem[{Pang {et~al.}(2021)Pang, Tews, Coughlin, Bulla, Broeck, \& Dietrich}]{pang_nuclear-physics_2021}
Pang, P. T.~H., Tews, I., Coughlin, M.~W., {et~al.} 2021, The Astrophysical Journal, 922, 14, \dodoi{10.3847/1538-4357/ac19ab}

\bibitem[{Peeples \& {The STScI Science Mission Office}(2025)}]{peeples_hst_2025}
Peeples, M., \& {The STScI Science Mission Office}. 2025, {{HST Primer}}: {{Scientific Instrument Comparisons}} - {{HST User Documentation}}, https://hst-docs.stsci.edu/hsp/the-hubble-space-telescope-primer-for-cycle-33/hst-primer-scientific-instrument-comparisons

\bibitem[{Petrov {et~al.}(2022)Petrov, Singer, Coughlin, Kumar, Almualla, Anand, Bulla, Dietrich, Foucart, \& Guessoum}]{petrov_data-driven_2022}
Petrov, P., Singer, L.~P., Coughlin, M.~W., {et~al.} 2022, The Astrophysical Journal, 924, 54, \dodoi{10.3847/1538-4357/ac366d}

\bibitem[{Pian {et~al.}(2017)Pian, D'Avanzo, Benetti, Branchesi, Brocato, Campana, Cappellaro, Covino, D'Elia, Fynbo, Getman, Ghirlanda, Ghisellini, Grado, Greco, Hjorth, Kouveliotou, Levan, Limatola, Malesani, Mazzali, Melandri, M{\o}ller, Nicastro, Palazzi, Piranomonte, Rossi, Salafia, Selsing, Stratta, Tanaka, Tanvir, Tomasella, Watson, Yang, Amati, Antonelli, Ascenzi, Bernardini, Bo{\"e}r, Bufano, Bulgarelli, Capaccioli, Casella, {Castro-Tirado}, {Chassande-Mottin}, Ciolfi, Copperwheat, Dadina, De~Cesare, Di~Paola, Fan, Gendre, Giuffrida, Giunta, Hunt, Israel, Jin, Kasliwal, Klose, Lisi, Longo, Maiorano, Mapelli, Masetti, Nava, Patricelli, Perley, Pescalli, Piran, Possenti, Pulone, Razzano, Salvaterra, Schipani, Spera, Stamerra, Stella, Tagliaferri, Testa, Troja, Turatto, Vergani, \& Vergani}]{pian_spectroscopic_2017}
Pian, E., D'Avanzo, P., Benetti, S., {et~al.} 2017, Nature, 551, 67, \dodoi{10.1038/nature24298}

\bibitem[{Piro \& Kollmeier(2018)}]{piro_evidence_2018}
Piro, A.~L., \& Kollmeier, J.~A. 2018, The Astrophysical Journal, 855, 103, \dodoi{10.3847/1538-4357/aaaab3}

\bibitem[{Raaijmakers {et~al.}(2020)Raaijmakers, Greif, Riley, Hinderer, Hebeler, Schwenk, Watts, Nissanke, Guillot, Lattimer, \& Ludlam}]{raaijmakers_constraining_2020}
Raaijmakers, G., Greif, S.~K., Riley, T.~E., {et~al.} 2020, The Astrophysical Journal, 893, L21, \dodoi{10.3847/2041-8213/ab822f}

\bibitem[{Raaijmakers {et~al.}(2021)Raaijmakers, Greif, Hebeler, Hinderer, Nissanke, Schwenk, Riley, Watts, Lattimer, \& Ho}]{raaijmakers_constraints_2021}
Raaijmakers, G., Greif, S.~K., Hebeler, K., {et~al.} 2021, arXiv:2105.06981 [astro-ph, physics:nucl-ex, physics:nucl-th].
\newblock \doarXiv{2105.06981}

\bibitem[{Robitaille {et~al.}(2013)Robitaille, Tollerud, Greenfield, Droettboom, Bray, Aldcroft, Davis, Ginsburg, {Price-Whelan}, Kerzendorf, Conley, Crighton, Barbary, Muna, Ferguson, Grollier, Parikh, Nair, G{\"u}nther, Deil, Woillez, Conseil, Kramer, Turner, Singer, Fox, Weaver, Zabalza, Edwards, Bostroem, Burke, Casey, Crawford, Dencheva, Ely, Jenness, Labrie, Lim, Pierfederici, Pontzen, Ptak, Refsdal, Servillat, \& Streicher}]{robitaille_astropy:_2013}
Robitaille, T.~P., Tollerud, E.~J., Greenfield, P., {et~al.} 2013, Astronomy \& Astrophysics, 558, A33, \dodoi{10.1051/0004-6361/201322068}

\bibitem[{Rosswog {et~al.}(2017)Rosswog, Feindt, Korobkin, Wu, Sollerman, Goobar, \& {Martinez-Pinedo}}]{rosswog_detectability_2017}
Rosswog, S., Feindt, U., Korobkin, O., {et~al.} 2017, Classical and Quantum Gravity, 34, 104001, \dodoi{10.1088/1361-6382/aa68a9}

\bibitem[{Rosswog {et~al.}(2018)Rosswog, Sollerman, Feindt, Goobar, Korobkin, Wollaeger, Fremling, \& Kasliwal}]{rosswog_the_2018}
Rosswog, S., Sollerman, J., Feindt, U., {et~al.} 2018, Astronomy \& Astrophysics, 615, A132, \dodoi{10.1051/0004-6361/201732117}

\bibitem[{Sagiv {et~al.}(2014)Sagiv, {Gal-Yam}, Ofek, Waxman, Aharonson, Kulkarni, Nakar, Maoz, Trakhtenbrot, Phinney, Topaz, Beichman, Murthy, \& Worden}]{sagiv_science_2014}
Sagiv, I., {Gal-Yam}, A., Ofek, E.~O., {et~al.} 2014, The Astronomical Journal, 147, 79, \dodoi{10.1088/0004-6256/147/4/79}

\bibitem[{Saleem {et~al.}(2020)Saleem, Resmi, Arun, \& Mohan}]{saleem_on_2020}
Saleem, M., Resmi, L., Arun, K.~G., \& Mohan, S. 2020, The Astrophysical Journal, 891, 130, \dodoi{10.3847/1538-4357/ab6731}

\bibitem[{Shvartzvald {et~al.}(2024)Shvartzvald, Waxman, {Gal-Yam}, Ofek, {Ben-Ami}, Berge, Kowalski, B{\"u}hler, Worm, Rhoads, Arcavi, Maoz, Polishook, Stone, Trakhtenbrot, Ackermann, Aharonson, Birnholtz, Chelouche, Guetta, Hallakoun, Horesh, Kushnir, Mazeh, Nordin, Ofir, Ohm, Parsons, Pe'er, Perets, Perdelwitz, Poznanski, Sadeh, Sagiv, Shahaf, Soumagnac, {Tal-Or}, Santen, Zackay, Guttman, Rekhi, Townsend, Weinstein, \& Wold}]{shvartzvald_ultrasat:_2024}
Shvartzvald, Y., Waxman, E., {Gal-Yam}, A., {et~al.} 2024, The Astrophysical Journal, 964, 74, \dodoi{10.3847/1538-4357/ad2704}

\bibitem[{Singer \& Price(2016)}]{singer_rapid_2016}
Singer, L.~P., \& Price, L.~R. 2016, Physical Review D, 93, 024013, \dodoi{10.1103/PhysRevD.93.024013}

\bibitem[{Singer {et~al.}(2016{\natexlab{a}})Singer, Chen, Holz, Farr, Price, Raymond, Cenko, Gehrels, Cannizzo, Kasliwal, Nissanke, Coughlin, Farr, Urban, Vitale, Veitch, Graff, Berry, Mohapatra, \& Mandel}]{singer_going_2016}
Singer, L.~P., Chen, H.-Y., Holz, D.~E., {et~al.} 2016{\natexlab{a}}, The Astrophysical Journal, 829, L15, \dodoi{10.3847/2041-8205/829/1/L15}

\bibitem[{Singer {et~al.}(2016{\natexlab{b}})Singer, Chen, Holz, Farr, Price, Raymond, Cenko, Gehrels, Cannizzo, Kasliwal, Nissanke, Coughlin, Farr, Urban, Vitale, Veitch, Graff, Berry, Mohapatra, \& Mandel}]{singer_supplement:_2016}
---. 2016{\natexlab{b}}, The Astrophysical Journal Supplement Series, 226, 10, \dodoi{10.3847/0067-0049/226/1/10}

\bibitem[{Singer {et~al.}(2025)Singer, Criswell, Leggio, Kiendrebeogo, Coughlin, Earnshaw, Gezari, Grefenstette, Harrison, Kasliwal, Morris, Tollerud, \& Cenko}]{singer_optimal_2025}
Singer, L.~P., Criswell, A.~W., Leggio, S.~C., {et~al.} 2025, Optimal {{Follow-Up}} of {{Gravitational-Wave Events}} with the {{UltraViolet EXplorer}} ({{UVEX}}),  arXiv, \dodoi{10.48550/arXiv.2502.17560}

\bibitem[{Smartt {et~al.}(2017)Smartt, Chen, Jerkstrand, Coughlin, Kankare, Sim, Fraser, Inserra, Maguire, Chambers, Huber, Kr{\"u}hler, Leloudas, Magee, Shingles, Smith, Young, Tonry, Kotak, {Gal-Yam}, Lyman, Homan, Agliozzo, Anderson, Angus, Ashall, Barbarino, Bauer, Berton, Botticella, Bulla, Bulger, Cannizzaro, Cano, Cartier, Cikota, Clark, De~Cia, Della~Valle, Denneau, Dennefeld, Dessart, Dimitriadis, {Elias-Rosa}, Firth, Flewelling, Fl{\"o}rs, Franckowiak, Frohmaier, Galbany, {Gonz{\'a}lez-Gait{\'a}n}, Greiner, Gromadzki, Guelbenzu, Guti{\'e}rrez, Hamanowicz, Hanlon, Harmanen, Heintz, Heinze, Hernandez, Hodgkin, Hook, Izzo, James, Jonker, Kerzendorf, Klose, {Kostrzewa-Rutkowska}, Kowalski, Kromer, Kuncarayakti, Lawrence, Lowe, Magnier, Manulis, {Martin-Carrillo}, Mattila, McBrien, M{\"u}ller, Nordin, O'Neill, Onori, Palmerio, Pastorello, Patat, Pignata, Podsiadlowski, Pumo, Prentice, Rau, Razza, Rest, Reynolds, Roy, Ruiter, Rybicki, Salmon, Schady, Schultz, Schweyer, Seitenzahl, Smith, Sollerman,
  Stalder, Stubbs, Sullivan, Szegedi, Taddia, Taubenberger, Terreran, {van Soelen}, Vos, Wainscoat, Walton, Waters, Weiland, Willman, Wiseman, Wright, Wyrzykowski, \& Yaron}]{smartt_a_2017}
Smartt, S.~J., Chen, T.-W., Jerkstrand, A., {et~al.} 2017, Nature, 551, 75, \dodoi{10.1038/nature24303}

\bibitem[{{Soares-Santos} {et~al.}(2017){Soares-Santos}, Holz, Annis, Chornock, Herner, Berger, Brout, Chen, Kessler, Sako, Allam, Tucker, Butler, Palmese, Doctor, Diehl, Frieman, Yanny, Lin, Scolnic, Cowperthwaite, Neilsen, Marriner, Kuropatkin, Hartley, {Paz-Chinch{\'o}n}, Alexander, Balbinot, Blanchard, Brown, Carlin, Conselice, Cook, {Drlica-Wagner}, Drout, Durret, Eftekhari, Farr, Finley, Foley, Fong, Fryer, {Garc{\'i}a-Bellido}, Gill, Gruendl, Hanna, Kasen, Li, Lopes, Louren{\c c}o, Margutti, Marshall, Matheson, Medina, Metzger, Mu{\~n}oz, Muir, Nicholl, Quataert, Rest, Sauseda, Schlegel, Secco, Sobreira, Stebbins, Villar, Vivas, Walker, Wester, Williams, Zenteno, Zhang, Abbott, Abdalla, Banerji, Bechtol, {Benoit-L{\'e}vy}, Bertin, Brooks, {Buckley-Geer}, Burke, Rosell, Kind, Carretero, Castander, Crocce, Cunha, D'Andrea, da~Costa, Davis, Desai, Dietrich, Doel, Eifler, Fernandez, Flaugher, Fosalba, Gaztanaga, Gerdes, Giannantonio, Goldstein, Gruen, Gschwend, Gutierrez, Honscheid, Jain, James, Jeltema,
  Johnson, Johnson, Kent, Krause, Kron, Kuehn, Kuhlmann, Lahav, Lima, Maia, March, McMahon, Menanteau, Miquel, Mohr, Nichol, Nord, Ogando, Petravick, Plazas, Romer, Roodman, Rykoff, Sanchez, Scarpine, Schubnell, {Sevilla-Noarbe}, Smith, Smith, Suchyta, Swanson, Tarle, Thomas, Thomas, Troxel, Vikram, Wechsler, \& {and}}]{soares-santos_the_2017}
{Soares-Santos}, M., Holz, D.~E., Annis, J., {et~al.} 2017, The Astrophysical Journal, 848, L16, \dodoi{10.3847/2041-8213/aa9059}

\bibitem[{Song {et~al.}(2019)Song, Ai, Wang, Xing, Gao, \& Zhang}]{song_viewing_2019}
Song, H.-R., Ai, S.-K., Wang, M.-H., {et~al.} 2019, The Astrophysical Journal Letters, 881, L40, \dodoi{10.3847/2041-8213/ab3921}

\bibitem[{{The Astropy Collaboration} {et~al.}(2018){The Astropy Collaboration}, {Price-Whelan}, Sip{\H o}cz, G{\"u}nther, Lim, Crawford, Conseil, Shupe, Craig, Dencheva, Ginsburg, VanderPlas, Bradley, {P{\'e}rez-Su{\'a}rez}, de~{Val-Borro}, Contributors), Aldcroft, Cruz, Robitaille, Tollerud, Committee), Ardelean, Babej, Bach, Bachetti, Bakanov, Bamford, Barentsen, Barmby, Baumbach, Berry, Biscani, Boquien, Bostroem, Bouma, Brammer, Bray, Breytenbach, Buddelmeijer, Burke, Calderone, Rodr{\'i}guez, Cara, Cardoso, Cheedella, Copin, Corrales, Crichton, D'Avella, Deil, Depagne, Dietrich, Donath, Droettboom, Earl, Erben, Fabbro, Ferreira, Finethy, Fox, Garrison, Gibbons, Goldstein, Gommers, Greco, Greenfield, Groener, Grollier, Hagen, Hirst, Homeier, Horton, Hosseinzadeh, Hu, Hunkeler, Ivezi{\'c}, Jain, Jenness, Kanarek, Kendrew, Kern, Kerzendorf, Khvalko, King, Kirkby, Kulkarni, Kumar, Lee, Lenz, Littlefair, Ma, Macleod, Mastropietro, McCully, Montagnac, Morris, Mueller, Mumford, Muna, Murphy, Nelson, Nguyen,
  Ninan, N{\"o}the, Ogaz, Oh, Parejko, Parley, Pascual, Patil, Patil, Plunkett, Prochaska, Rastogi, Janga, Sabater, Sakurikar, Seifert, Sherbert, {Sherwood-Taylor}, Shih, Sick, Silbiger, Singanamalla, Singer, Sladen, Sooley, Sornarajah, Streicher, Teuben, Thomas, Tremblay, Turner, Terr{\'o}n, van Kerkwijk, de~la Vega, Watkins, Weaver, Whitmore, Woillez, Zabalza, \& Contributors)}]{theastropycollaboration_the_2018}
{The Astropy Collaboration}, {Price-Whelan}, A.~M., Sip{\H o}cz, B.~M., {et~al.} 2018, The Astronomical Journal, 156, 123, \dodoi{10.3847/1538-3881/aabc4f}

\bibitem[{Tohuvavohu {et~al.}(2024)Tohuvavohu, Kennea, Roberts, DeLaunay, Ronchini, Cenko, Ewing, Magee, Messick, Sachdev, \& Singer}]{tohuvavohu_swiftly_2024}
Tohuvavohu, A., Kennea, J.~A., Roberts, C.~J., {et~al.} 2024, Swiftly Chasing Gravitational Waves across the Sky in Real-Time,  arXiv, \dodoi{10.48550/arXiv.2410.05720}

\bibitem[{Van~Rossum \& Drake(2009)}]{vanrossum_python_2009}
Van~Rossum, G., \& Drake, F.~L. 2009, Python 3 Reference Manual (Scotts Valley, CA: CreateSpace)

\bibitem[{Virtanen {et~al.}(2020)Virtanen, Gommers, Oliphant, Haberland, Reddy, Cournapeau, Burovski, Peterson, Weckesser, Bright, {van der Walt}, Brett, Wilson, Millman, Mayorov, Nelson, Jones, Kern, Larson, Carey, Polat, Feng, Moore, VanderPlas, Laxalde, Perktold, Cimrman, Henriksen, Quintero, Harris, Archibald, Ribeiro, Pedregosa, \& {van Mulbregt}}]{virtanen_scipy_2020a}
Virtanen, P., Gommers, R., Oliphant, T.~E., {et~al.} 2020, Nature Methods, 17, 261, \dodoi{10.1038/s41592-019-0686-2}

\bibitem[{Wang \& Giannios(2021)}]{wang_multimessenger_2021}
Wang, H., \& Giannios, D. 2021, The Astrophysical Journal, 908, 200, \dodoi{10.3847/1538-4357/abd39c}

\bibitem[{Watson {et~al.}(2019)Watson, Hansen, Selsing, Koch, Malesani, Andersen, Fynbo, Arcones, Bauswein, Covino, Grado, Heintz, Hunt, Kouveliotou, Leloudas, Levan, Mazzali, \& Pian}]{watson_identification_2019}
Watson, D., Hansen, C.~J., Selsing, J., {et~al.} 2019, Nature, 574, 497, \dodoi{10.1038/s41586-019-1676-3}

\bibitem[{Yu {et~al.}(2018)Yu, Liu, \& Dai}]{yu_a_2018}
Yu, Y.-W., Liu, L.-D., \& Dai, Z.-G. 2018, The Astrophysical Journal, 861, 114, \dodoi{10.3847/1538-4357/aac6e5}

\bibitem[{Zonca {et~al.}(2019)Zonca, Singer, Lenz, Reinecke, Rosset, Hivon, \& Gorski}]{zonca_healpy:_2019}
Zonca, A., Singer, L., Lenz, D., {et~al.} 2019, Journal of Open Source Software, 4, 1298, \dodoi{10.21105/joss.01298}

\end{thebibliography}
\bibliographystyle{aasjournal}



\end{document}